\newcommand{\ev}{\hbox{ eV}}

\newcommand{\mev}{\hbox{ MeV}}
\newcommand{\gev}{\hbox{ GeV}}
\newcommand{\tev}{\hbox{ TeV}}
\newcommand{\cm}{\hbox{ cm}}
\newcommand{\sr}{\hbox{ sr}}
\newcommand{\s}{\hbox{ s}}
\newcommand{\km}{\hbox{ km}}
\newcommand{\kms}{\hbox{ km}/\hbox{s}}
\newcommand{\mpc}{\hbox{ Mpc}}
\newcommand{\K}{\hbox{ K}}
\newcommand{\Y}{\ensuremath{\mathcal{Y}}}
\newcommand{\cfrac}[2]{\textstyle{\frac{#1}{#2}}}
\def\bentarrow{\:\raisebox{1.2ex}{\rlap{$\vert$}}\!\rightarrow}

\def\dk#1#2#3{
	\begin{displaymath}
	\begin{array}{r c l}
	#1 & \rightarrow & #2 \\
	 & & \bentarrow #3
	\end{array}
	\end{displaymath}
		}

\documentclass[aps,prd,twocolumn,groupedaddress,showpacs,nofootinbib,
nobibnotes,twoside]{revtex4}
\usepackage{graphicx} 
\usepackage{hyperref}
\usepackage{latexsym}
\preprint{FERMILAB--PUB--04/nnn--T}

\catcode`\@=11
\def\ps@fnal{\def\@oddhead{\textsf{FERMILAB--Pub--04/379--T \hfil \thepage}}
\def\@evenhead{\thepage \hfil \textsf{FERMILAB--Pub--04/379--T}}}
\relax

\begin{document}

\pagestyle{fnal} %%% FOR PREPRINT

%\preprint{}

\title{Diagnostic Potential of Cosmic-Neutrino Absorption Spectroscopy}


\author{Gabriela Barenboim}
\email[]{Gabriela.Barenboim@uv.es}
%\thanks{}
\affiliation{Departament de F\'{\i}sica Te\`{o}rica,
 Universitat de Val\`{e}ncia, 
Carrer Dr.~Moliner 50, E-46100 Burjassot (Val\`{e}ncia),  ÊÊSpainÊ}

\author{Olga Mena Requejo}
\email[]{omena@fnal.gov}
\affiliation{Theoretical Physics Department, Fermi National 
Accelerator Laboratory,\\ P.O.\ Box 500, Batavia, Illinois 60510 USA}

\author{Chris Quigg}
\email[]{quigg@fnal.gov}
\affiliation{Theoretical Physics Department, Fermi National 
Accelerator Laboratory,\\ P.O.\ Box 500, Batavia, Illinois 60510 USA}


\date{\today}

\begin{abstract}
Annihilation of extremely energetic cosmic neutrinos on the relic-neutrino
background can give rise to absorption lines at energies 
corresponding to  formation of the electroweak gauge boson 
$Z^{0}$. The positions of the absorption dips are set by the masses of 
the relic neutrinos. Suitably intense sources of extremely energetic 
($10^{21}$ -- $10^{25}$-eV) cosmic neutrinos might therefore enable the 
determination of the absolute neutrino masses and the flavor composition of 
the mass eigenstates. Several factors---other than neutrino mass and 
composition---distort the absorption lines, however. We analyze the 
influence of the time-evolution of the relic-neutrino density and the 
consequences of neutrino decay. We consider the sensitivity of the 
lineshape to the age and character of extremely energetic neutrino 
sources, and to the thermal history of the Universe, reflected in the 
expansion rate. We take into account Fermi motion arising from the 
thermal distribution of the relic-neutrino gas. We also note the 
implications of Dirac \textit{vs.} Majorana relics, and briefly 
consider unconventional neutrino histories. We ask what kinds of 
external information would enhance the potential of cosmic-neutrino absorption 
spectroscopy, and estimate the sensitivity required to make the 
technique a reality.
\end{abstract}

\pacs{96.40.Tv, 14.60.Pq, 13.35.Hb,  95.35.+d, 95.85.Ry,  98.70.Sa \hfill 
\fbox{FERMILAB--PUB--04--379--T}} 
%\keywords{}

\maketitle

\section{Introduction \label{intro}}
\subsection{A Brief History of Relic Neutrinos}
According to the standard cosmology, neutrinos should be the most
abundant particles in the Universe, after the photons of the cosmic
microwave background, provided that they are stable over cosmological
times.\footnote{For compact summaries of the canonical thermal history
of the Universe, see Weinberg~\cite{GravCos}, \S15.6, the review
article by Steigman~\cite{Steigman:1979kw},  \S19--23 of the \textit{Review of 
Particle Physics}~\cite{Eidelman:2004wy}, or a standard 
textbook~\cite{KT,EH,AL,SD}.} Because they interact only
weakly, neutrinos decoupled when the age of the Universe was $\approx
0.1\hbox{ s}$ and the temperature of the Universe was $\hbox{a
few}\mev$.  Accordingly, relic neutrinos have been present---as witnesses
or participants---for landmark events in the history of the Universe: the
era of big-bang nucleosynthesis, a few minutes into the life of the
Universe; the decoupling era around $379\,000\hbox{
y}$~\cite{Bennett:2003bz}, when the cosmic microwave background was
imprinted on the surface of last scattering; and the era of large-scale
structure formation, when the Universe was only a few percent  of its
current age.\footnote{For a recent quantitative assessment
of evidence that neutrinos were 
present at these times, see Ref.~\cite{Barger:2003zg}.}

Some of the earliest cosmological bounds on neutrino masses 
followed from the requirement that massive relic neutrinos, present 
today in the expected numbers, do not saturate the critical density of the 
Universe~\cite{Gershtein:1966gg,Cowsik:1972gh}. Refined analyses, 
incorporating constraints from a suite of cosmological measurements, 
sharpen the bounds on the sum of light-neutrino masses~\cite{Fogli:2004as}. 
The discovery of neutrino 
oscillations~\cite{Fukuda:1998mi,Ahmad:2002jz,Eguchi:2002dm} implies 
that neutrinos have mass, but we cannot reliably compute the contribution 
of relic neutrinos to the dark matter of the Universe until we establish 
the absolute scale of neutrino masses. Current estimates for the 
neutrino fraction of the Universe's mass--energy density lie in the 
range $0.1\% \lesssim \Omega_{\nu} \lesssim 1.5\%$.

The neutrino gas that we believe permeates the present Universe has never been 
detected directly. Imaginative schemes have been proposed to 
record the elastic scattering of the 1.95-K relic neutrinos, but all appear to require 
significant further technological development before they can 
approach the needed sensitivity~\cite{Duda:2001hd,Ringwald:2004np}. 
In this paper, we elaborate a complementary approach: detecting relic 
neutrinos by observing the resonant annihilation of extremely-high-energy 
cosmic neutrinos on the background neutrinos through the reaction 
$\nu\bar{\nu} \to 
Z^{0}$~\cite{Weiler:1982qy,Weiler:1983xx,Roulet:1992pz,Gondolo:1991rn,Yoshida:1996ie,Eberle:2004ua}.
By observing $Z$-bursts or absorption lines, one may hope to
determine the absolute neutrino masses and the flavor composition of 
the neutrino mass eigenstates.\footnote{See Ref.~\cite{Han:2004kq} for 
a general review of other aspects of particle physics at neutrino 
observatories.}

As a \textit{Gedankenexperiment,} the prospect of cosmic-neutrino absorption 
spectroscopy has great clarity and appeal. Reality is more 
complicated, and it is our purpose---building on earlier work---to analyze all the important 
effects that will influence the execution and interpretation of  
neutrino-absorption experiments. We are encouraged in this effort by 
the imminent construction and operation of neutrino observatories and 
by imaginative efforts to develop new techniques to detect 
super-high-energy neutrinos. A novel aspect of the analysis 
presented here is our attention to the thermal motion of the relics. 
We also raise the possibility that cosmic-neutrino absorption spectroscopy 
might open a new vista on the thermal history of the universe, as 
well as extending or validating our understanding of neutrino 
properties.

In the body of this introductory section, we develop the pieces that 
enter the analysis of neutrino absorption spectra: our expectations 
for the relic neutrino background now and in the past, details of the 
annihilation cross section, and possible sources of extremely 
energetic cosmic neutrinos. We also survey experiments that aim to 
detect ultrahigh-energy neutrinos. In \S\ref{sec:toy}, we describe the 
idealized situation of a super-high-energy neutrino beam incident on 
a (very long) uniform column of relic neutrinos at today's 
density, but with negligible temperature. We describe the 
information that could be extracted from absorption 
dips, assuming perfect energy resolution and flavor tagging. 

The extremely long interaction length for neutrinos traversing the relic 
background means that we must integrate over cosmic time, or redshift, 
and this takes up \S\ref{sec:reds}. There we discuss the mechanisms 
that distort absorption lines and how the distortions compromise the 
dream of determining the absolute neutrino masses. We also remark on 
the sensitivity of the line-shape to the thermal history of the 
Universe.

We include Fermi motion due to the relic-neutrino temperature---which 
evolves with redshift---in \S\ref{sec:temp}. The mean relic-neutrino 
momentum at the present epoch acts as a rough lower bound on the 
effective target mass. Section~\ref{sec:loose} 
is devoted to the implications of unconventional neutrino histories, 
including neutrino decay and the consequences of a lepton asymmetry in 
the early Universe. We summarize what we have learned, and assess the 
prospects for experimental realization of these ideas in \S\ref{sec:summa}. 
Looking forward to the experiments, we consider how external 
information could enhance the potential of cosmic-neutrino absorption 
spectroscopy, and we estimate the sensitivity required to make the 
technique a reality.

\subsection{Character of the Relic Neutrino Background 
\label{subsec:char}}
The cosmic microwave background is characterized by a Bose--Einstein blackbody 
distribution of photons (per unit volume)\footnote{We adopt units such 
that $\hbar = 1 = c$, and we will measure temperature in kelvins or 
electron volts, as appropriate to the situation. The conversion factor 
is Boltzmann's constant, $k = 8.617343 \times 10^{-5}\ev\K^{-1}$.}
\begin{equation}
    \frac{dn_{\gamma}(T)}{d^{3}p} = \frac{1}{(2\pi)^{3}} \,
    \frac{1}{\exp{(p/T)} - 1}\; ,
    \label{eq:bbphotons}
\end{equation}
where $p$ is the relic momentum and $T$ is the 
temperature of the photon ensemble. The number density of photons throughout 
the Universe is
\begin{equation}
    n_{\gamma}(T) = \frac{1}{(2\pi)^{3}} \!\!\int \!\!d^{3}p \;
    \frac{1}{\exp{(p/T)} - 1} = \frac{2\zeta(3)}{\pi^{2}}\, 
    T^{3},
    \label{eq:bbgamnum}
\end{equation}
where $\zeta(3) \approx 1.20205$ is Riemann's zeta function. In the 
present Universe, with a photon temperature $T_{0} = 
(2.725 \pm 0.002)\K$~\cite{Bennett:2003bz}, the photon density is
\begin{equation}
    n_{\gamma0} \equiv n_{\gamma}(T_{0}) \approx 410\cm^{-3}\;.
    \label{eq:numgam}
\end{equation}

The present photon density provides a reference for other big-bang 
relics. The essential observation is that neutrinos decoupled when 
the cosmic soup cooled to around $1\mev$, so did not share in the 
energy released when electrons and positrons annihilated at $T \approx 
m_{e}$, the electron mass. Applying entropy conservation and counting interacting degrees 
of freedom, it follows that the ratio of neutrino and photon 
temperatures (below $m_{e}$) is
\begin{equation}
    T_{\nu}/T = \left(\cfrac{4}{11}\right)^{\!1/3}\;,
    \label{eq:nutogam}
\end{equation}
so that the present neutrino temperature is 
\begin{equation}
    T_{\nu0} = \left(\cfrac{4}{11}\right)^{\!1/3}T_{0} = 1.945\K 
    \leadsto 1.697 \times 10^{-4}\ev\;.
    \label{eq:nutemp}
\end{equation}

The momentum distribution of relic neutrinos follows the Fermi--Dirac 
distribution (with zero chemical potential),
\begin{equation}
    \frac{dn_{\nu_{i}}(T_{\nu})}{d^{3}p} =
    \frac{dn_{\nu^{c}_{i}}(T_{\nu})}{d^{3}p} = \frac{1}{(2\pi)^{3}} \,
    \frac{1}{\exp{(p/T_{\nu})} + 1}\; .
    \label{eq:nuFD}
\end{equation}
The number distribution of relic neutrinos is therefore
\begin{eqnarray}
    n_{\nu_{i}}(T_{\nu}) = & n_{\nu^{c}_{i}}(T_{\nu}) = & \frac{1}{(2\pi)^{3}} \!\!\int \!\!d^{3}p \;
    \frac{1}{\exp{(p/T_{\nu})} + 1} \nonumber\\ & & = \frac{3\zeta(3)}{4\pi^{2}}  \, 
    T_{\nu}^{3},
    \label{eq:numbnu} \\
    & & = \cfrac{3}{22} n_{\gamma}(T)\;. \nonumber
\end{eqnarray}
In the present Universe, the number density of each (active) neutrino 
species is~\footnote{The unconventional neutrino histories described 
in \S\ref{sec:loose} can alter this expectation.}
\begin{equation}
    n_{\nu_{i}0} = n_{\nu^{c}_{i}0} \equiv n_{\nu_{i}}(T_{\nu0}) 
    \approx 56\cm^{-3}\;, 
    \label{eq:numbnunow}
\end{equation}
and the mean momentum of relic neutrinos today is
\begin{equation}
    \langle p_{\nu0} \rangle = \frac{7}{2} \, 
    \frac{\zeta(4)}{\zeta(3)} \cdot T_{\nu0} \approx 3.151 T_{\nu0}
    \approx 5.314 \times 10^{-4}\ev\;,
    \label{eq:numom}
\end{equation}
where we have used $\zeta(4) = \pi^{4}/90 = 1.08232$. In the same way, 
the mean-squared neutrino momentum is given by
\begin{equation}
\label{eqn:p2mean}
\langle p^{2}_{\nu 0} \rangle = 15\,\frac{\zeta(5)}{\zeta(3)} \cdot 
T^{2}_{\nu 0} \approx
12.94\, T^{2}_{\nu 0}\;,
\end{equation}
so that
\begin{equation}
    \langle p_{\nu0}^{2} \rangle^{\cfrac{1}{2}} \approx 
    3.597\,T_{\nu0} \approx 6.044 \times 10^{-3}\ev\;.
    \label{eqn:prms}
\end{equation}

\begin{figure}
\includegraphics[width=8cm]{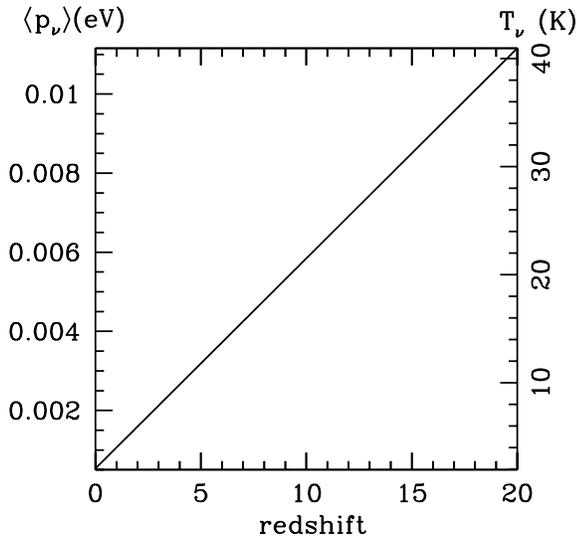}%{figs/thermalbis.eps}%
\caption{Neutrino temperature $T_{\nu}$ (right-hand scale) and mean 
momentum $p_{\nu}$ of relic neutrinos (left-hand scale) as a function of the 
redshift $z$.}
\label{fig:thermal}
\end{figure}

\begin{figure}
\includegraphics[width=8cm]{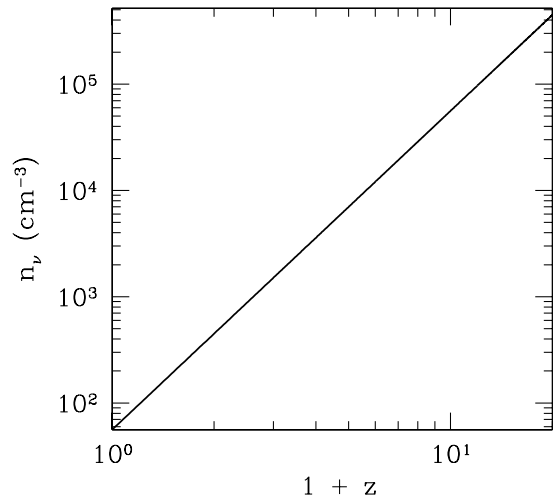}%{figs/nunumtemp.eps}
\caption{Relic neutrino number density versus redshift.}
\label{fig:nnu}
\end{figure}

Neutrinos decoupled very early in the history of the Universe, at
redshift $z = \mathcal{O}(10^{10})$.  The temperature of a massless
decoupled species scales as $T\sim (1+z)$, as shown in Figure~\ref{fig:thermal} for  redshifts 
since the formation of the first stars~\cite{Bennett:2003bz}.  
By Eq.~(\ref{eq:numbnu}), this means that
the relic neutrino number densities will be redshifted as
$n_{\nu_{i}}(z) = n_{\nu0}\ (1+z)^3$.
The dependence of the relic neutrino number on the redshift is
depicted in Figure~\ref{fig:nnu}.  

\begin{figure}
\includegraphics[width=8cm]{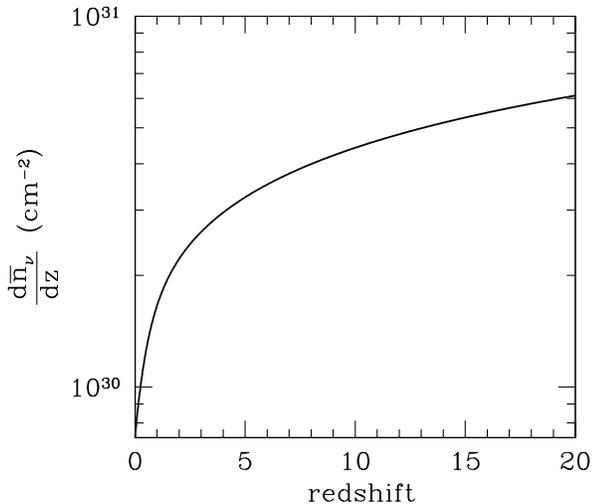}%{figs/eff.eps}
\caption{Column density (\ref{eqn:fermi3}) versus redshift for the $\Lambda$CDM model, with 
$\Omega_{m} = 0.27 \pm 0.04$, $\Omega_{\Lambda} = 0.73 \pm 0.04$, and 
$h = 0.71^{+0.04}_{-0.03}$~\cite{Eidelman:2004wy}.}
\label{fig:nnueff}
\end{figure}

The effective relic neutrino density that
an extremely-high-energy neutrino would encounter while traversing 
the expanding Universe is the neutrino density per unit 
redshift. The  propagation distance $r$ is related to the 
redshift $z$ through 
\begin{equation}
    dr = dz/(1+z)H(z)\;,
    \label{eq:linescale}
\end{equation}
where $H(z)$ is the Hubble parameter. In a flat Universe with 
negligible radiation component, we may write
\begin{equation}
    H^{2}(z) = H_{0}^{2}\left[\Omega_{m}(1+z)^{3} + 
    \Omega_{\Lambda}\right]\;,
    \label{eq:Hubble}
\end{equation}
where $\Omega_{m}$ is the matter density, $\Omega_{\Lambda}$ is the 
cosmological constant, and $H_{0} = h \cdot 100\kms\mpc^{-1}$ is the 
present value of the Hubble constant. The neutrino density per unit 
redshift, sometimes called  the \textit{column density,} is then
\begin{eqnarray}
\label{eqn:fermi3}
\bar{n}_{\nu_{i}}(z) &=&n_{\nu0}\ (1+z)^3 \ dr\\ \nonumber 
&=&\ \frac{ n_{\nu0}\ (1+z)^3 dz}{(1+z)\  H(z)} = \frac{n_{\nu0}\ 
(1+z)^2 dz}{H(z)}\;.
\end{eqnarray}
In Figure~\ref{fig:nnueff} 
we show this
column density as a function of the redshift for the $\Lambda$CDM
model,\footnote{For three complementary views of today's concordance 
cosmology, see 
Refs.~\cite{Peebles:2002gy,Freedman:2003ys,Trodden:2004st}.}
that is, a flat universe with a cosmological constant and cold dark
matter.  

The appearance of the Hubble parameter in Eq.~(\ref{eqn:fermi3}) 
raises the possibility that careful observation of  neutrino absorption lines 
could reveal something of the thermal history of the Universe. We 
shall see in \S\ref{sec:reds} examples of how the lineshapes differ 
in different cosmologies.

\subsection{The Neutrino Mass Spectrum \label{subsec:numass}}
Through the past few years, experiments have adduced robust 
evidence for  flavor change involving solar, atmospheric, and
reactor neutrinos, as well as neutrinos produced by  accelerator 
beams~\cite{ichep2004}. Putting aside exotic interpretations, these 
results establish that neutrinos have nonvanishing masses and that 
neutrino species mix. The most economical description of the new 
phenomena is given in terms of the $3\times3$ neutrino mixing 
matrix that relates flavor eigentstates to mass 
eigenstates,\footnote{We consider here the canonical picture of three 
neutrino families. The ongoing MiniBooNE
experiment~\cite{miniboone} is expected to explore all the 
parameter space of the $\nu_{\mu} \leftrightarrow \nu_{e}$ mutation 
claimed by the LSND experiment~\cite{Athanassopoulos:1996jb}, with its 
implication of additional (presumably sterile) neutrinos. 
The neutrino mixing matrix is sometimes
called the Pontecorvo--Maki--Nakagawa--Sakata (PMNS) matrix.} the 
analogue of the (Cabibbo--Kobayashi--Maskawa) quark mixing matrix. 
The standard neutrino-mixing phenomenology 
entails six parameters: three real mixing angles ($\theta_{12}$, $ \theta_{23}$,
$\theta_{13}$),  one Dirac \textsf{CP} phase ($\delta$), 
and two independent
mass-squared differences\footnote{We define $\Delta m_{ij}^{2} \equiv m_j^2
-m_i^2$, where $m_{i}$ is the mass of $\nu_{i}$.} ($\Delta m_{12}^2$ and $\Delta
m_{23}^2$). 

\begin{figure*}
    \includegraphics[width=12cm]{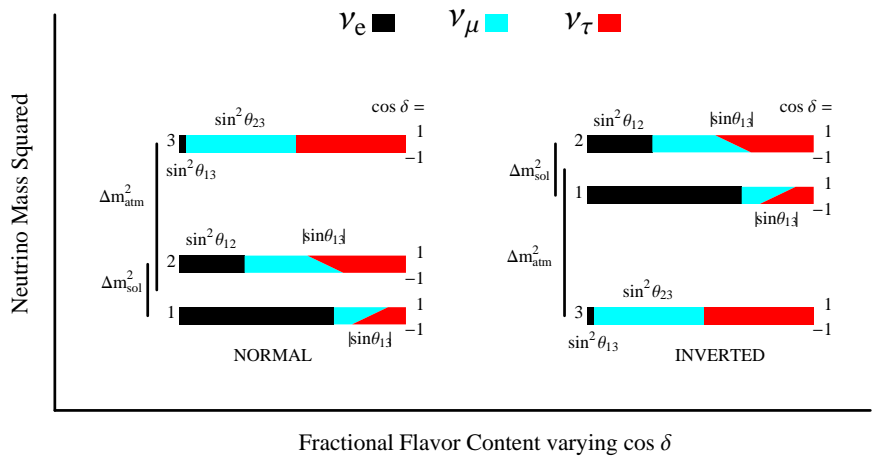}
 \caption{Composition of the neutrino mass eigenstates 
($\nu_{1},\nu_{2},\nu_{3}$) 
in terms of the flavor eigenstates ($\nu_{e},\nu_{\mu},\nu_{\tau}$) for 
the normal (left panel) and inverted (right panel) hierarchies [from 
Ref.~\cite{Mena:2003ug}]. Each bar shows the flavor mixture of a mass 
eigenstate
as the \textsf{CP}-violating phase, $\delta$, varies from $\cos \delta = 
-1$ (bottom) to    $\cos \delta = 1$ (top).
    The other mixing parameters are held fixed at the representative 
    values
    $\sin^2 \theta_{12} = 0.30$, $\sin^2 \theta_{13} = 0.03$ and
    $\sin^2 \theta_{23} = 0.50$. To an excellent approximation, 
    $\Delta m_{12}^{2} = \Delta m_{\mathrm{sol}}^{2}$ and 
    $\Delta m_{23}^{2} = \Delta m_{\mathrm{atm}}^{2}$.}
\label{fig:hierarchy2}
\end{figure*}

To connect the solar, atmospheric, reactor, and accelerator 
observations with neutrino parameters, we follow convention~\cite{boris} in 
identifying the mass splittings and mixing angles that drive the solar 
and atmospheric transitions as $(\Delta m_{12}^2,\theta_{12})$ and 
$(|\Delta m_{23}^2|,\theta_{23})$. The sign of the mass splitting 
between the atmospheric state $\nu_3$ and the solar doublet 
($\nu_1,\nu_2$) is not yet known.  Both the normal hierarchy ($m_{3} > 
m_{2} > m_{1}$) and the inverted hierarchy ($m_{3} < m_{1} 
\lesssim m_{2}$) are illustrated in Figure~\ref{fig:hierarchy2}, where 
the colored bars represent the flavor content of the mass eigenstates.

The best
fit point for the combined analysis of solar neutrino data together
with KamLAND reactor data~\cite{Araki:2004mb} is at $\Delta 
m_{12}^2=8.2^{+0.6}_{-0.5}
\times 10^{-5}\ev^2$ and $\tan^{2}\theta_{12}=0.40^{+0.09}_{-0.07}$.  In the
atmospheric neutrino sector, the most recent analysis of K2K
accelerator data and atmospheric neutrino data constrains 
$1.9 \times 10^{-3}\ev^{2} < |\Delta m_{23}^2|<3.0 \times 
10^{-3}\ev^2$, $\sin^{2}2\theta_{23}> 0.90$~\cite{paris1,paris2}. 

Only an upper bound exists on the mixing angle $\theta_{13}$ (which connects the solar and
atmospheric neutrino realms), and   the \textsf{CP}-violating phase $\delta$ is
unobservable in current neutrino oscillation experiments, so we allow in Figure~\ref{fig:hierarchy2}
for the variation $-1 < \cos\delta < +1$. The CHOOZ 
reactor experiment bounds $\sin^{2} 2 \theta_{13} < 0.1$
(at $90\%$ CL) for a value of the atmospheric mass gap close to the 
current central value~\cite{chooz}.

Several oscillation experiments that exploit neutrino beams from
nuclear reactors and accelerators are taking data, and similar
experiments will take data over the next few years.  In particular,
future reactor neutrino experiments could set the value of
$\theta_{13}$, as explored in detail in Ref.~\cite{Anderson:2004pk}.  For a
recent study of the measurement of leptonic \textsf{CP} violation and the
pattern of the neutrino mass spectrum at the future T2K 
(Tokai-to-Kamioka)~\cite{Itow:2001ee} and
NO$\nu$A~\cite{nova} long-baseline experiments, see Ref.~\cite{Mena:2004sa}.  If
the value of $\theta_{13}$ turns out to be very small, the ultimate
high-precision measurements may require 
super-beams~\cite{Gomez-Cadenas:2001eu}, neutrino
factories~\cite{Cervera:2000kp,Albright:2000xi}, beta 
beams~\cite{Burguet-Castell:2003vv}, or a 
combination~\cite{Burguet-Castell:2002qx,Donini:2004hu}.

 Despite the great promise of planned and future
 neutrino oscillation experiments,  two essential neutrino
 properties would still remain unknown. Is the neutrino  
 distinct from its antiparticle ($\nu \ne \bar{\nu}$, Dirac case), or 
 identical to it ($\nu \equiv \bar{\nu}$, Majorana 
 case)?\footnote{Prevailing theoretical opinion favors the Majorana 
 character, if only because we know no principle that forbids 
 Majorana mass terms. The see-saw mechanism~\cite{Yanagida:1979as,Gell-Mann:1980vs,Mohapatra:1979ia} offers a 
 natural interpretation of the smallness of neutrino masses, and 
 points to a new scale associated with the heavy Majorana neutrino.
 The see-saw mechanism  implies lepton-number--violating processes that 
 could have happened in the early Universe.  The decay of the heavy, 
 weak-isoscalar, Majorana neutrinos,
 together with  $B+L$-violating sphaleron transitions can give rise to
 the baryon asymmetry of the Universe~\cite{Fukugita:1986hr}.} What are the 
 absolute  values of the neutrino masses? 

 Searches for neutrinoless double beta decay, a rare---and hitherto 
 unobserved---transition between two nuclei with the same mass number 
 ($A$) that changes the nuclear charge ($Z$) by two units, %$\beta\beta(0\nu)$,
\begin{equation}
    (Z,A) \to (Z+2,A) + e^{-}_{1} + e^{-}_{2}\;,
    \label{eq:bb0nu}
\end{equation}
 are for now our only probe for Majorana neutrinos.  Observational upper 
 limits on $\beta\beta(0\nu)$ rates provide an
 upper bound on the so-called \emph{``effective Majorana mass''} of the
 electron neutrino~\cite{Eidelman:2004wy},
 \begin{equation} 
     \langle m_\mathrm{eff} \rangle <
 0.3-1.0 \ev\;.
 \end{equation}
Forthcoming $\beta\beta(0\nu)$ experiments that aim for sensitivity 
 approaching $\langle m_\mathrm{eff} \rangle \lesssim 0.05\ev$
 could well establish neutrinos as Majorana particles~\cite{Elliott:2002xe,Bahcall:2004ip}.

 Direct information on the absolute scale of neutrino masses can be
extracted from kinematical studies.  The present upper bound on the 
electron-neutrino mass from tritium beta-decay experiments is $2.2\ev$ ($95 \%$
CL)~\cite{Farzan:2001cj,Farzan:2002zq}, and in the future the KATRIN experiment is expected to
be sensitive to electron-neutrino masses approaching $0.2\ev$ ($90\%$
CL)~\cite{katrin}.

The evolution of the Universe is sensitive to the absolute neutrino
mass scale, independent of mixing parameters or \textsf{CP}-violating
phases~\cite{Dolgov:2002wy,Hannestad:2004nb}, for example, through the
influence of neutrinos on the matter power spectrum~\cite{Crotty:2004gm,Lesgourgues:2004ps}.  
Combining WMAP observations of the cosmic microwave background with 
large-scale structure data from  the 2dF Galaxy Redshift 
Survey~\cite{aussies} or the Sloan Digital Sky Survey~\cite{sloan} 
yields  impressive constraints on neutrino mass $\sum_i m_{\nu_i} \lesssim 
1\ev$, the precise value depending on the priors and the data 
set~\cite{Fogli:2004as,Seljak:2004xh}.  However, the neutrino mass 
limits arising from existing large-scale structure can be evaded if new 
interactions (such as the coupling of neutrinos to a light boson) 
enable the relic neutrinos to annihilate at late 
times~\cite{Beacom:2004yd}. 

\begin{figure*}
  \includegraphics[width=7cm]{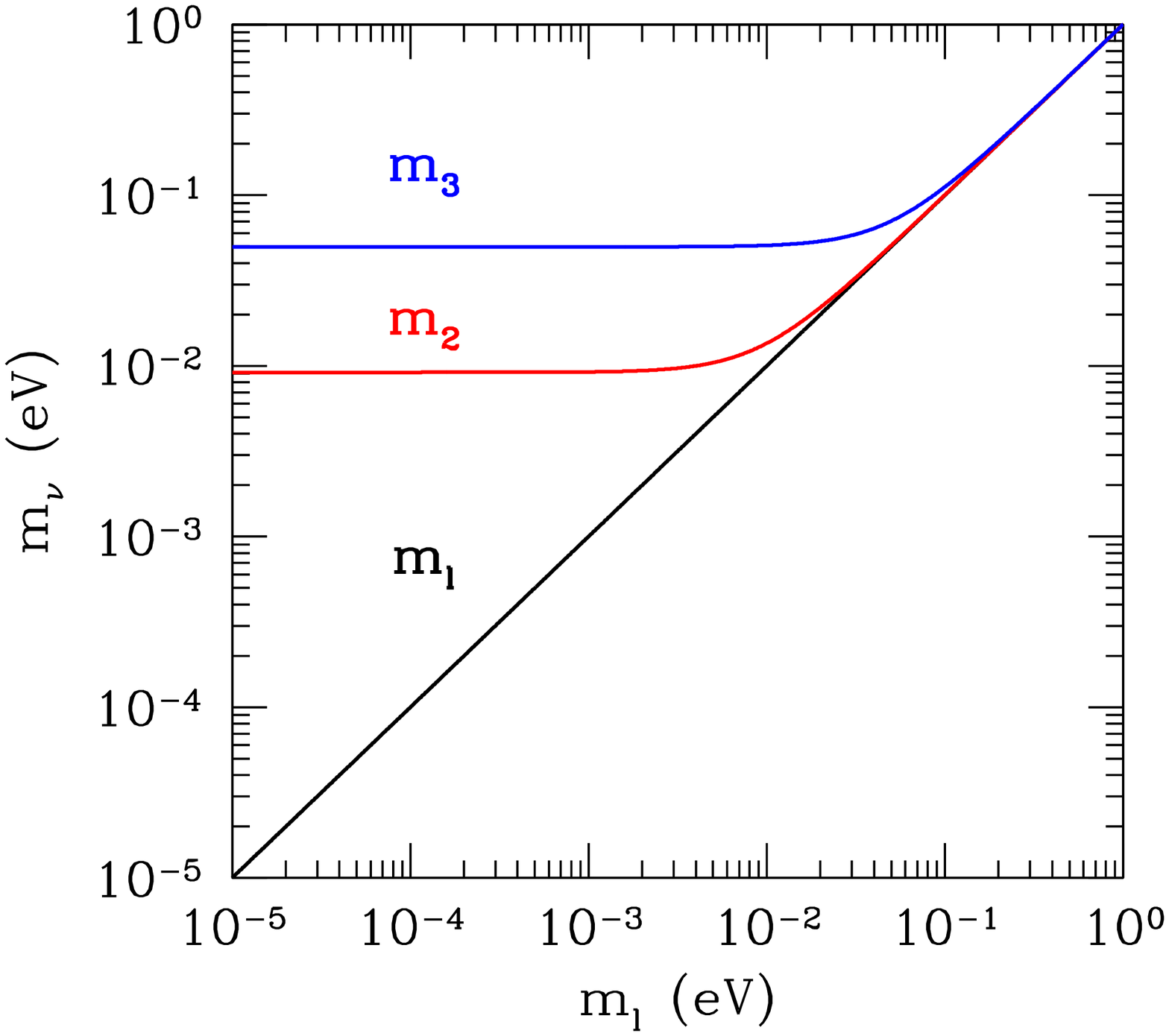}%{figs/normal_x.eps} \quad %&
 \includegraphics[width=7cm]{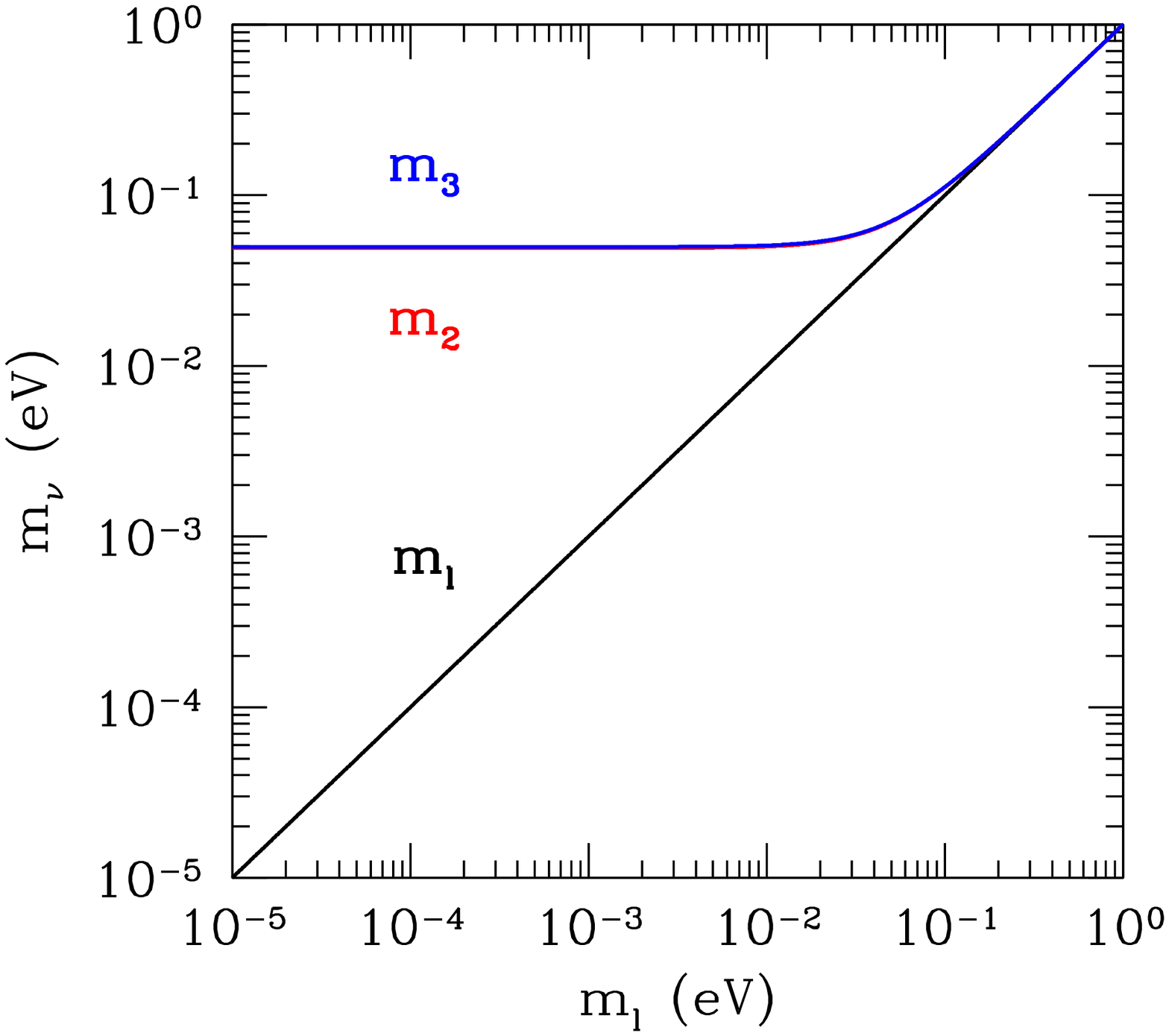}%{figs/inverted_x.eps}  %\\
 \caption{Favored values for the neutrino masses as functions of the lightest 
 neutrino mass, $m_\ell$, in the three neutrino scenario for normal 
 hierarchy (left panel, $m_{\ell} = m_{1}$) and the inverted hierarchy
  (right panel, $m_{\ell} = m_{3}$). [After 
  Ref.~\cite{Beacom:2002cb}.] }
 \label{fig:hierarchy}
 \end{figure*}

In Figure~\ref{fig:hierarchy}
 we  show the allowed ranges of neutrino masses
 ($m_{1}, m_2$, $m_3$)  in terms of the lightest neutrino mass
 $m_\ell$ for the normal and inverted hierarchies, using the best-fit values of
 the solar and atmospheric mass gaps reported above. For the entire 
 range of permitted masses, at least two relic species are 
 nonrelativistic ($m_{\nu} \gg \langle p_{\nu0}\rangle$) in the 
 present Universe. Consult Ref.~\cite{Paes:2001nd,Bilenky:2002aw} for
 recent surveys of the prospects for 
 determining the absolute mass scale of the neutrinos. 

\subsection{The Absorption Cross Section \label{subsec:abs}}
In terrestrial experiments that seek to detect neutrinos  
originating in particle accelerators or astrophysical sources, the 
reactions of interest are usually the deeply inelastic scattering 
processes $\nu N \to \mu + \hbox{anything}$ or $\nu N \to \nu + 
\hbox{anything}$.  Nucleons are so rare throughout the Universe at 
large ($n_{B} = (2.5 \pm 0.1)\times 10^{-7}\cm^{-3}$ in the current 
Universe~\cite{Eidelman:2004wy}) that neutrino--nucleon scattering is 
a negligible source of attenuation, even over cosmological distances. 
A path length of $8\times 10^{5}\mpc$ in today's Universe corresponds 
to $1\cm$ water equivalent (cmwe).

It is convenient to define the interaction length,
\begin{equation}
    \mathcal{L}_{\mathrm{int}}^{\nu N} = 1/\sigma_{\nu N}(E_{\nu})N_{A},
    \label{eqn:Nintlength}
\end{equation}
where $\sigma_{\nu N}$ is the appropriate neutrino--nucleon cross 
section and $N_{A} = 6.022 \times 10^{23}\hbox{ mol}^{-1} = 6.022 \times 
10^{23}\cm^{-3}$ (water equivalent) is Avogadro's number. For neutrino 
energies in the range $10^{16}\ev \lesssim E_{\nu} \lesssim 
10^{21}\ev$, a recent calculation~\cite{Gandhi:1998ri} yields 
 $(\nu,\bar{\nu})N$ total cross sections 
 \begin{equation}
     \sigma_{\nu N \to {\mathrm{all}}} \approx 0.78 \times 
     10^{-35}\cm^{2} \left( \frac{E_{\nu}}{1\gev} \right)^{\!0.363}\;.
     \label{eq:signun}
 \end{equation}
 It is not unreasonable to extrapolate this form a few orders of 
 magnitude higher in energy.\footnote{For an examination of different 
 extrapolations in energy and of the influence of exotic mechanisms, 
 see Ref.~\cite{hallsie}.}
 
\begin{figure}
\includegraphics[width=8.5cm]{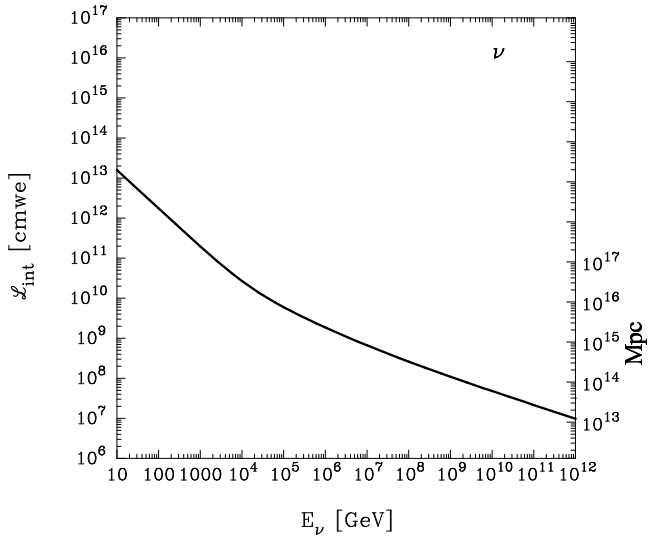} 
\caption{Interaction length defined in Eq.~(\ref{eqn:Nintlength}) for the reactions $\nu N \to 
\hbox{anything}$ as a function of the incident neutrino energy. The 
left-hand scale, in cmwe, is appropriate for terrestrial 
applications; the right-hand scale, in Mpc for the current Universe, 
is appropriate for transport over cosmological distances. [After~\cite{Gandhi:1998ri}.]}
\label{fig:Nintlength}
\end{figure}

 We plot the resulting interaction length in Figure~\ref{fig:Nintlength}. 
For the energies that will be of interest to us, the interaction 
length lies in the range $10^{6}\hbox{ -- }10^{9}\cm$we, or 
$10^{12}\hbox{ -- }10^{15}\mpc$ in the current Universe. These 
distances are extraordinarily vast, in view of the expectation that 
astrophysical sources of ultrahigh-energy neutrinos such as active 
galactic nuclei lie within 
$100\mpc$ of Earth. At earlier---but not too early---epochs, the 
nucleon density scales with redshift as $n_{B}(z) = n_{B}(0)(1 + z)^{3}$. 
Even so, back to $z \approx 20$ when the first astrophysical neutrino sources 
began to shine, the interaction length is far too long for $\nu N$ 
scattering to be a significant mechanism for reducing the flux of 
neutrinos incident on Earth.\footnote{A complete treatment, taking 
into account redshifting of the neutrino energy, does not change the 
conclusions~\cite{Gondolo:1991rn}.}

Over the energy range of interest for neutrino astronomy, the 
interactions of $\nu_{e}, \nu_{\mu}, \nu_{\tau}, \bar{\nu}_{\mu}, 
\bar{\nu}_{\tau}$ with electrons can generally be neglected compared 
to interactions with nucleons. The case of $\bar{\nu}_{e}e$ 
interactions is exceptional, because of the intermediate-boson 
resonance formed in the neighborhood of $E_{\nu}^{W \mathrm{res}} = 
M_{W}^{2}/2m_{e} \approx 6.3 \times 10^{15}\ev$. The peak cross section,
$\sigma(\bar{\nu}_{e}e \to W^{-} \to \hbox{anything}) \approx 5 \times 
10^{-31}\cm^{2}$, corresponds to an interaction length at resonance 
of $6 \times 10^{6}\cm$we~\cite{Gandhi:1995tf}. Assuming that the density of 
electrons throughout the current Universe is comparable to the density of 
baryons, even resonant $\bar{\nu}_{e}e$ scattering contributes a negligible 
attenuation of astrophysical neutrinos \textit{en route} to Earth.

The number density of relic neutrinos (of each species) in the 
current Universe is some 2.2 million times the number density of 
baryons. The thicker relic neutrino target (!) combined with an 
appreciable cross section for $\nu\bar{\nu} \to Z^{0}$ annihilation 
accounts for the importance of resonant absorption as an attenuator 
of extremely high energy neutrinos. As we will see shortly, the 
interaction length for resonant annihilation in the current Universe 
is $\mathcal{L}_{\mathrm{int}}^{\nu\bar{\nu}} \approx 1.2 \times 
10^{4}\mpc$, some six orders of magnitude shorter than the 
interaction lengths for $\nu N$ or resonant $\bar{\nu}_{e}e$ 
scattering.

The cross section for neutrino-antineutrino annihilation into fermion 
pairs through the 
$Z^{0}$ is given by
\begin{eqnarray}
    \sigma^{s:Z}(\nu_{\alpha}\bar{\nu}_{\alpha} \to f\bar{f}) & = & 
    \frac{G_{{F}}^{2}m_{i}E_{\nu}N_{c}^{(f)}}{3\pi}     \label{eq:Zann}\\ & \times &
    \frac{L_{f}^{2} + R_{f}^{2}}{(1 - \Y)^{2} + 
    \Gamma_{Z}^{2}/M_{Z}^{2}} \nonumber \;,
\end{eqnarray}
where $\Y = 2m_{i}E_{\nu}/M_{Z}^{2}$, $G_{{F}} = 1.166\,37 \times 10^{-5}\gev^{-2}$ is the Fermi 
constant, $m_{i}$ is the mass of the target (relic) neutrino, 
$E_{\nu}$ is the incident neutrino energy, and $N_{c}^{(f)}$ is the 
number of colors of the fermion $f$: 1 for leptons and 3 for quarks. 
The chiral couplings of $f$ are $L_{f} = \tau_{3}^{(f)} - 
2Q_{f}\sin^{2}\theta_{W}$ and $R_{f} = - 2Q_{f}\sin^{2}\theta_{W}$, 
where $Q_{f}$ is the fermion's electric charge and $\tau_{3}^{(f)} = 
\pm 1$ 
is the third component of its (left-handed) weak isospin.\footnote{We take $M_{Z} = 
91.1876\gev$, $\Gamma_{Z} = 2.4952\gev$, $\sin^{2}\theta_{W} = 
{0.231}$; we have taken account of fermion masses in our 
numerical studies.} 

We have written 
Eq.~(\ref{eq:Zann}) for the annihilation of a neutrino of flavor 
$\alpha$ on its antineutrino counterpart. In our application, the 
cross section must be weighted by the  probability for the 
mass eigenstate $\nu_{i}$ to contain the flavor component 
$\nu_{\alpha}$, which is to say, by the absolute square of the 
appropriate neutrino mixing matrix element. We have also assumed that 
the neutrinos are Majorana particles; the difference between Majorana 
and Dirac particles is explained in the following \S\ref{subsec:DorM}.

When summed over the kinematically accessible decay products of 
$Z^{0}$, namely the charged leptons $e,\mu,\tau$, the neutrinos 
$\nu_{e}, \nu_{\mu}, \nu_{\tau}$, and the quarks $u, d, s, c, b$, 
(\ref{eq:Zann}) leads to the branching fractions collected in Table~\ref{tab:Zdecay}.
\begin{table}%[H] add [H] placement to break table across pages
\caption{Decay modes and branching fractions of the $Z^{0}$~\cite{Eidelman:2004wy}.
\label{tab:Zdecay}}
\begin{ruledtabular}
\begin{tabular}{cc}
    Decay mode & Branching fraction \\
    \hline
    hadrons ($u\bar{u}+d\bar{d}+s\bar{s}+c\bar{c}+b\bar{b}$) & $70\%$ \\
    charged leptons ($e^{+}e^{-} + \mu^{+}\mu^{-} + 
    \tau^{+}\tau^{-}$) & $10\%$ \\
    invisible ($\nu_{e}\bar{\nu}_{e}+ \nu_{\mu}\bar{\nu}_{\mu}+ 
    \nu_{\tau}\bar{\nu}_{\tau}$) & $20\%$
\end{tabular}
\end{ruledtabular}
\end{table}
In our study of absorption lines, we shall regard the 20\% of $Z^{0} 
\to \nu\bar{\nu}$ decays as removed from the incident beam. In a 
detailed study of particular experimental circumstances, one might 
choose to improve this approximation.

\begin{figure}
\includegraphics[width=8.75cm]{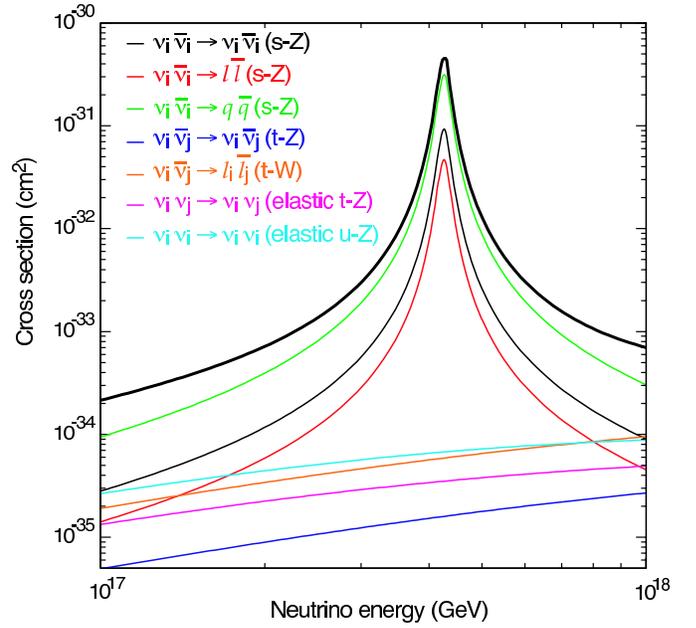}
\vspace*{-12pt}
\caption{Total neutrino annihilation cross section and the different
contibution channels as a function of the ultra-high neutrino energy
assuming a relic neutrino mass of $m_\nu =10^{-5}\ev$ and zero
redshift.}
\label{fig:cross}
\end{figure}

Depending on the incident beam and the relic target, other processes 
may contribute. A complete catalogue was given by 
Roulet~\cite{Roulet:1992pz}, whose notation we emulate here.  
We present the components of the neutrino-(anti)neutrino cross sections in Figure~\ref{fig:cross}.
Neutrino-antineutrino scattering in general receives a contribution from $t$-channel $Z$-exchange,
\begin{eqnarray} \sigma^{t:Z}(\nu_\alpha \bar{\nu}_\beta\rightarrow 
\nu_\alpha\bar{\nu}_\beta)& =&\frac{G^2_F m_{i}E_{\nu}}{\pi} F_{1}(\Y),
\label{eqn:resonantt}
\end{eqnarray} 
where $ F_{1} (\Y)=[\Y^2 +2 \Y -2
(1+\Y)\ln(1+\Y)]/\Y^3$.  The $s$-$t$ interference term is
\begin{eqnarray} \sigma^{st:Z}(\nu_\alpha \bar{\nu}_\beta \rightarrow 
\nu_\alpha
\bar{\nu}_\beta)& =& \delta_{\alpha\beta}\frac{G^2_F m_{i}E_{\nu}}{2 \pi} F_{2}(\Y)  \\
& \times &  \frac{(\Y - 1)}{(1 - \Y)^{2} +\Gamma^2_Z / M^2_Z} ~,\nonumber 
\label{eqn:resonantts}
\end{eqnarray}
where $ F_{2} (\Y)=[3\Y^2 +2 \Y -2 (1+\Y)^2\ln(1+\Y)]/\Y^3$. 

Neutrino-antineutrino scattering to a pair of charged leptons may 
proceed by $W$ exchange in the $t$ channel,
\begin{eqnarray}
\sigma^{t:W}(\nu_\alpha \bar{\nu}_\beta\rightarrow \ell_\alpha 
\bar{\ell}_\beta)& =&\frac{4 G^2_F m_{i}E_{\nu}}{\pi} F_{1}(\Y) .
\label{eqn:wt}
\end{eqnarray}
For charged-lepton pair production, the interference between
the $s$-channel $Z$ exchange and the $t$-channel $W$ exchange is
\begin{eqnarray}
\sigma^{WZ}(\nu_\alpha \bar{\nu}_\beta \rightarrow \ell_\alpha 
\bar{\ell}_\beta)& =& 
\delta_{\alpha\beta}\frac{4 
G^2_F m_{i}E_{\nu}}{\pi}  F_{2}(\Y)(\sin^{2}\theta_{W} - \cfrac{1}{2})
 \nonumber \\
& \times & \frac{(\Y - 1)}{(1 - \Y)^2 +\Gamma^2_Z/ M^2_Z}\;.
\label{eqn:resonantzw}
\end{eqnarray} 

Neutrino-neutrino (or antineutrino-neutrino) elastic scattering is 
mediated by $t$-channel $Z$ exchange, with a cross section
\begin{eqnarray}
\sigma^{t:Z}(\nu_\alpha \nu_\beta \rightarrow \nu_\alpha \nu_\beta) & =&\frac{G^2_F 
m_{i}E_{\nu}}{\pi}
\frac{1}{1 + \Y}\;,
\label{eqn:elt}
\end{eqnarray}
that is accompanied, for identical species, by the $u$-channel 
contribution
\begin{eqnarray}
\sigma^{u:Z}(\nu_\alpha \nu_\beta \rightarrow \nu_\alpha \nu_\beta)& =& 
\delta_{\alpha\beta}\frac{G^2_F 
m_{i}E_{\nu}}{\pi}  \\ & & \times
\left[\frac{1}{1+\Y}+\frac{\ln(1+\Y)}{\Y(1 + \cfrac{1}{2}\Y)}\right] .
\nonumber 
\label{eqn:elu}
\end{eqnarray}  

Above the thresholds for $W^{+}W^{-}$ and $Z^{0}Z^{0}$ pair production,
we include the $\nu\bar{\nu} \to$ vector-boson-pair cross sections in our numerical
analysis.  The effect of these processes on neutrino attenuation is
minor; the relevant formulas may be found in Ref.~\cite{Roulet:1992pz}.

\subsection{Dirac \textit{versus }Majorana Relics \label{subsec:DorM}}
The interaction cross section may depend on whether the relic (target) 
neutrinos are Dirac or Majorana particles. If the relic neutrinos are 
extremely relativistic, the Dirac and Majorana characters are indistinguishable: 
relativistic neutrinos are pure left-handed chirality states, because 
only such states are produced in the weak interactions. Chirality and 
helicity coincide, and the right-handed chirality is absent. 
Nonrelativistic Dirac and Majorana neutrinos exhibit distinctive 
behavior. In the static limit, Dirac neutrinos are left-handed helicity
eigenstates with equal populations of left- and right-handed
chiralities; Dirac antineutrinos are right-handed helicity eigenstates,
also with equal populations of left- and right-handed chiralities.
Since only the left-handed neutrino (right-handed antineutrino)
chiralities interact, the other two components of the Dirac neutrino
field are sterile.  Because Majorana neutrinos are their own
antiparticles, both chiralities interact.  Accordingly, in the static limit, 
the interaction cross section on a Majorana target is twice the cross 
section on a Dirac target.

The  interactions with a Majorana neutrino's ``wrong-chirality'' population 
enter with weight $m_{\nu}^{2}/(\varepsilon_\nu \ +\ p_\nu)^2$,
where $p_{\nu}$ is the relic neutrino momentum and 
$\varepsilon_{\nu} = \sqrt{p_{\nu}^{2}+m_{\nu}^{2}}$, so that 
\begin{equation}
    \frac{\sigma_{\mathrm{Majorana}}}{\sigma_{\mathrm{Dirac}}} =  1 + 
    \frac{m_\nu^{2}}{(\varepsilon_\nu \ +\ 
    p_\nu)^2}   = 
    \frac{2} {1 + p_{\nu}/\varepsilon_{\nu}}\;.
    \label{eq:MoverD}
\end{equation}

\begin{figure*}
\includegraphics[width=5.75cm]{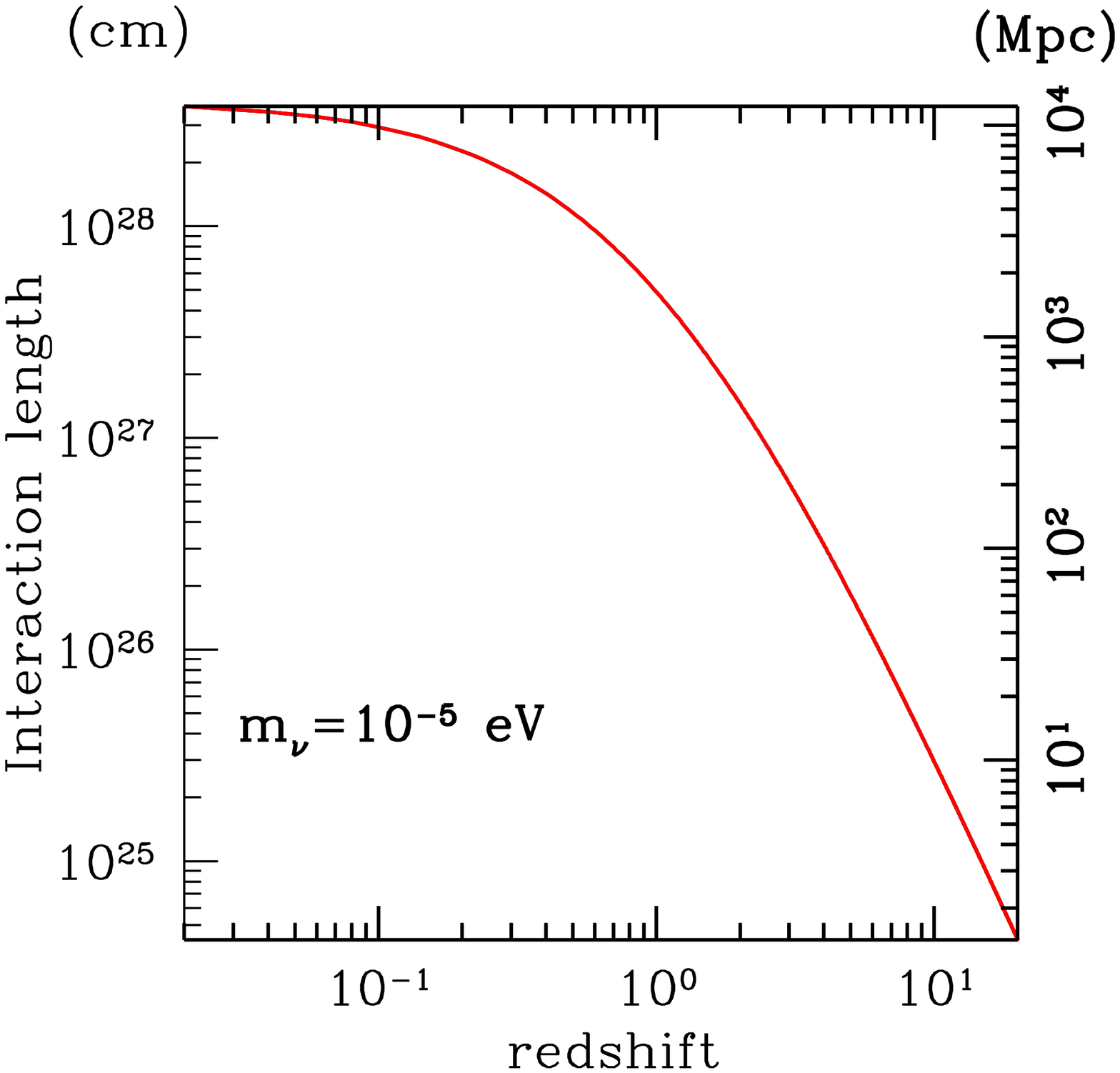}%{figs/5n.eps} 
\includegraphics[width=5.75cm]{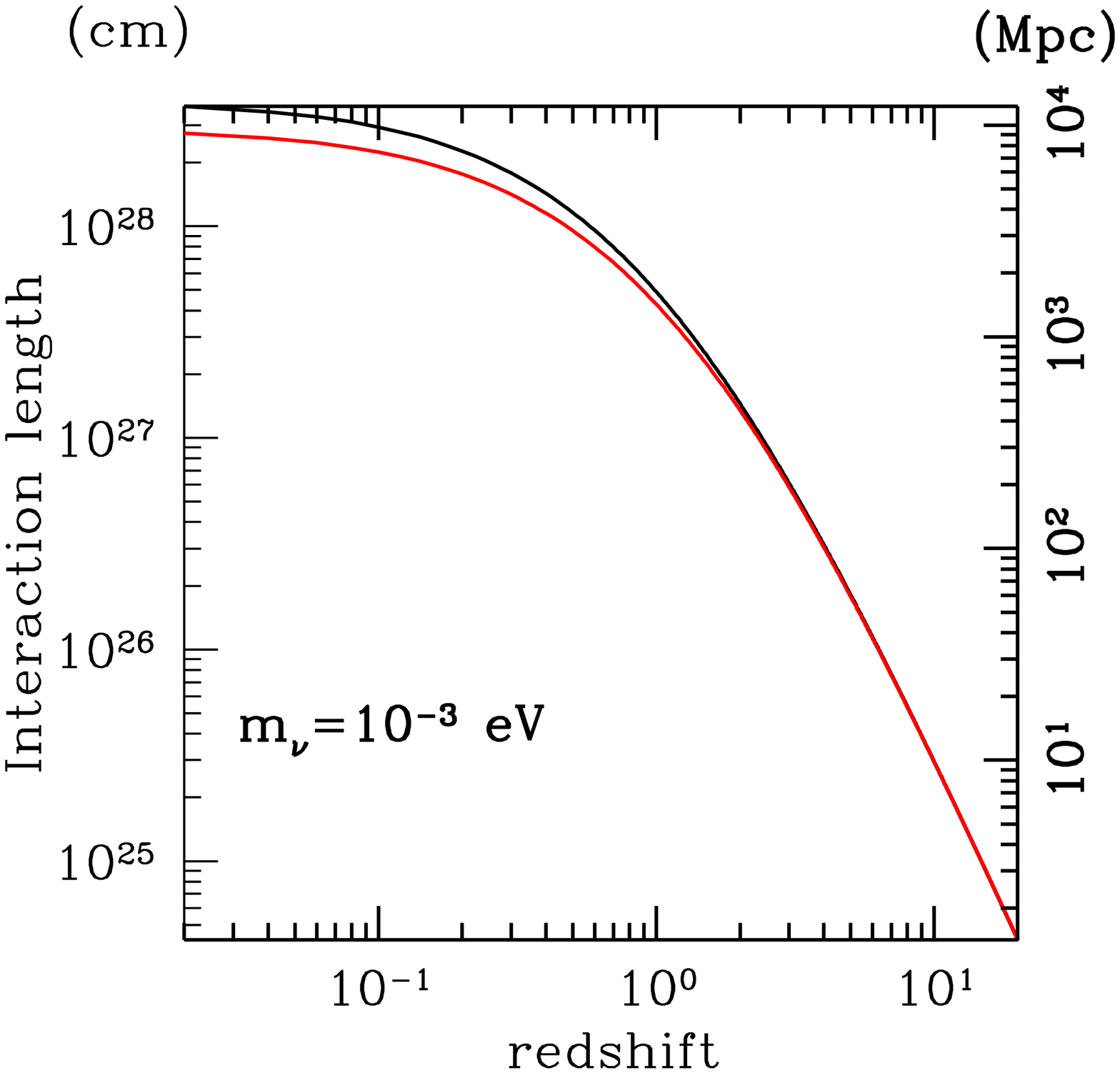}%{figs/3n.eps}
\includegraphics[width=5.75cm]{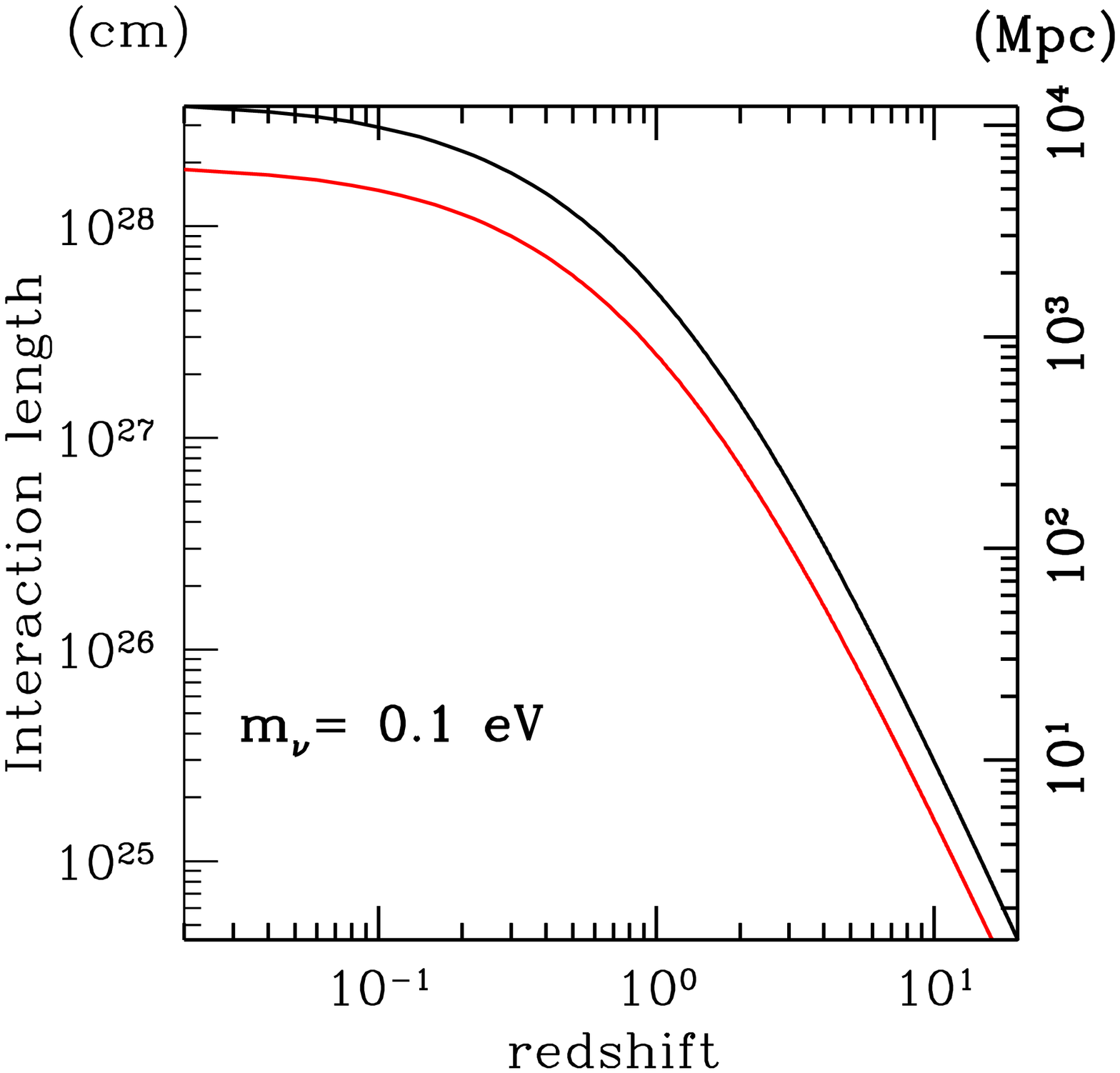}%{figs/1n.eps}
\vspace*{-12pt}
\caption{Interaction lengths defined in Eq.~(\ref{eqn:intlength})
versus redshift at the $Z^{0}$ resonance 
for neutrino masses $m_\nu =10^{-5}, 10^{-3}, 10^{-1}\ev$ (left, center, 
and right panels). The left-hand scales are in centimeters, the 
right-hand scales in megaparsecs ($1\mpc = 3.085678 \times 
10^{24}\cm$).
In the center and right panels, the lower (black) 
line is for the  Dirac-neutrino case; the  upper (red) line applies to Majorana 
neutrinos.}
\label{fig:5int}
\end{figure*}

The interaction length for annihilation on relic neutrinos 
illustrates the difference between the Majorana and
Dirac cases and the transition from static to extreme relativistic
regimes.  We define
\begin{equation}
\mathcal{L}^{\nu\bar{\nu}}_{\mathrm{int:M,D}}= 
1/\sigma_{\mathrm{M,D}}(E_{\nu}^{Z\mathrm{res}})n_{\nu_{i}}(z)\;,
\label{eqn:intlength}
\end{equation}
where $E_{\nu}^{Z\mathrm{res}} = M_{Z}^{2}/2m_{\nu}$, and evaluate the
ratio~(\ref{eq:MoverD}) using the mean momentum $\langle p_{\nu}
\rangle$ and energy $\langle \varepsilon_{\nu} \rangle = (\langle
p_{\nu}^{2} \rangle + m_{\nu}^{2})^{1/2}$ from the Fermi-Dirac
distribution~(\ref{eq:nuFD}).  In Figure~\ref{fig:5int},
we depict the Majorana and Dirac
interaction lengths 
over the range of redshifts considered in this study, 
for three illustrative values of the
relic neutrino mass.

Contrary to common wisdom, the distinction between Majorana and Dirac
relics is not readily observable in neutrino absorption lines, unless
the relic neutrino mass approaches $\approx 0.1\ev$, a value close to
the present cosmological upper bounds for a quasi-degenerate neutrino 
spectrum.  Appreciably lighter relics are relativistic over much or 
all of the redshift range we consider.

\subsection{Super-High-Energy Neutrino Sources \label{subsec:sources}}

\subsubsection{General orientation \label{subsubsec:go}}
Observing  absorption lines on the relic background requires 
an adequate neutrino flux at the resonant energies $E^{Z \mathrm{res}}_{\nu}= 
M_{Z}^{2}/2m_{\nu_{i}} \approx 4.2 \times 10^{21}\ev/m_{\nu_{i}}$. For 
the sub-eV relics that current information on the neutrino 
spectrum leads us to expect, the cosmic neutrinos must have energies 
no less than those of the highest-energy cosmic rays ever 
observed~\cite{Takeda:1998ps,Abbasi:2002ta}.   
Indeed, the required cosmic neutrinos have been named \textit{Super-GZK 
neutrinos}~\cite{Berezinsky:2003iv}, as their energies lie above 
the Greisen--Zatsepin--Kuzmin~\cite{Greisen:1966jv,Zatsepin:1966jv} cutoff 
in the cosmic ray spectrum. 

It is worth taking a moment to review the GZK argument, because it
implies the existence of so-called cosmogenic
neutrinos~\cite{Stecker:1978ah}.  Extremely high energy cosmic
rays---let us take protons, to be concrete---can lose energy by
interacting with the cosmic microwave background whose properties were
recalled in \S\ref{subsec:char}.  The key energy-loss mechanism is pion
photoproduction, $p + \gamma \to \pi + N$.  Ultrahigh-energy $
\nu_{\mu}, \gamma,\hbox{ and }\bar{\nu}_{\mu}$ arise from the decays of
$\pi^{+}\!, \pi^{0}\!, \pi^{-}\!$.

Taking the energy of a typical CMB photon as
\begin{equation}
    \langle p_{\gamma\,0} \rangle =3 \, 
    \frac{\zeta(4)}{\zeta(3)} \cdot T_{0} \approx 2.701 T_{0}
    \approx 6.341 \times 10^{-4}\ev\;,
    \label{eq:gammom}
\end{equation}
we estimate the threshold for pion production to be
\begin{equation}
    E_{p}^{\gamma\to\pi} \approx \frac{m_{\pi}(m_{\pi}+2M_{p})}
    {4\langle p_{\gamma\,0} \rangle} \approx 1.1 \times 10^{20}\ev\;,
\end{equation}
where $m_{\pi}$ is the pion mass and $M_{p}$ is the proton mass.
Accordingly, any proton with energy $\gtrsim 10^{20}\ev$ that traverses 
a long path in the current Universe will suffer energy loss through 
pion photoproduction. The interaction length, determined by 
scattering at the $\Delta(1232)$ resonance, is approximately 
$10\mpc$, short compared with the 100-Mpc distance to active 
galaxies. At earlier epochs, the interaction length (at redshifted 
energy) scales with the number density of CMB photons.

No experiment has yet detected neutrinos with energies above $1\tev$
that originate outside Earth's atmosphere.  To discuss possible
sources\footnote{See the extensive review of cosmic-ray sources in
Ref.~\cite{Stecker:2003wm}.} of cosmic neutrinos that might be useful
for absorption spectroscopy we enter a largely unexplored realm of
upper limits and models not disciplined by extensive data sets.

Both acceleration mechanisms and top-down (decay) phenomena may be at
the origin of super-high-energy neutrinos.  We consider these two 
classes of sources briefly in turn.

Extragalactic objects such as active galactic nuclei (AGNs) and gamma-ray 
bursters (GRBs) are generally regarded as promising sites for the 
production of ultrahigh-energy neutrinos. Protons accelerated to 
extreme energies may collide with the surrounding matter or the bath 
of photons to produce pions through the inclusive reactions
\begin{equation}
    p + (N,\gamma) \to \pi + \hbox{anything.}
    \label{eq:piprod}
\end{equation}
If $\pi^{+}, \pi^{0}, \pi^{-}$ are produced in equal numbers, then 
the decay chains $\pi^{0} \to \gamma\gamma$ and
\dk{\pi^{+}}{\mu^{+}{\nu}_{\mu}}{e^{+}\nu_{e}\bar{\nu}_{\mu}}
(and similarly for $\pi^{-}$) imply products in the 
proportions $\gamma:e^{+}:e^{-}:\nu_{\mu}:\bar{\nu}_{\mu}:\nu_{e}:\bar{\nu}_{e}::2:1:1:2:2:1:1$.
If such processes were responsible for the flux of ultrahigh-energy gamma 
rays, then a similar flux of ultrahigh-energy neutrinos would be 
essentially guaranteed.\footnote{A rather restrictive upper bound 
(Waxman--Bahcall) follows if energetic protons escape freely 
from such a source~\cite{Waxman:1998yy}. A more permissive (by $\sim 
40\times$) \textit{cascade limit}~\cite{cascade} relates the neutrino 
flux to the $\gamma$-ray flux observed by the EGRET 
instrument~\cite{Sreekumar:1997un} aboard the Compton Gamma-Ray 
Observatory, assuming the photons are not obscured. Future 
$\gamma$-ray detectors, such as GLAST~\cite{glast}, will improve the 
photon-flux baseline. In the case of \textit{hidden sources,} from 
which neither nucleons nor photons escape, there is no way to bound 
the neutrino flux from above.}

The top-down scenarios---superheavy relic particles or topological
defects formed in symmetry-breaking phase transitions predicted by
unified theories, for example---do not require regions in which
astrophysical processes can accelerate particles to super-high
energies, but they depend on physics beyond the standard model that has
not been established~\cite{Bhattacharjee:1991zm}.  They might populate
energies beyond the reach of even the most extreme astrophysical
processes, conceivably exceeding the scale on which the
$\mathrm{SU(3)}_c\otimes \mathrm{SU(2)}_{\mathrm{L}}\otimes
\mathrm{U(1)}_Y$ standard-model interactions are unified. 

In this class of models, ultrahigh-energy cosmic rays can be decay
products of some supermassive $X$-particles with masses $M_{X}$ close
to the GUT scale.  The supermassive $X$-particles could be long-lived
relics of the early universe or could themselves arise from the
collapse of topological defects.  The $X$ particles can decay into
nucleons, gamma rays, and neutrinos with energies approaching
$M_{X}$.\footnote{See Ref.~\cite{Semikoz:2003wv} and the works cited
there for a general discussion.} In the simplest models, the fluxes of
neutrinos that arise in this manner are bounded by the cascade limit.
Hidden-sector topological defects, which may arise in multi-brane
scenarios, evade the cascade limit and so might provide the largest
(which is to say, least constrained) flux of Super-GZK
neutrinos~\cite{Berezinsky:1999az}.

\subsubsection{Parametrizations of neutrino spectra 
\label{subsubsec:param}}
Calculation of the cosmogenic neutrino flux is accomplished by
propagating an assumed primary proton flux through the cosmic microwave
background over the relevant history of the universe, by means of
transport codes.  A standard
\textit{Ansatz}~\cite{Semikoz:2003wv,help2} is a power-law shape for
the injection spectrum per unit comoving volume,
\begin{equation}
    \varphi_{p}(E,z) = \mathcal{N} \cdot E^{-\alpha}\,
    (1+z)^m\,\Theta(E_{\mathrm{max}}-E)\;,
    \label{eqn:cosmogenic}
\end{equation}
during the era characterized by $z_{\mathrm{min}} \le z \le
z_{\mathrm{max}}$.  Here $E_{\mathrm{max}}$ represents the maximum
energy to which protons can be accelerated by astrophysical processes,
$\alpha$ is the spectral index, $m$ is the redshift-evolution index,
and $\mathcal{N}$ specifies the normalization.  

\begin{figure}[tb]
    \includegraphics[width=7.5cm]{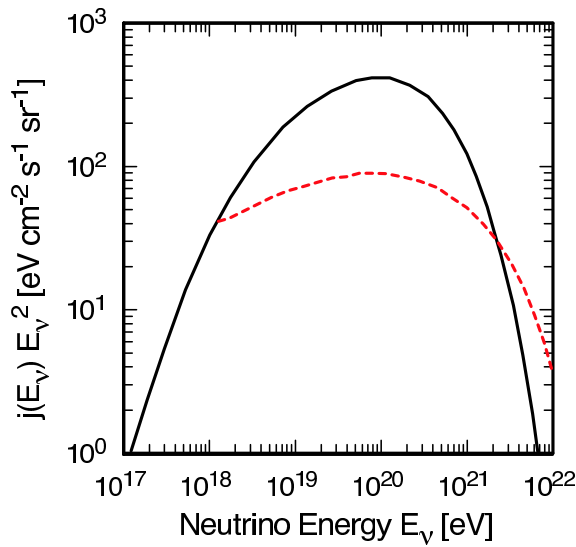}  
\vspace*{-6pt}
\caption{Solid line: Maximal cosmogenic neutrino fluxes \textit{per flavor} 
computed for an injection spectrum characterized by 
$E_{\mathrm{max}}=2 \times 10^{13}\gev$, $\alpha=1$,  $m=3$, for $0 
\le z \le 2$ (from Ref.~\cite{Semikoz:2003wv}). Dashed (red) line:
Diffuse neutrino flux from hidden topological defects in the 
form of necklaces,
as calculated in  Ref.~\cite{Aloisio:2004ap} for a superheavy particle of mass 
$M_X=10^{14}\gev$.}
\label{fig:presentfuture}
\end{figure}

As an example, we show %as the solid line 
in Figure~\ref{fig:presentfuture} ($E_{\nu}^{2}\times$) the cosmogenic
neutrino flux $j(E_{\nu})$ tuned~\cite{Semikoz:2003wv} to saturate the
cascade limit, $j(E_{\nu})\,E_{\nu}^{2} \lesssim 
450\ev\cm^{-2}\s^{-1}\sr^{-1}$, derived from a recent analysis of the EGRET
data~\cite{Strong:2003ex}. 
As anticipated, the cosmogenic neutrinos lie squarely in the domain 
of the GZK cutoff on the cosmic-ray spectrum.
The flux depicted in Figure~\ref{fig:presentfuture} is consistent 
with direct limits on the neutrino flux, which are summarized in 
Refs.~\cite{Eberle:2004ua,Semikoz:2003wv}. Equal fluxes (at Earth) of 
all neutrino flavors is implied by the pattern of neutrino mixing, as 
we elaborate in the opening paragraphs of \S\ref{sec:toy}.
We comment briefly on event rates for planned detectors in \S\ref{subsec:events}.

To illustrate top-down scenarios, we cite a calculation of the diffuse
neutrino flux arising from one species of topological defect, a
necklace of monopoles and antimonopoles strung on a cosmic string.
Once formed in symmetry-breaking phase transitions in the early
universe, topological defects can survive indefinitely, unless they
collapse or annihilate~\cite{hillm}, producing massive quanta
generically labelled $X$ particles.  As monopoles and antimonopoles on
a necklace annihilate, they produce $X$ particles in the form of heavy
Higgs bosons and gauge bosons.  These in turn are the source of
ultrahigh-energy cosmic rays and neutrinos; analytic expressions for 
the fluxes in several cases are computed in Ref.~\cite{Bhattacharjee:1991zm}.

The dashed (red) curve in Figure~\ref{fig:presentfuture}
shows the diffuse neutrino spectrum in the present universe that arises for 
monopole-antimonopole annihilations into particles with mass
$M_X=10^{14}\gev$~\cite{Aloisio:2004ap}. In top-down scenarios, 
energetic neutrinos might have been generated at very early times; in 
that case our computation of the absorption lines will entail an 
integration over redshift back to those early times.

\subsection{Detectors for Cosmic Neutrinos \label{subsec:detect}}
Neutrino astronomy is moving into a new, and much anticipated, era. The value 
of neutrino observatories has been clearly demonstrated in the MeV 
range through the detection of neutrinos from Supernova 1987A and from 
the Sun~\cite{Davis:2003kh,Koshiba:2003xy}. The detailed observation 
of atmospheric neutrinos with GeV energies was crucial to 
establishing neutrino mixing~\cite{Fukuda:1998mi}, and the AMANDA 
experiment has detected atmospheric neutrinos up to about 
$10^{5}\gev$~\cite{Ahrens:2004qq}.  Current exploration is dedicated 
to the search for extraterrestrial neutrinos---either from the diffuse 
background of active galactic nuclei or from point sources such as 
gamma-ray bursters---with energies between $10^{5}$ and $10^{9}\gev$. 
In addition to the ice-Cherenkov detector AMANDA~II, dedicated 
neutrino telescopes include the BAIKAL water-Cherenkov 
array~\cite{Spiering:2004dt} and the antenna array 
RICE~\cite{Kravchenko:2003tc}, which aims to detect radio pulses 
emitted by neutrino-induced showers in the Antarctic ice. The 
Fly's Eye~\cite{Baltrusaitis:1985mt}, HiRes~\cite{wonyong}, and AGASA~\cite{Hayashida:1998qb} 
air-shower arrays  are sensitive to   horizontal air showers initiated by neutrino
interactions deep in the atmosphere.

The Goldstone Lunar Ultrahigh-energy neutrino Experiment
(GLUE)~\cite{Gorham:2001aj} has begun to search  for radio emission from
ultrahigh-energy cascades induced by neutrinos or cosmic rays skimming
the moon surface.  The FORT\'{E} (Fast On-orbit Recording of Transient Events)
satellite~\cite{Lehtinen:2003xv} has set upper limits on
the UHE neutrino fluxes at energies beyond the GZK cutoff, looking 
for radio pulses generated by neutrino interactions in the Greenland 
ice sheet.

Over the next few years, significantly increased
sensitivities will be attained in IceCube~\cite{icecube}, a cubic-kilometer-scale
ice-Cherenkov detector evolved from the AMANDA experience that is beginning
construction at the South Pole.  The ANTARES~\cite{antares},
NEMO~\cite{nemo}, and NESTOR~\cite{nestor} projects are developing
techniques for a cubic-kilometer water-Cherenkov array in the
Mediterranean Sea~\cite{mediterranean}.  It is highly desirable that
the next generation of neutrino telescopes not only characterize the
incident neutrino energy and direction, but also tag the neutrino
flavor by identifying the outgoing charged lepton.  An optimistic
assessment of prospects for flavor tagging in detectors such as IceCube
can be found in Ref.~\cite{Beacom:2002vi}.

ANITA~\cite{anita}, a balloon-borne array of radio antennas, will
circle the Antarctic continent at an altitude of $\sim 35\km$ to record
radio bursts from neutrino interactions in the polar ice cap.  The
Pierre Auger Observatory, a $3000\hbox{-km}^{2}$ hybrid detector for
air showers in Argentina's Mendoza province, will be sensitive to the
interactions of $\gtrsim10^{9}$-GeV neutrinos in the
atmosphere~\cite{auger}.  Space-based instruments such as the Extreme 
Universe Space Observatory (EUSO)~\cite{euso} and the Orbiting Wide-angle 
Light-collectors (OWL)~\cite{Cline:1999ez}  would have an energy threshold 
near $10^{10}\gev$.

\section{A Toy Experiment \label{sec:toy}}
As a prelude to our investigation of neutrino absorption spectroscopy 
in the physical Universe, we describe a highly idealized situation in 
which an extremely high-energy neutrino beam traverses a
very long column with the relic-neutrino properties of the 
current Universe. We neglect for now the expansion of the Universe 
and the thermal motion of the relic neutrinos. The ``cosmic neutrino attenuator'' is 
thus a column of length $L$ with uniform neutrino density $n_{\nu0} = 
56\cm^{-3}$ of each neutrino species, $\nu_{e}, \bar{\nu}_{e}, \nu_{\mu}, 
\bar{\nu}_{\mu}, \nu_{\tau}, \bar{\nu}_{\tau}$.

We imagine a terrestrial detector capable of 
distinguishing arriving neutrino species through charged-current 
interactions producing electrons, muons, or tau leptons, and of 
inferring the energies of the neutrinos that initiated those 
interactions.  A detector of the required scale is unlikely to measure 
the charge of the outgoing lepton, and so would not distinguish 
neutrinos from antineutrinos in the beam. Our aim here is to identify 
the sorts of observations that might be made, should neutrino 
absorption line spectroscopy become practical, and to point out 
how various neutrino properties would manifest themselves.

We assume that the incoming neutrino beam originates at 
least $100\mpc$ from Earth, that it contains $\nu_{e}, \bar{\nu}_{e}, \nu_{\mu}, 
\bar{\nu}_{\mu}, \nu_{\tau}, \bar{\nu}_{\tau}$ in sufficient numbers 
to allow a measurement of the energy spectra of the neutrinos arriving 
at Earth, and that the neutrino energy spectrum at the source is 
reasonably smooth.\footnote{It is a reasonable bet, though not 
essential to our analysis, that the beam contains an equal mix of 
neutrinos and antineutrinos.} Neutrino oscillations tend to produce a 
``beam'' with roughly equal mixtures of $\nu_{e},\nu_{\mu},\nu_{\tau}$,
whatever the flavor mixture of the extremely high-energy neutrinos at 
the source~\cite{Athar:2000yw,Barenboim:2003jm}. The vacuum oscillation length,
$L_{\mathrm{osc}} = 4\pi E_{\nu}/|\Delta m^2|$, is typically short compared with
the intergalactic distances we contemplate.  For $|\Delta m^2| = 
\Delta m_{12}^{2} \approx
8.2 \times 10^{-5}\ev^2$, for example, the oscillation length is $L_{\mathrm{osc}}\approx 3 \times
10^{-25}\mpc\cdot (E_{\nu}/1\ev)$. At the resonance energy 
$E_{\nu}^{Z \mathrm{res}} = M_{Z}^{2}/2m_{\nu}$, the (solar) %vacuum 
oscillation length is thus $L_{\mathrm{osc}}\approx 1.2 \times 
10^{-3}\mpc/(m_{\nu_{i}}/1\ev)$. Only for the normal hierarchy with 
$m_{\ell} \lesssim 10^{-4}\ev$ is the oscillation length not 
a negligible distance. The decoherence length for neutrinos above 
$10^{18}\ev$ is many orders of magnitude greater than the Hubble 
distance, $D_{H} \equiv c/H_{0} = 4200\mpc 
(0.7/h)$~\cite{Eberle:2004ua}.

\subsection{An Idealized Experiment \label{subsec:ideal}}
As we reviewed in \S\ref{subsec:numass}, the mass $m_{\ell}$
of the lightest neutrino, which corresponds to $m_1$ for the normal 
spectrum or to $m_{3}$ for the inverted spectrum, lies between 
$10^{-5}\ev$ and $0.1\ev$.  We present
here studies for three example values: $m_{\ell} =
10^{-5}, 10^{-3}, \hbox{ and }0.1\ev$. For a given mass of the 
lightest neutrino, the graphs in Figure~\ref{fig:hierarchy} show our current
expectations for the masses of the other neutrinos.

If the column of relic neutrinos is thick enough to attenuate 
neutrinos appreciably through resonant absorption at the $Z^{0}$ gauge 
boson, the energies that display absorption dips point to the neutrino 
masses through the condition $m_{\nu} = M_{Z}^{2}/2E_{\nu}^{Z 
\mathrm{res}}$.\footnote{See Ref.~\cite{Gelmini:2004zb} for a related strategy to 
determine the neutrino masses.} Evidently the heaviest neutrino corresponds to the 
lowest-energy absorption line. For a normal hierarchy, we expect 
three or two or one absorption lines, as $m_{\ell}$ increases; for an 
inverted hierarchy, two or one absorption lines.

\begin{figure*}
\includegraphics[width=8.75cm]{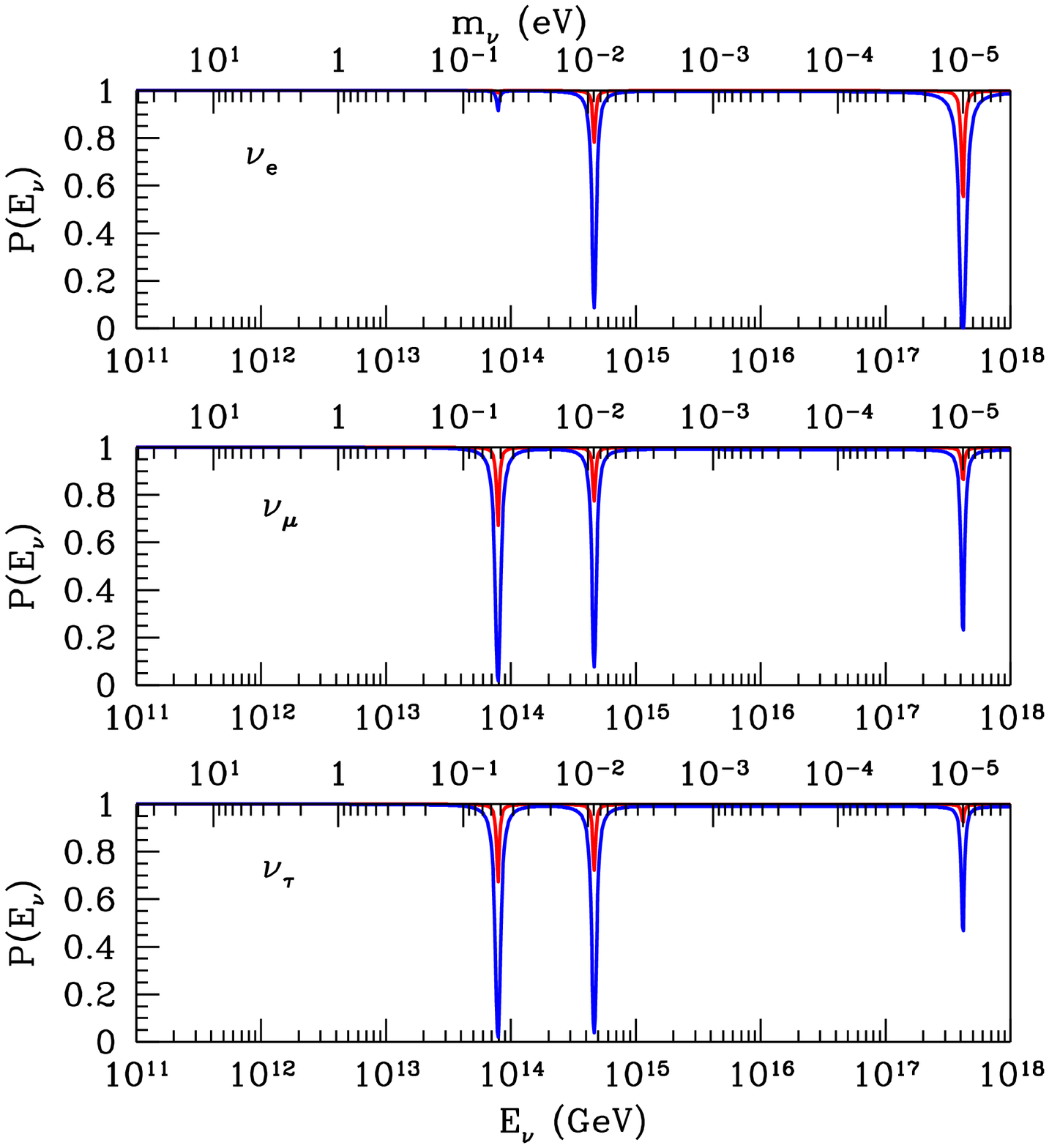}%{figs/maj_light_normalxx.eps} 
\includegraphics[width=8.75cm]{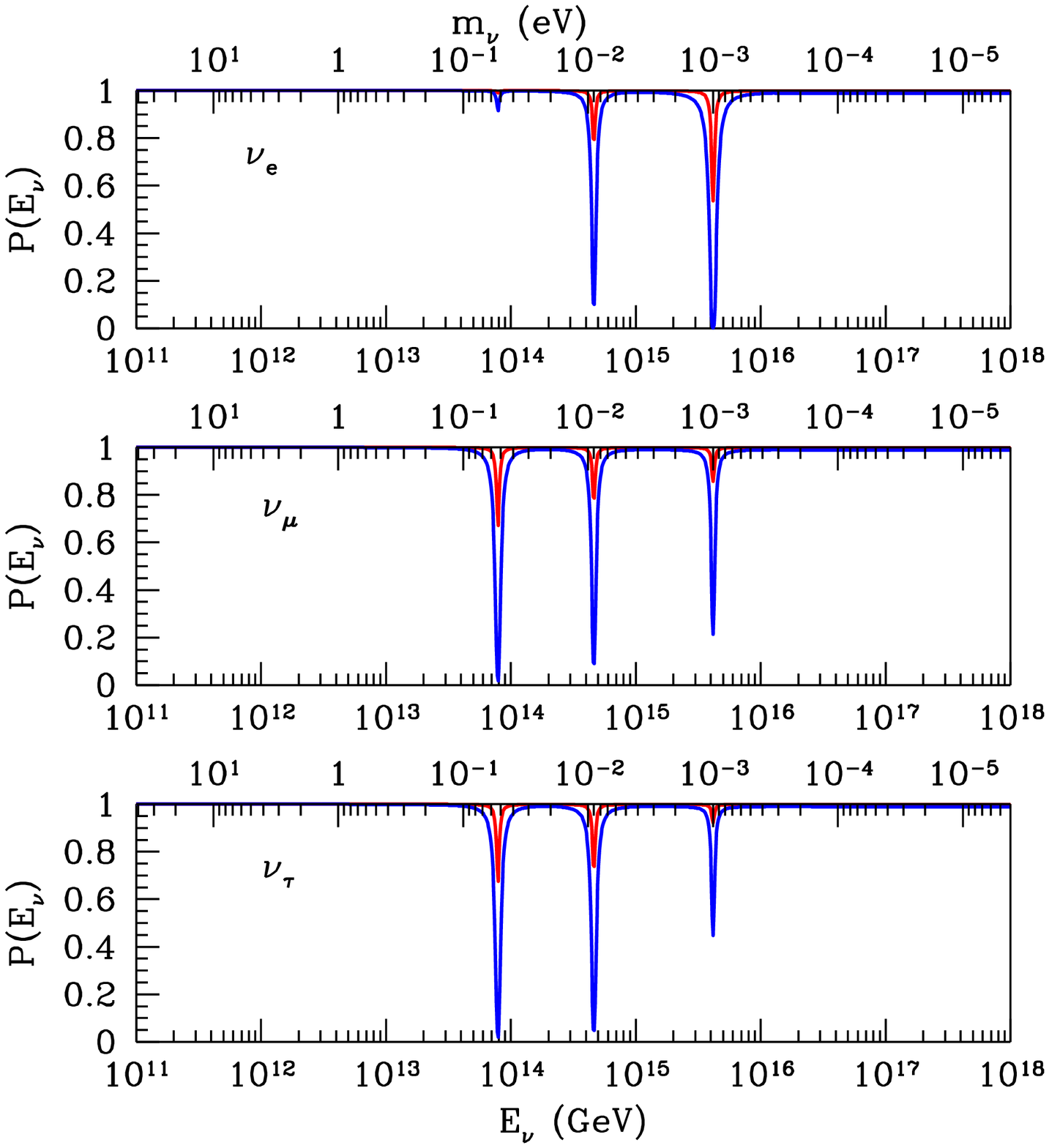}%{figs/maj_medium_normalxx.eps} 
\caption{Survival probabilities for $\nu_e$, $\nu_\mu$, and
$\nu_\tau$ in the ideal experimental scenario, for a normal hierarchy 
with $m_{\ell} = 10^{-5}\ev$ (left panel) or $m_{\ell} = 10^{-3}\ev$ 
(right panel).  The red and blue curves correspond to $L = 10^{4}\hbox{ and 
}10^{5}\mpc$.}
\label{fig:5normal}
\end{figure*}

We illustrate this prospect in Figure~\ref{fig:5normal}, where we plot 
the survival probabilities for $\nu_{e}, \nu_{\mu}, \nu_{\tau}$ over 
the relevant range of neutrino energies, for $m_{\ell} = 10^{-5}$ and 
$10^{-3}\ev$. We imagine that the survival probability can be 
estimated reliably by fitting the shape of the energy spectrum away 
from the dips, in much the same spirit as backgrounds are estimated 
in experiments that observe peaks.
The amount of attenuation is governed by the length of 
the column and by the flavor composition of the neutrino mass 
eigenstates, which was indicated in Figure~\ref{fig:hierarchy2}. 

\begin{figure*}
    \includegraphics[width=8.75cm]{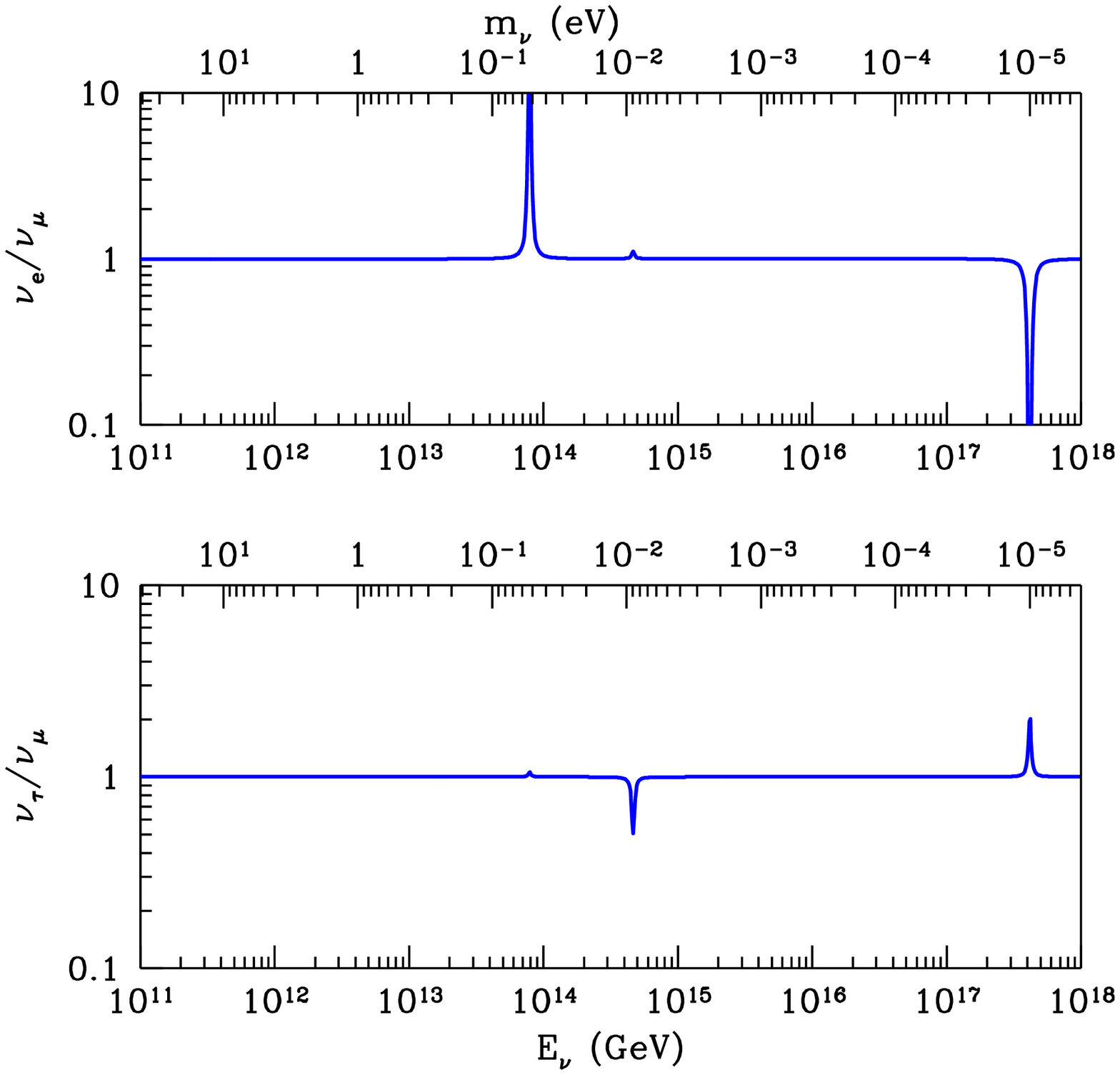}%{figs/ratios_testn.eps}  
    \includegraphics[width=8.75cm]{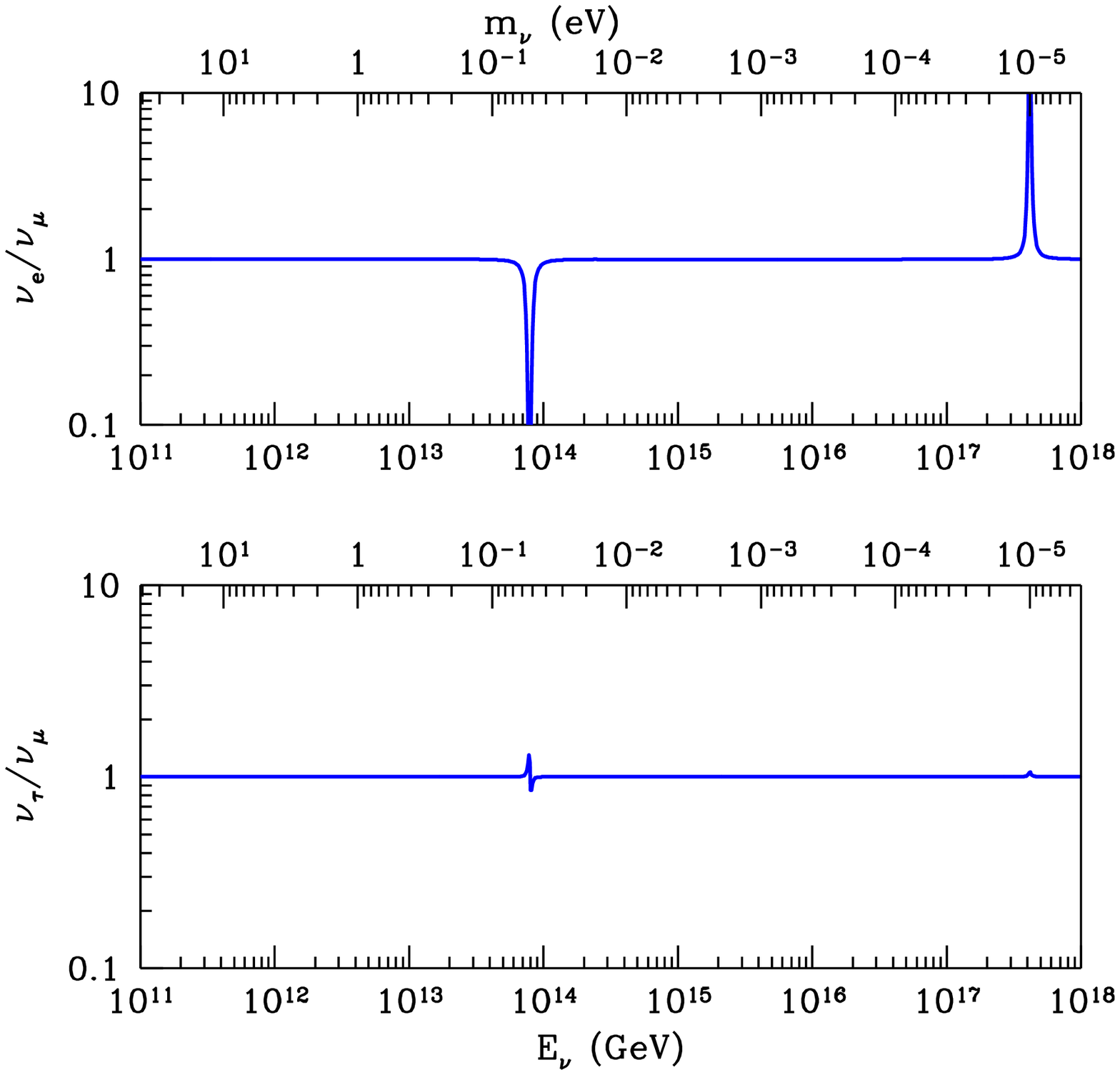}%{figs/ratios_itestn.eps}  
    \caption{Flux ratios $\nu_{e}/\nu_{\mu}$ and 
    $\nu_{\tau}/\nu_{\mu}$ at Earth, for normal (left panel) and 
    inverted (right panel) mass hierarchies with $m_{\ell} = 
    10^{-5}\ev$. The column length is $L = 10^{5}\mpc$.}
    \label{fig:rats}
\end{figure*}

As in 
other aspects of ultrahigh-energy neutrino studies, flavor tagging at 
the detector greatly enriches the scientific 
program~\cite{Beacom:2003nh,Barenboim:2003jm}. 
With flavor tagging, 
simply looking for variations in the relative fluxes 
$\nu_{e}/\hbox{all}$, $\nu_{\mu}/\hbox{all}$, $\nu_{\tau}/\hbox{all}$
with energy may be highly revealing. We plot in Figure~\ref{fig:rats} 
the flux ratios $\nu_{e}/\nu_{\mu}$ and $\nu_{\tau}/\nu_{\mu}$ for 
normal and inverted hierarchies with $m_{\ell}= 
10^{-5}\ev$.\footnote{We assume the natural mix,  
$\nu_{e}:\nu_{\mu}:\nu_{\tau}::1:1:1$.} The 
flavor ratios are a powerful discriminant between the normal and 
inverted hierarchies, because the $\nu_{e}/\nu_{\mu}$ ratio is an 
excellent diagnostic for the mass eigenstates---even if only the 
lowest-energy (highest-mass) dip is visible. In the normal hierarchy, 
muon neutrinos are depleted with respect to electron neutrinos, while 
for the inverted hierarchy, $\nu_{e}$ is depleted with respect to 
$\nu_{\mu}$. This distinction can be essential 
in case the incident neutrino flux runs out before all of the 
absorption dips are revealed. 

\begin{figure*}
    \includegraphics[width=8.75cm]{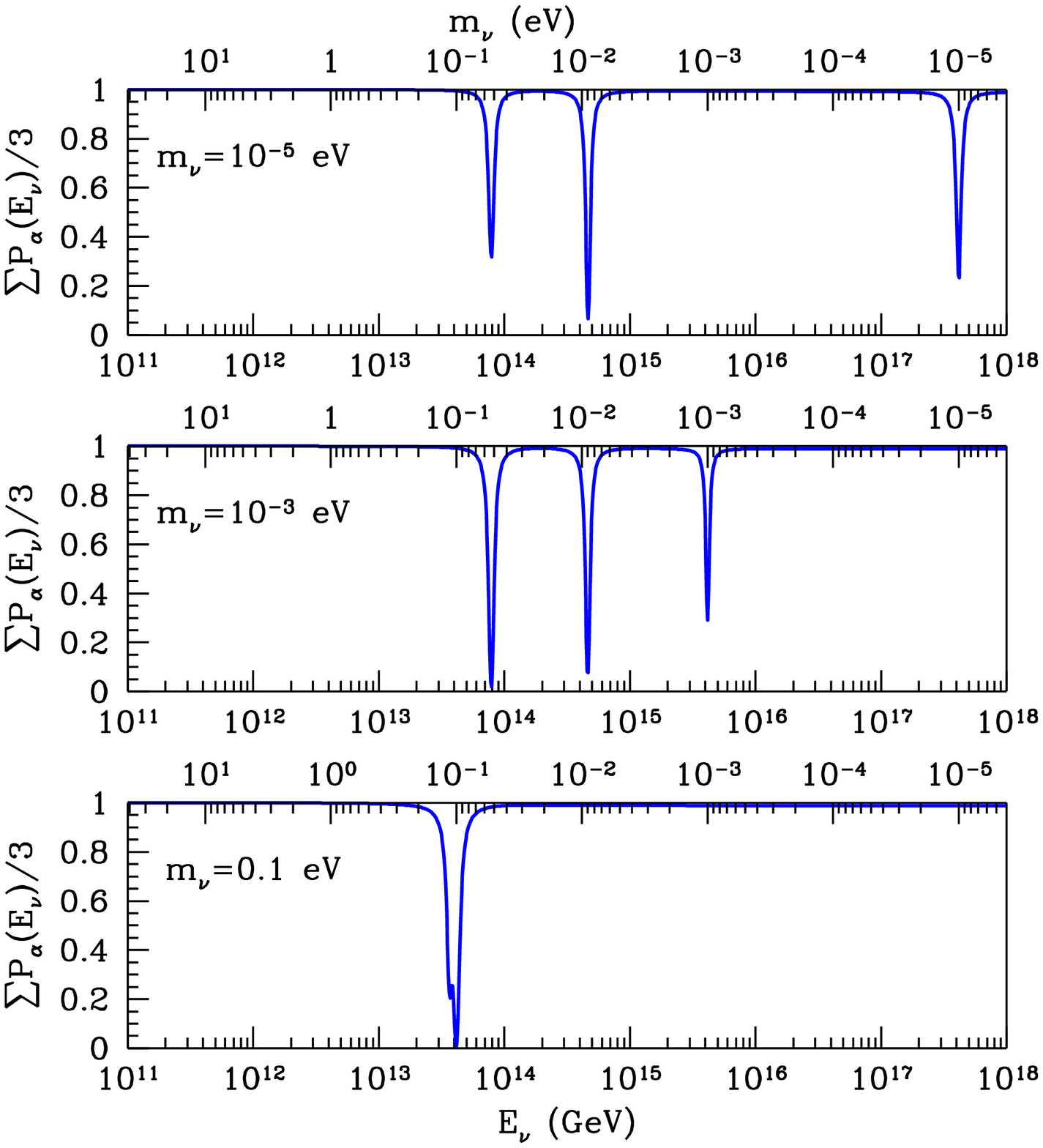}%{figs/averaged_normal.eps}  
    \includegraphics[width=8.75cm]{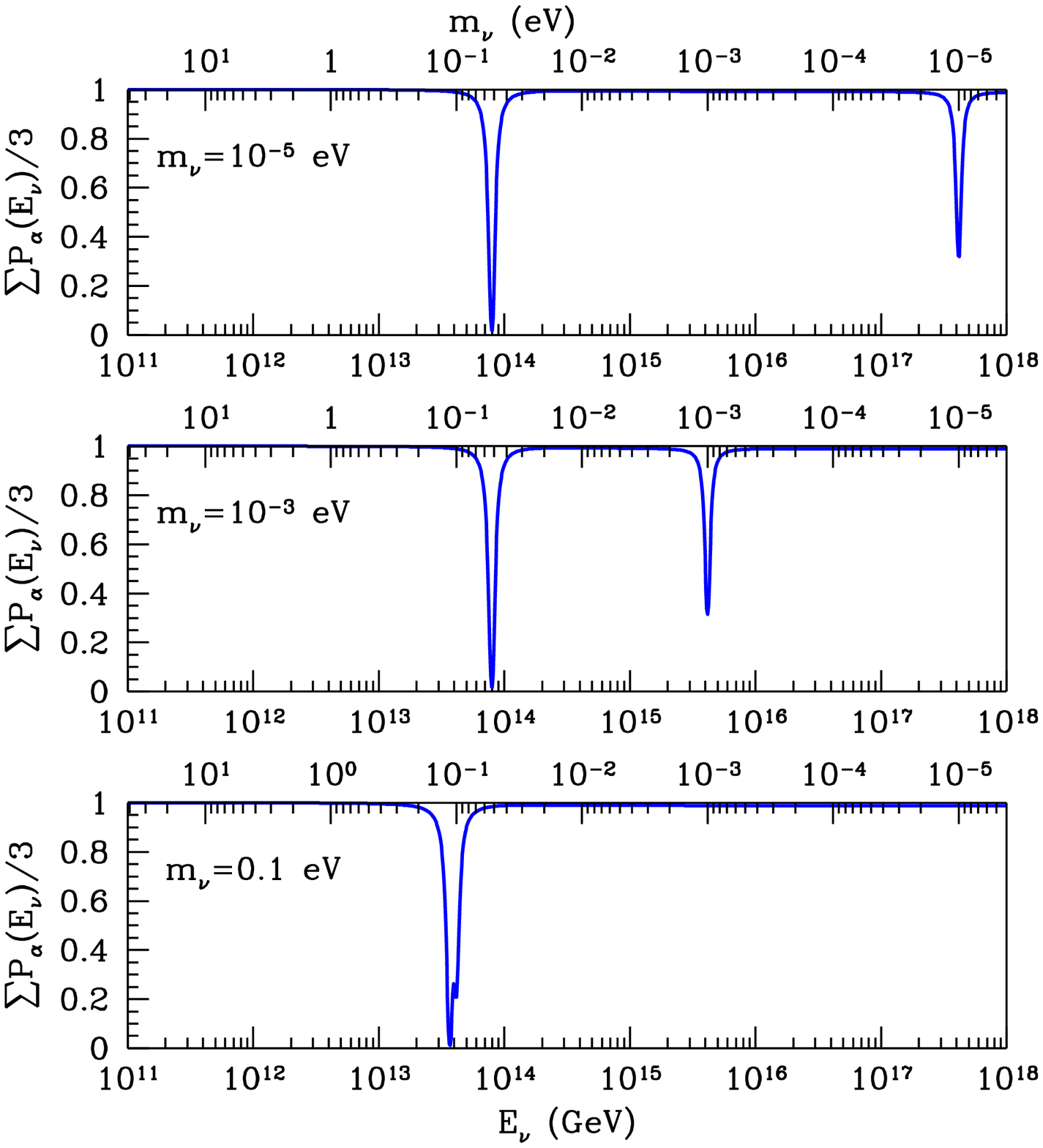}%{figs/averaged_inverted.eps} 
    \caption{Survival probabilities summed over neutrino flavors in the ideal
    experimental scenario, for  $m_{\ell} = 10^{-5}, 10^{-3}, 
    10^{-1}\ev$ (top, middle, bottom) in the normal (left panel) and 
    inverted (right panel) hierarchy. The length of the relic 
    neutrino column is $L = 10^{5}\mpc$. }
    \label{fig:in}
\end{figure*}

If the neutrino masses exhibit a distinct hierarchy---as is the case 
for $m_{\ell} \lesssim 10^{-3}\ev$---then it is possible, with 
adequate flux, to distinguish the normal and inverted hierarchies, 
even without flavor identification.  
In Figure~\ref{fig:in} we show the all-flavor survival probabilities,
for a representative relic neutrino column length of $10^{5}\mpc$,
for the normal and inverted spectra with $m_{\ell} = 
10^{-5}, 10^{-3}, 10^{-1}\ev$. For the normal hierarchy, we observe 
three, three, and one absorption lines; for the inverted hierarchy, 
two, two, and one dips.

\subsection{Majorana or Dirac Relics? \label{subsec:MDR}}
We reviewed in \S\ref{subsec:DorM} the implications of Majorana 
\textit{versus} Dirac character for the annihilation cross sections, 
and showed in Figure~\ref{fig:5int} how the distinction depends on neutrino 
mass and temperature. Absent a calibration of the length of the 
relic-neutrino column, we do not see how to exploit the factor-of-two 
difference between Dirac and Majorana cross sections in the static 
limit to distinguish the two cases. Even in the idealized (zero 
neutrino temperature) situation we treat in this Section, it is hard 
to imagine inferring the neutrino character from the depth of a single 
absorption line. If the lightest neutrino were  relativistic and the 
heaviest neutrino nonrelativistic, it is remotely possible to imagine 
calibrating the effective column density on the highest-energy 
(lowest neutrino mass) dip, and then noticing a smaller apparent depth for the 
lowest-energy (highest neutrino mass) dip, in the case of Dirac relics. In the more realistic 
circumstances we shall describe below, with integration over redshift 
and attention to the thermal motion of the relics, we do not see how 
to distinguish Dirac from Majorana relics.

\subsection{Unstable Relics \label{subsec:reldecay}}
To this point, we have considered neutrinos as stable particles. 
``Invisible'' decays, such as the decay of a heavy neutrino into a 
lighter neutrino plus a very light---or massless--(pseudo)scalar 
boson such as the majoron~\cite{Valle:1983ua,Gelmini:1983ea} are not 
very well constrained by observations~\cite{Beacom:2002vi}.\footnote{A majoron too massive 
to serve as a neutrino decay product can nevertheless have important 
consequences for cosmology, including deviations from the standard 
expectations for the radiation energy density and changes in the 
positions of peaks in the cosmic microwave background power 
spectrum~\cite{Chacko:2003dt}.} 
If \textsf{CPT} invariance holds, SN1987a data set an upper limit on the lifetime of
the lightest neutrino of $\tau_{\ell}/m_{\ell} \gtrsim 10^{5}$ s/eV. 
Observations of solar neutrinos lead to $\tau_{2}/m_{2}\gtrsim 10^{-4}$ s/eV. Finally, if the neutrino
spectrum is normal, the data on  atmospheric neutrinos coming upward 
through the Earth, imply $\tau_{3}/m_{3}\gtrsim 10^{-10}$ s/eV. All of 
these bounds leave open the possibility that heavy relic neutrinos 
might have long since decayed away, so that the current Universe is 
filled not with the primordial neutrinos, but with their decay 
products.\footnote{As we noted in \S\ref{subsec:numass}, majoron interactions
that would mediate neutrino decays can also mediate (co)annihilations
that might lead to a vanishing relic neutrino density today.}

\begin{figure}
    \includegraphics[width=8.75cm]{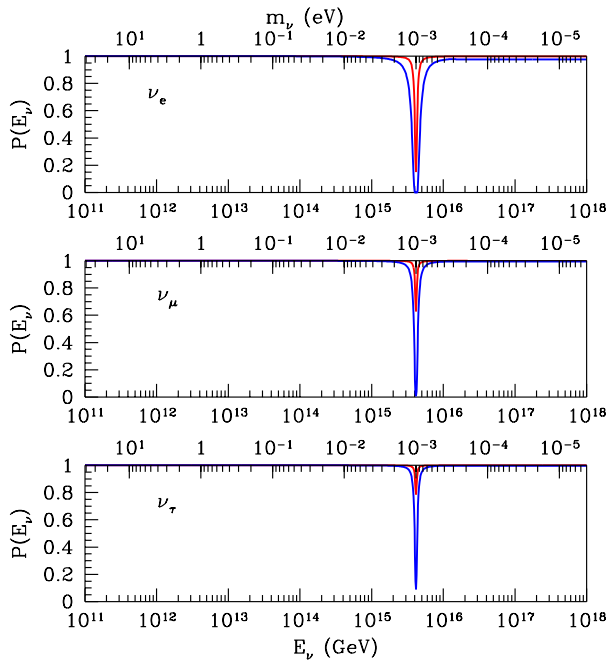}%{figs/maj_decays.eps} %maj_medium_normal_d23.eps}
\caption{Survival probabilities for $\nu_{e}, \nu_{\mu}$, and $\nu_{\tau}$ in 
the idealized experimental scenario, for the normal hierarchy in which 
all the heavy relics ($\nu_{2}$ and $\nu_{3}$) have 
decayed. The mass of the surviving (lightest) relic is  $m_{\ell} = 
m_{1} = 10^{-3}\ev$.  The curves correspond to column lengths $L = 
10^{4}\mpc$ (red) and $10^{5}\mpc$ (blue). }
\label{fig:3nd}
\end{figure}

Let us consider, for illustration, the simplest case of unstable 
neutrinos, in which the two heavier neutrinos have decayed into the 
lightest neutrino plus invisible products. In that event, the density 
of the heavier neutrinos is now zero, but the density of the lightest 
neutrino is three times $n_{\nu0}$. We further assume that the 
neutrinos in our beam have not decayed in flight.
We show in Figure~\ref{fig:3nd} the neutrino survival probability for 
the idealized experiment, for the normal hierarchy with
$m_{\ell} = m_1=10^{-3}\ev$. In contrast to what we see  in the right-hand 
panel of Figure~\ref{fig:5normal},  there is only one
absorption dip located at the resonant energy that corresponds to the
lightest neutrino---the surviving relic.  The dip is more pronounced for 
fixed column density than in the standard picture, reflecting the 
threefold increase in the population of 
$\nu_{\ell}$.\footnote{We suppose that all the $\nu_{2}$ and 
$\nu_{3}$ relics have decayed to $\nu_{1}$, and neglect cosmic expansion.}

If only the lightest neutrino survives as a relic, the absorption line 
is necessarily at a higher energy than in the standard case of three 
surviving flavors. In the extreme case that $m_{\ell} = 10^{-5}\ev$, 
the single absorption line would occur at $E_{\nu} \approx 
10^{18}\gev$, so the demands for adequate flux at the highest energies 
are very great.

We note that, with flavor identification, even the single absorption 
line could signal the normal or inverted hierarchy. In the normal 
hierarchy, the $\nu_{e}$ flux is strongly absorbed, whereas in the 
inverted hierarchy it is attenuated only slightly. Without flavor 
tagging, there is no prospect of unmasking the hierarchy, if the 
heavy neutrinos are absent as relics.

\section{Neutrino Absorption Lines in an Evolving Universe\label{sec:reds}}
Even if we had the means to prepare neutrino beams of the requisite
energy, the Universe would not hold still for us to perform the
idealized measurements described in \S\ref{sec:toy}.  The time required
to traverse one interaction length for $\nu\bar{\nu} \to Z^{0}$
annihilation on the relic background in the current Universe ($1.2
\times 10^{4}\mpc = 39\hbox{ Gly}$) exceeds the age of the
Universe,\footnote{Some $13\hbox{ Gly}$ according to the current best
estimates.} not to mention the human attention span.  If we are ever to
detect the attenuation of neutrinos on the relic-neutrino background,
we shall have to make use of astrophysical or cosmological neutrinos
sources traversing the Universe over cosmic time scales.  The expansion
of the Universe over the propagation time of the neutrinos entails two
important effects: the evolution of relic-neutrino density and the
redshift of the incident neutrino energy.  We consider both effects in
this Section.

\subsection{The Influence of Expansion and Redshift 
\label{subsec:expred}}
We displayed the redshift dependence of the relic-neutrino number 
density in Figure~\ref{fig:nnu} in  \S\ref{subsec:char}. There we also introduced the {column
density} defined in Eq.~(\ref{eqn:fermi3}) and plotted in 
Figure~\ref{fig:nnueff}, which characterizes the neutrino number
density per unit of redshift.

As we look back in time, the present energy $E_{\nu0}$ of an 
ultrahigh-energy neutrino increases with redshift as $E_{\nu}(z) = 
(1+z)E_{\nu0}$. This scaling is easily understood: the energy of an 
ultrarelativistic particle is inversely proportional to its 
wavelength, which stretches out in an expanding Universe as 
$(1+z)^{-1}$. 

We evaluate the survival probabilities as
\begin{equation}
\mathcal{P} (E_{\nu0}) =\exp\left[{-\int^{z}_{0} 
\!\!\!dz\,\sigma_{\nu\nu}((1+z)E_{\nu0})\, \frac{d\bar{n}_{\nu_i}
(z)}{dz} }\right],
\label{eqn:prob1}
\end{equation}
where $\sigma_{\nu\nu}$ is the interaction cross section of 
\S\ref{subsec:abs} weighted with the appropriate
mass-flavor mixing factor  and $d\bar{n}_{\nu_i}/dz$ is the {column
density}, Eq.~(\ref{eqn:fermi3}).\footnote{We choose not to smear the 
cross section over a neutrino spectrum, to make our analysis as 
generally informative as possible.  Equation~(\ref{eqn:prob1}) would 
result for any power-law neutrino 
spectrum.} The attenuation process is thus
summed over redshift, with the simplifying assumption that the 
neutrino sources in question switched on at a time in the past 
characterized by a single redshift, which we take to be $z_{\nu} 
\lesssim 20$.  

Plausible acceleration sources of ultrahigh-energy 
neutrinos, including active galactic nuclei and gamma-ray bursters, 
lie at redshifts of a few~\cite{Semikoz:2003wv}. 
Extremely energetic neutrinos might have been generated in the decays of super-heavy 
particles or topological 
defects~\cite{Semikoz:2003wv,Bhattacharjee:1991zm,Hill:1986mn,Aloisio:2004ap} long before the 
first stars began to shine. Their interaction with the relic neutrino 
background could reach as far back as the epoch of light neutrino decoupling, 
$z \sim \mathcal{O}(10^{10})$~\cite{Gondolo:1991rn}.

\begin{figure*}%[t]
    \includegraphics[width=8.75cm]{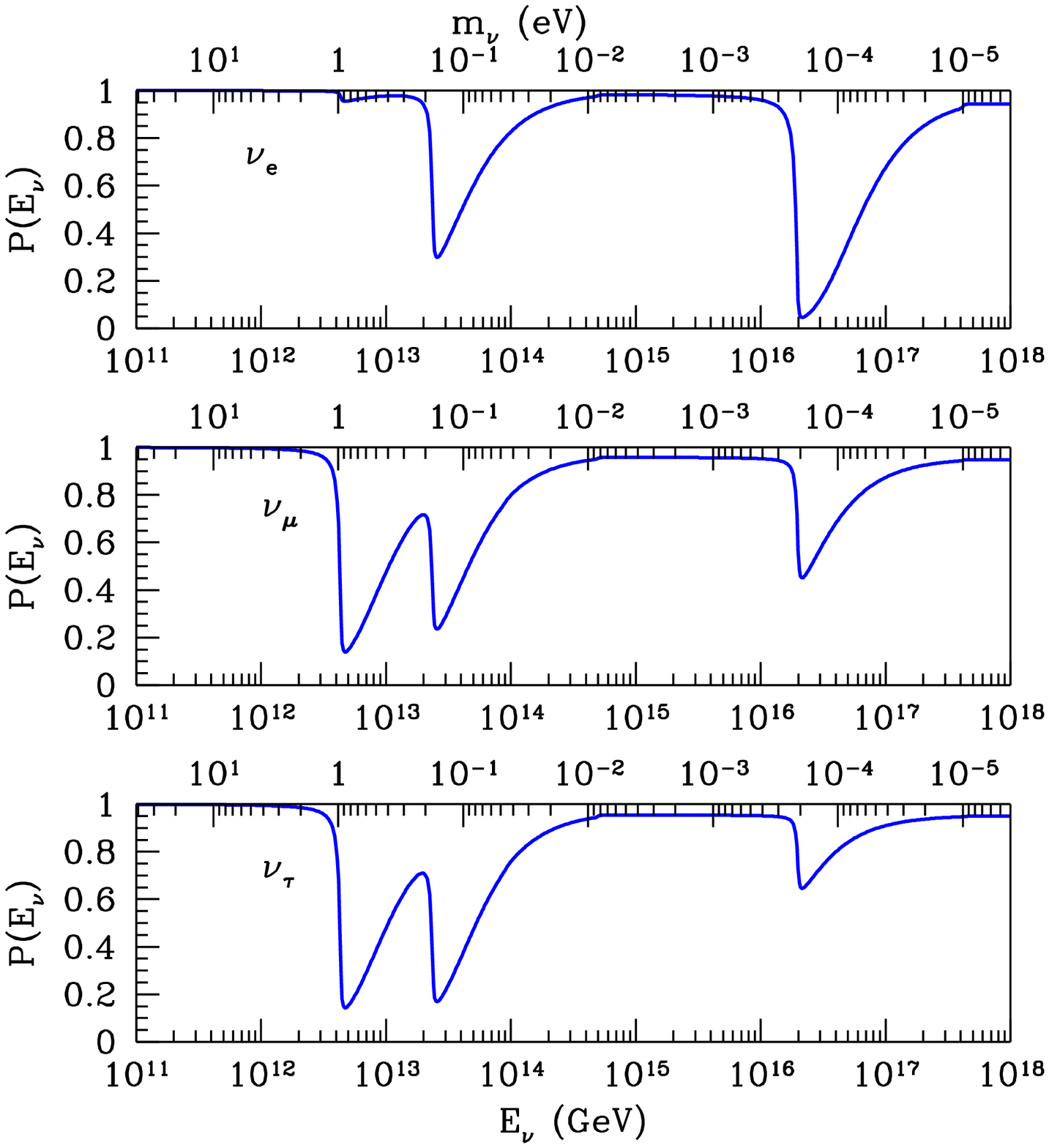}%{figs/maj_light_normalz.eps} 
    \includegraphics[width=8.75cm]{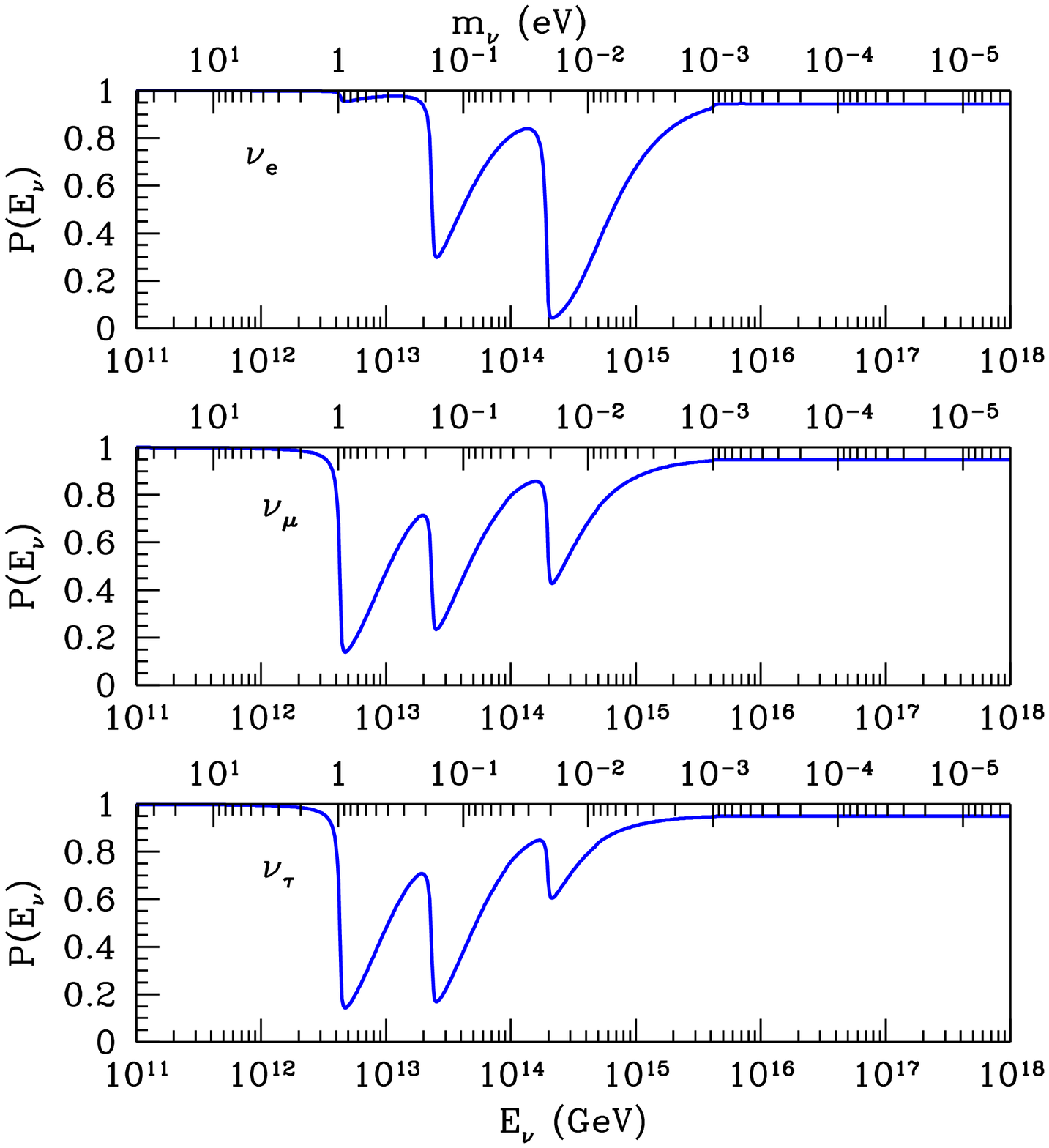}%{figs/maj_medium_normalz.eps} 
\caption{Survival probabilities for $\nu_e$, $\nu_\mu$ and
$\nu_\tau$ after an integration back to redshift $z = 20$, for a normal
hierarchy with $m_{\ell}=10^{-5}$ eV (left panel) or $m_{\ell}=10^{-3}$ eV
(right panel), as a function of the energy of the UHE neutrino. 
The scale at the top shows the neutrino mass 
defined as $m_{\nu} = M_{Z}^{2}/2E_{\nu}$ that would be inferred if 
neutrino energies were not redshifted.}
\label{fig:5znormal}
\end{figure*}

We show in Figure~\ref{fig:5znormal} the survival probabilities for 
$\nu_{e}, \nu_{\mu}, \nu_{\tau}$ integrated from the present back to 
redshift $z = 20$, for $m_{\ell} = 10^{-5}\hbox{ and }10^{-3}\ev$. 
Comparing with the analogous idealized ``toy experiment'' whose 
outcome was depicted in Figure~\ref{fig:5normal}, we find that the 
absorption dips are distorted by the redshift of neutrino energies. 
Neutrino-antineutrino annihilation into $Z^{0}$ at redshift $z$ 
implies a depletion of neutrino flux now at an energy
\begin{equation}
    E_{\nu0}^{Z \mathrm{res:}z} = \frac{M_{Z}^{2}}{2m_{\nu_{i}}(1+z)}\;.
    \label{eq:zscale}
\end{equation}
The correspondence between neutrino mass and energy of an absorption 
line is compromised. No longer does the un-redshifted relation
$m_{\nu} = M_{Z}^{2}/2E_{\nu}^{Z \mathrm{res}}$ allow  the neutrino mass 
to be inferred reliably. In these examples, that simple relation would 
overestimate neutrino masses by about an order of magnitude. In 
compensation, moving the dips to lower energies may be an important 
advantage for an experiment that is almost sure to be flux-limited.

If terrestrial experiments determine the neutrino masses---or if 
other considerations place increasingly stringent bounds on the 
neutrino masses---then the location of the neutrino absorption lines 
might provide new information about the neutrino sources. If, 
for example, the location of the dip lies much more than one order of 
magnitude below the anticipated energy, that would be strong evidence 
that  sources of extremely high energy neutrinos existed before the 
stars---a surprising and important conclusion.

\begin{figure}%[h]
\includegraphics[width=9.25cm]{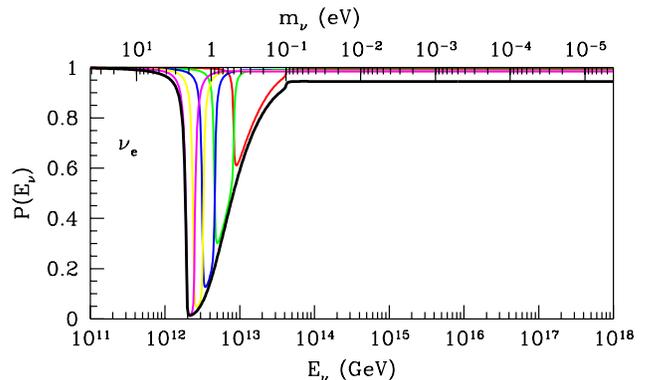}%{figs/zbin.eps} 
\vspace*{-12pt}
\caption{Electron-neutrino survival probabilities integrated back to 
$z = 20$ for a
degenerate neutrino spectrum with $m_\ell=0.1\ev$.  The thick black curve 
shows the distorted absorption line after integration back to 
redshift $z = 20$, and the colored lines (from right to left) show the contributions from 
bins in redshift: red, $0 \le z \le 4$; green, $4 \le z \le 8$; 
blue, $8 \le z \le 12$; yellow, $12 \le z \le 16$; magenta, $16 \le z 
\le 20$.}
\label{fig:explz}
\end{figure}

The morphology of the distortion is 
illustrated in Figure~\ref{fig:explz}, where we plot the $\nu_{e}$ 
survival probability binned in redshift for the simple (one-dip) case of a degenerate neutrino 
spectrum with $m_{\ell} = 0.1\ev$. Evidently binning in redshift is 
not an observational possibility, but we can impose it on our 
calculation to deconstruct the origin of the shift and broadening of the 
absorption line. As expected, we observe that the downward shift in the 
attenuated energy in the present Universe grows with the redshift at 
which the annihilation occurred. Moreover, the depth of the absorption 
dips is greater at higher redshifts, because of the increased column 
density of relic neutrinos, Eq.~(\ref{eqn:fermi3}).

\begin{figure}%[t]
    \includegraphics[width=8.75cm]{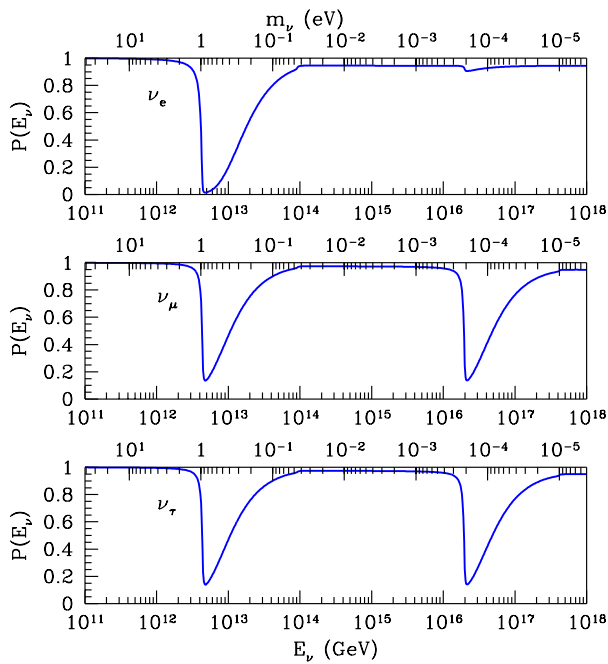}%{figs/maj_light_invertedz.eps} 
\caption{Survival probabilities for $\nu_e$, $\nu_\mu$ and
$\nu_\tau$ after an integration back to redshift $z = 20$, for an 
inverted
hierarchy with $m_{\ell}=10^{-5}$ eV, as a function of the energy of the UHE neutrino. 
The scale at the top shows the neutrino mass 
defined as $m_{\nu} = M_{Z}^{2}/2E_{\nu}$ that would be inferred if 
neutrino energies were not redshifted.}
\label{fig:5zinvert}
\end{figure}

\begin{figure*}
    \includegraphics[width=8.75cm]{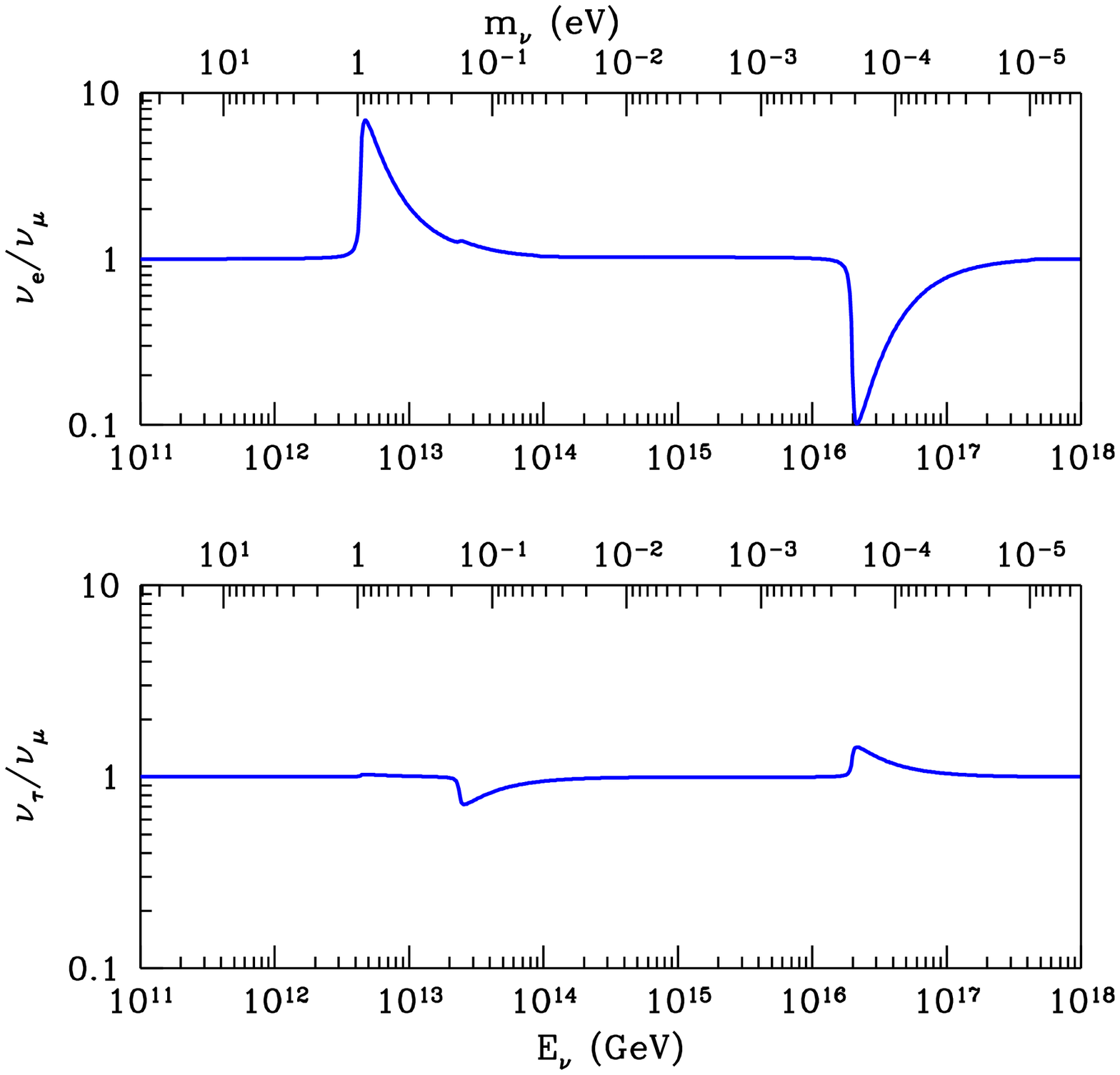}%{figs/ratios_testzn.eps}  
    \includegraphics[width=8.75cm]{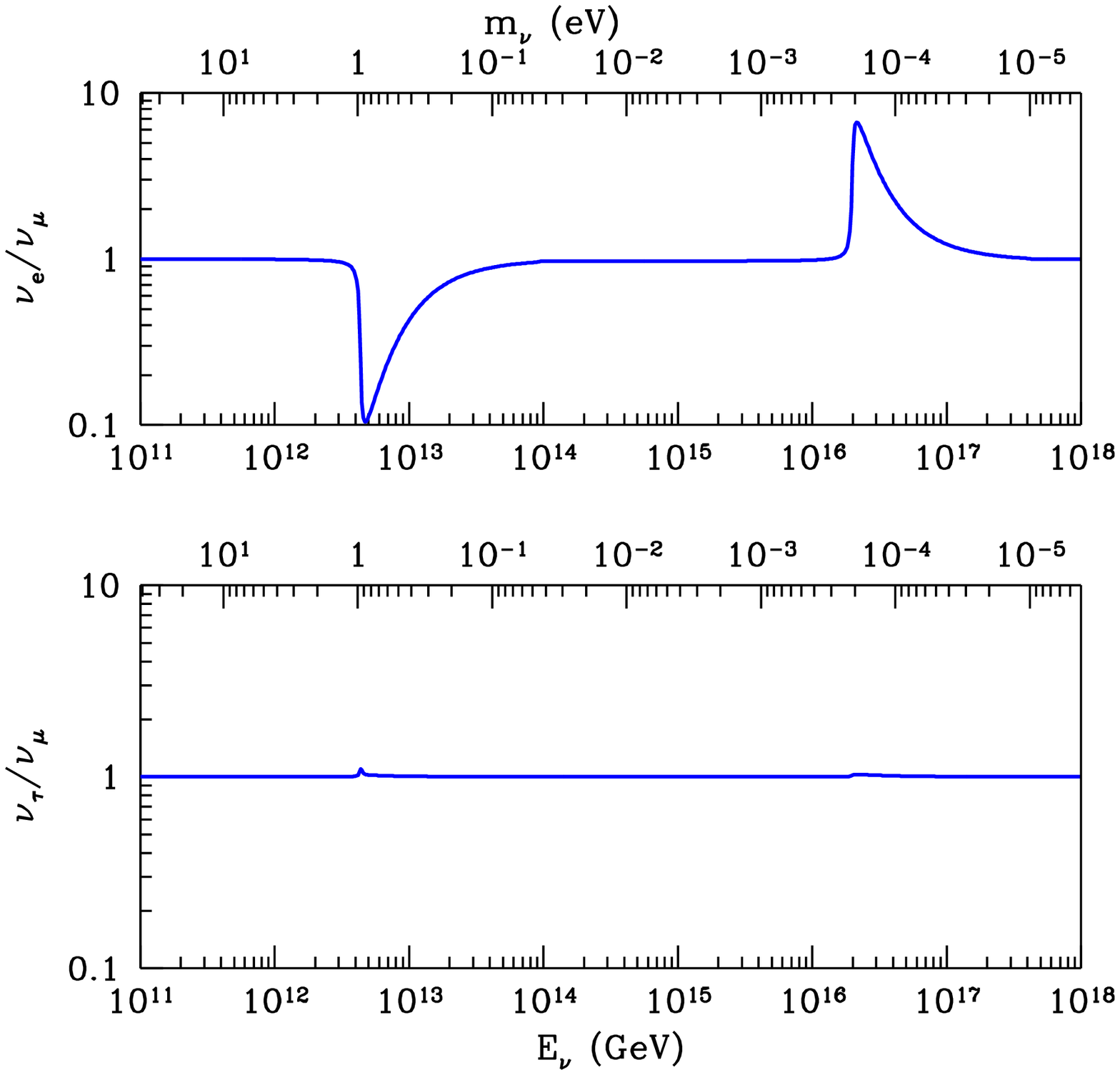}%{figs/ratios_testi.eps}  %ratios_itestzn.eps 
    \caption{Flux ratios $\nu_{e}/\nu_{\mu}$ and 
    $\nu_{\tau}/\nu_{\mu}$ at Earth, for normal (left panel) and 
    inverted (right panel) mass hierarchies with $m_{\ell} = 
    10^{-5}\ev$, after integration back to redshift $z=20$.
    The scale at the top shows the neutrino mass 
    defined as $m_{\nu} = M_{Z}^{2}/2E_{\nu}$ that would be inferred if 
    neutrino energies were not redshifted.}
    \label{fig:ratsz}
\end{figure*}

In an evolving Universe, the neutrino flux ratios remain an effective 
discriminant of the mass hierarchy. Figure~\ref{fig:5zinvert} shows 
the survival probabilities for an inverted spectrum with $m_{\ell} = 
10^{-5}\ev$.
We plot in Figure~\ref{fig:ratsz} 
the flux ratios $\nu_{e}/\nu_{\mu}$ and $\nu_{\tau}/\nu_{\mu}$ for 
normal and inverted hierarchies with $m_{\ell}= 
10^{-5}\ev$, integrated back from the present to redshift $z = 20$. 
Compared with the toy-model ratios plotted in Figure~\ref{fig:rats}, 
the features are broadened and displaced toward lower energies, but 
the essential message is unchanged. In the normal hierarchy, 
$\nu_{e}/\nu_{\mu} > 1$ at the lowest-energy line, whereas in the 
inverted hierarchy  $\nu_{e}/\nu_{\mu} < 1$.

\subsection{Alternative Thermal Histories \label{subsec:althist}}
Since the redshift dependence of the relic-neutrino column density is
imprinted on the neutrino absorption lines, it might be possible to 
infer some information about the column density from the absorption 
lineshape. As we saw in the discussion surrounding 
Eqs.~(\ref{eq:linescale}) -- (\ref{eqn:fermi3}), each particular 
cosmology influences the relic-neutrino column density through the 
Hubble parameter (\ref{eq:Hubble}). 

We have adopted the $\Lambda$CDM 
model as our standard reference cosmology. Let us compare its 
implications for  neutrino absorption lines with those of its 
predecessor, the SCDM (for standard cold dark matter) model 
characterized by matter density $\Omega_{m} = 1$ and zero cosmological 
constant.

\begin{figure}
\includegraphics[width=8cm]{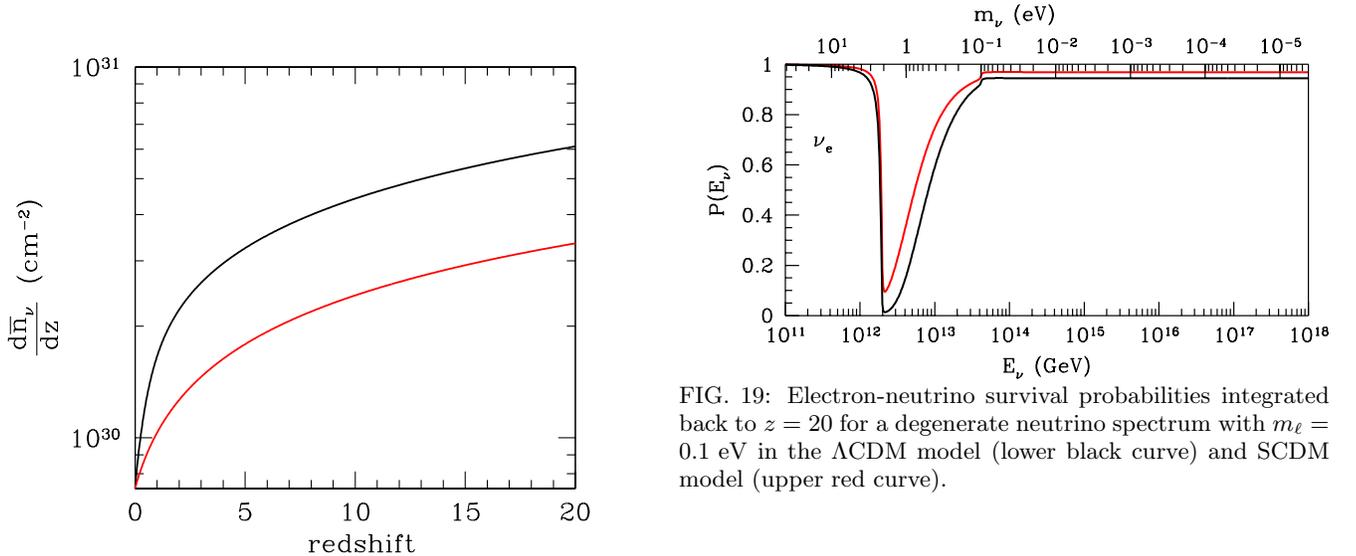}%{figs/eff_bis_2.eps} 
\caption{Relic neutrino column densities versus redshift for the
 $\Lambda$CDM model (upper black curve) and SCDM
model (lower red curve).}
\label{fig:effbis}
\end{figure}

\begin{figure}
    \includegraphics[width=9.25cm]{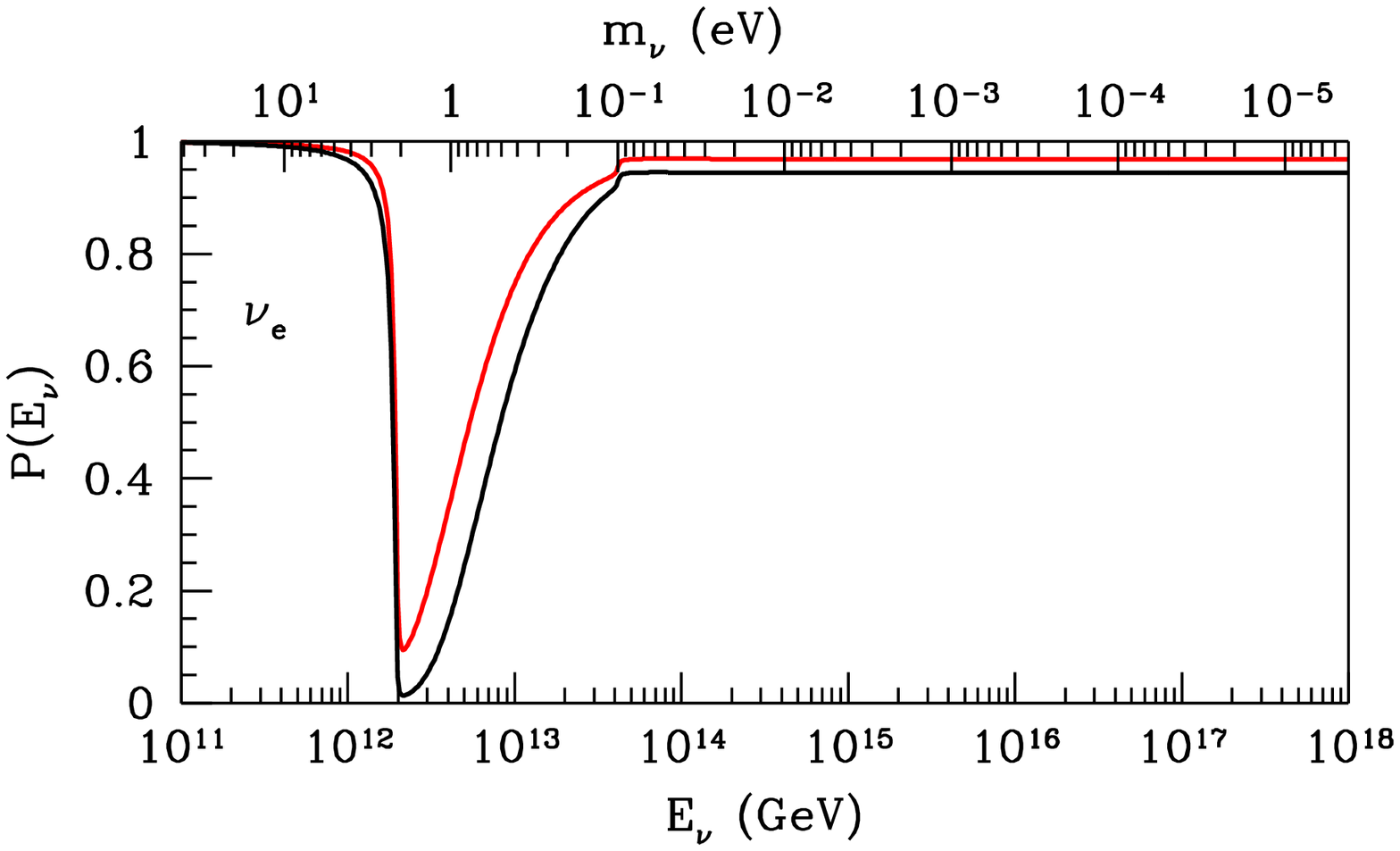}%{figs/history_2.eps} 
\vspace*{-12pt}
    \caption{Electron-neutrino survival probabilities integrated back to 
$z = 20$ for a
degenerate neutrino spectrum with $m_\ell=0.1\ev$
in the $\Lambda$CDM model (lower black curve) and SCDM
model (upper red curve).}
\label{fig:thermalhistory}
\end{figure}

In Figure~\ref{fig:effbis} we compare the relic neutrino
column densities for these two cosmological models. The column density 
is sytematically larger in the $\Lambda$CDM model, 
because flat-Universe models with a cosmological constant imply a larger physical 
volume associated with unit redshift than flat-Universe models 
dominated by matter. As Figure~\ref{fig:thermalhistory} shows, neutrino attenuation is more efficient in the 
$\Lambda$CDM model than in the SCDM model.
There, as in Figure~\ref{fig:explz}, we show our expectations for 
the $\nu_{e}$ survival probabilities integrated back to $z=20$, for a 
degenerate neutrino spectrum with $m_{\ell} = 0.1\ev$.

If we had a ``standard candle'' for neutrinos analogous to the Type 
Ia Supernovae for photons, we might imagine discriminating between 
these two cosmological models, because the difference we compute here 
is similar to what is revealed by the supernova luminosity distances. 
Just as distant supernovae appear fainter in a dark-energy dominated 
Universe, so are the extremely high energy neutrinos more thoroughly 
absorbed. 

The discrimination among cosmological models would be more acute if 
extremely high energy neutrino creation---and attenuation---were 
initiated at much earlier times, or much larger redshifts, than we 
consider in this example. If supermassive particle decays are an 
additional---early---neutrino source, then different thermal 
histories might yield dramatically different neutrino absorption lines.

\subsection{Event rates \label{subsec:events}}

Since experiments have just begun to explore the spectrum of
ultrahigh-energy neutrinos, quantifying the challenge of establishing
neutrino absorption lines is a very uncertain undertaking.  We refer to
\S III of Ref.~\cite{Eberle:2004ua} for a useful assessment.  We concur
that observing the cosmic-neutrino absorption lines with planned
detectors will require extended exposures, exotic sources of
ultrahigh-energy neutrinos, or both.  Here we briefly examine the case
of quasidegenerate neutrino masses, which minimizes the energy required
of the incoming neutrinos.

Neutrino absorption lines in the flux of cosmogenic neutrinos will
appear at neutrino energies $E_{\nu} \gtrsim 10^{12}\gev$, whereas
those in the flux of neutrinos generated much earlier from topological
defects may be redshifted to energies $E_{\nu} \lesssim 10^{10}\gev$.
Such energies lie in the domain of radio-Cherenkov and air-shower
detectors~\cite{Spiering:2003xm}.  IceCube is optimized for neutrinos
in the TeV-to-PeV range; it is sensitive to neutrinos of much higher
energies, though the ability to characterize the neutrino energy
diminishes progressively.  IceCube and possible
extensions~\cite{Halzen:2003fi} hold promise for identifying the flavor
of the ultrahigh-energy neutrinos.  We explore here the expected
sensitivities to the absorption lines in the future ANITA
mission~\cite{anita}.

Suppose now that super-high-energy neutrinos arise from the decay 
of superheavy particles created in the collapse (or annihilation) of 
topological defects. If these neutrinos are produced at very early 
times, detecting their annihilation on the relic-neutrino background 
may give us a rare glimpse of processes that prevailed in the very 
early universe. For example, the monopole-antimonopole boundstates 
dubbed ``monopolonium'' would have formed at the time of helium 
synthesis~\cite{hillm}. Neutrinos released in the decay of $X$ 
particles, as described in \S\ref{subsubsec:param}, have traversed 
many different epochs; the expansion of the universe during the 
neutrino's propagation will be imprinted on the neutrino spectrum. It 
remains to be seen whether we can learn to read the imprints in the 
absorption lines.

\begin{figure}[tb]
\includegraphics[width=8.5cm]{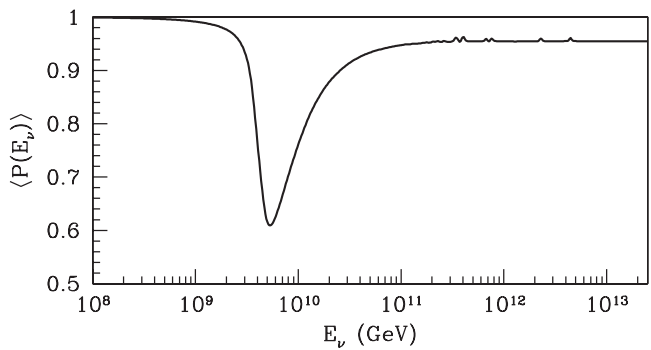}%{figs/topdef_ap.eps}  
%\vspace*{-12pt}
\caption{Survival probability for electron neutrinos created in the 
decay of a superheavy particle, $X$, with $M_X=10^{14}\gev$, 
for the case of $m_{\ell}=0.1$ eV.}
\label{fig:flux2}
\end{figure}

We show in Figure~\ref{fig:flux2}  the survival probability for 
neutrinos emitted in the decay of $10^{14}$-GeV $X$ particles,
following Ref.~\cite{Bhattacharjee:1991zm}.  For a degenerate neutrino
spectrum, the absorption line occurs at $E_{\nu} \lesssim 10^{10}\gev$,
well within the range to be covered by future experiments, including
ANITA and EUSO. In the most optimistic scenario for the TD neutrino
fluxes, a 5-$\sigma$ significance level ($164$ events) could be
achieved after a single fifteen-day flight of ANITA~\cite{anita}.  For
these redshifted, early-time neutrinos, we must face an additional
complication: the presence of relatively late-time neutrino sources,
such as the cosmogenic mechanism, that provide a pedestal under the
signal.  If the foreground sources are too bright, compared to the flux
from top-down sources, more events will be needed to establish the
absorption lines.

\section{Effect of the Relic-Neutrino Temperature \label{sec:temp}}
\begin{figure}
     \includegraphics[width=8.5cm]{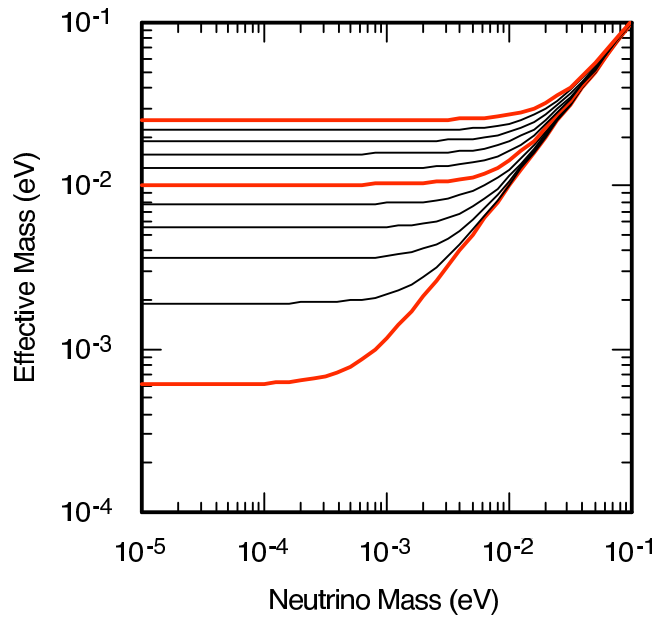}%                                             
 \caption{Effective masses of relic neutrinos, estimated by 
 $\varepsilon_{\nu} = [\langle p_{\nu}^{2}\rangle + 
 m_{\nu}^{2}]^{1/2}$, as functions of the neutrino mass $m_{\nu}$. 
 The mean-squared neutrino momentum in the current Universe is given by 
 Eq.~(\ref{eqn:p2mean}).
 From bottom to top, the curves refer to redshifts $z$ from 0 to 20, 
 in steps of 2.}
 \label{fig:effmass}
 \end{figure}

Relic neutrinos are moving targets, with their momentum distribution 
characterized in the present Universe by Eq.~(\ref{eq:nuFD}). 
The thermal motion of the neutrinos gives rise to a Fermi (momentum) 
smearing of the UHE-$\nu$--relic-$\nu$ cross section. The resonant 
incident-neutrino energy for a relic neutrino in motion is given 
 by
\begin{equation}
    {E}_{\nu}^{Z \mathrm{res}} = 
    \frac{M_{Z}^{2}}{2(\varepsilon_{\nu} - p_{\nu}\cos\theta)}\;,
    \label{eq:smres}
\end{equation}
where $p_{\nu}$ and $\varepsilon_{\nu}$ are the relic-neutrino 
momentum and energy. The angle $\theta$ characterizes the direction of 
the relic neutrino with respect to the line of flight of the incident 
UHE neutrino. Accordingly, the resonant energy will be displaced 
downward from $M_{Z}^{2}/2m_{\nu}$ to approximately
\begin{equation}
    \widetilde{E}_{\nu}^{Z \mathrm{res}} = 
    \frac{M_{Z}^{2}}{2\langle\varepsilon_{\nu}\rangle}\;,
    \label{eq:smresavg}
\end{equation}
where $\langle\varepsilon_{\nu}\rangle = [\langle p_{\nu}^{2}\rangle + 
 m_{\nu}^{2}]^{1/2}$ plays the role of an \textit{effective relic-neutrino 
 mass.} We plot the effective mass in Figure~\ref{fig:effmass}, for the 
interesting range of neutrino masses and redshifts.

The root-mean-square relic-neutrino momentum, which ranges from $6 
\times 10^{-4}\ev$ in the present Universe to $2.5 \times 10^{-2}\ev$ 
at redshift $z = 20$, thus serves as a rough 
lower bound on the effective neutrino mass. At a given redshift, the 
resonance peak for scattering 
from any neutrino with $m_{\nu} \lesssim \langle\varepsilon_{\nu}\rangle$ 
will be changed significantly.

\begin{turnpage}
\begin{figure*}
\includegraphics[width=4.5cm]{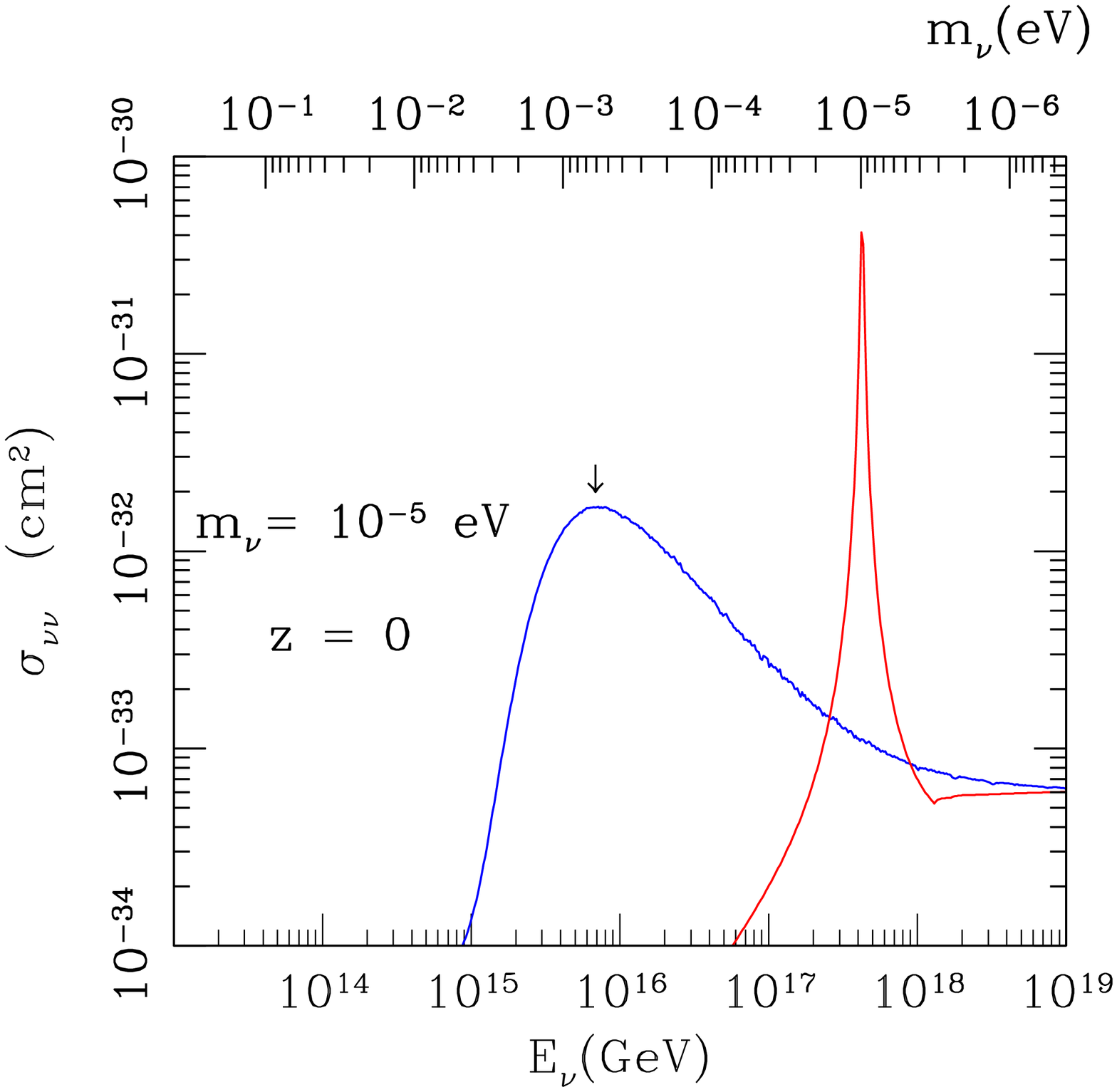}%
\includegraphics[width=4.5cm]{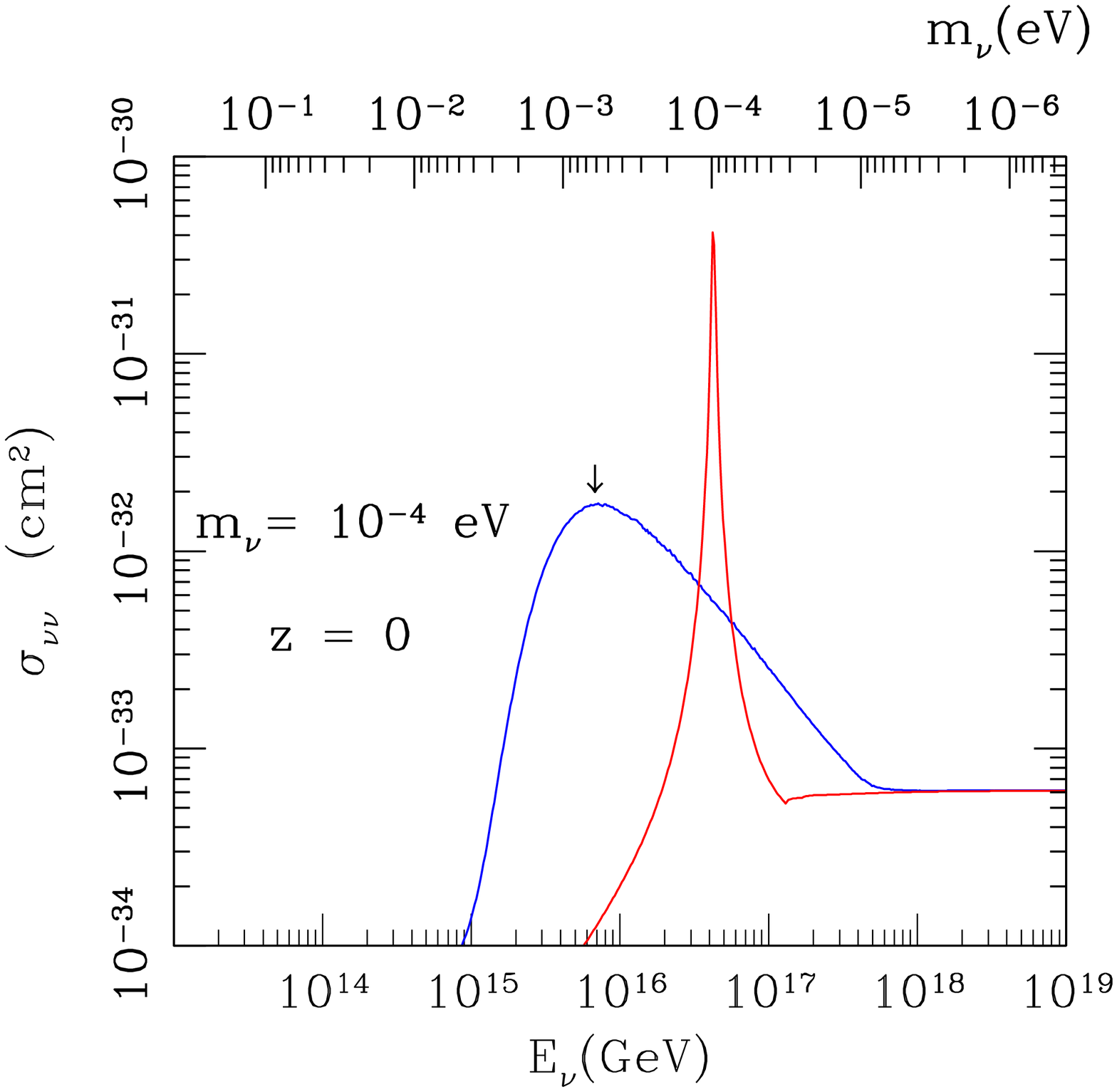}%
\includegraphics[width=4.5cm]{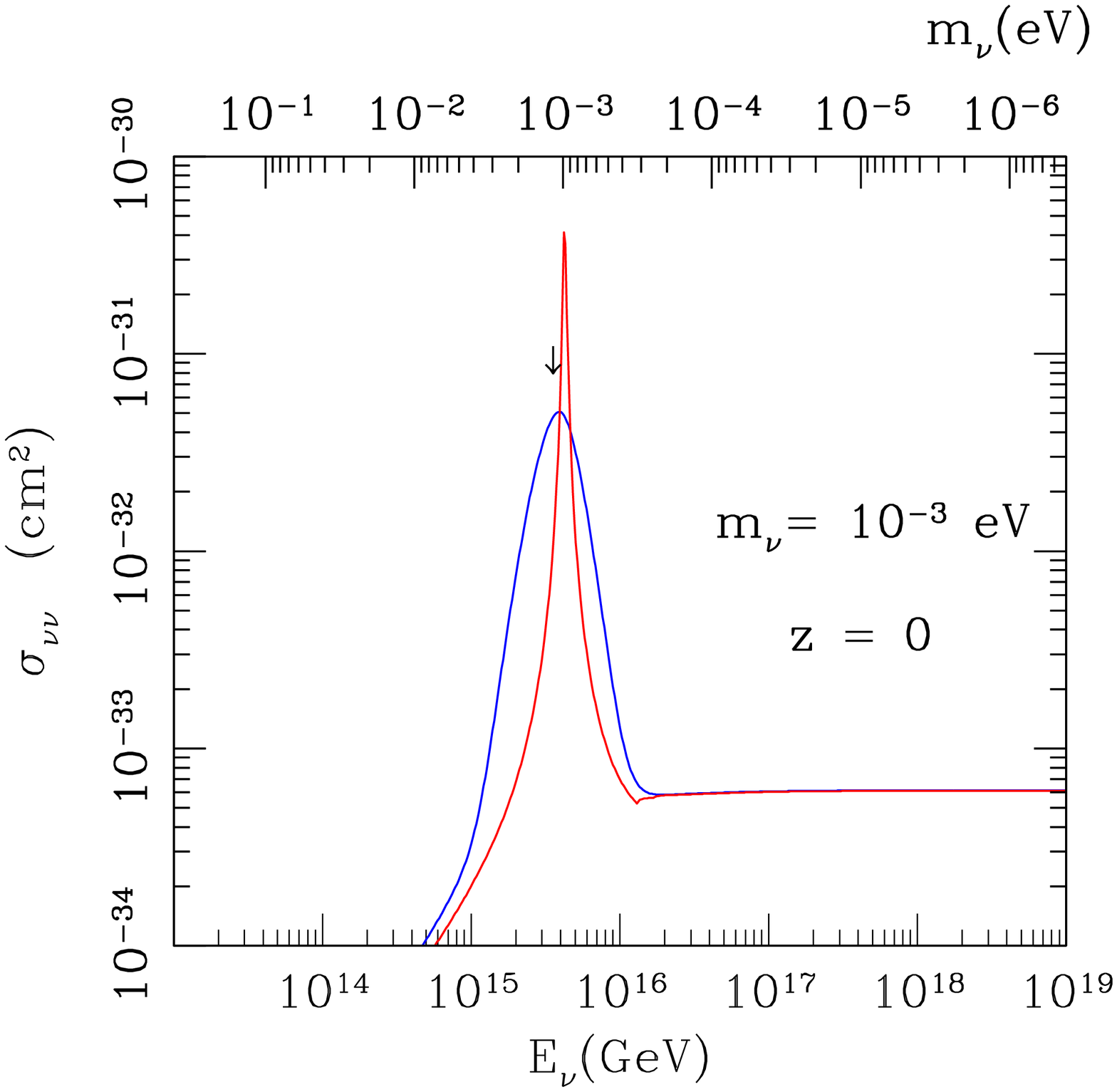}%
\includegraphics[width=4.5cm]{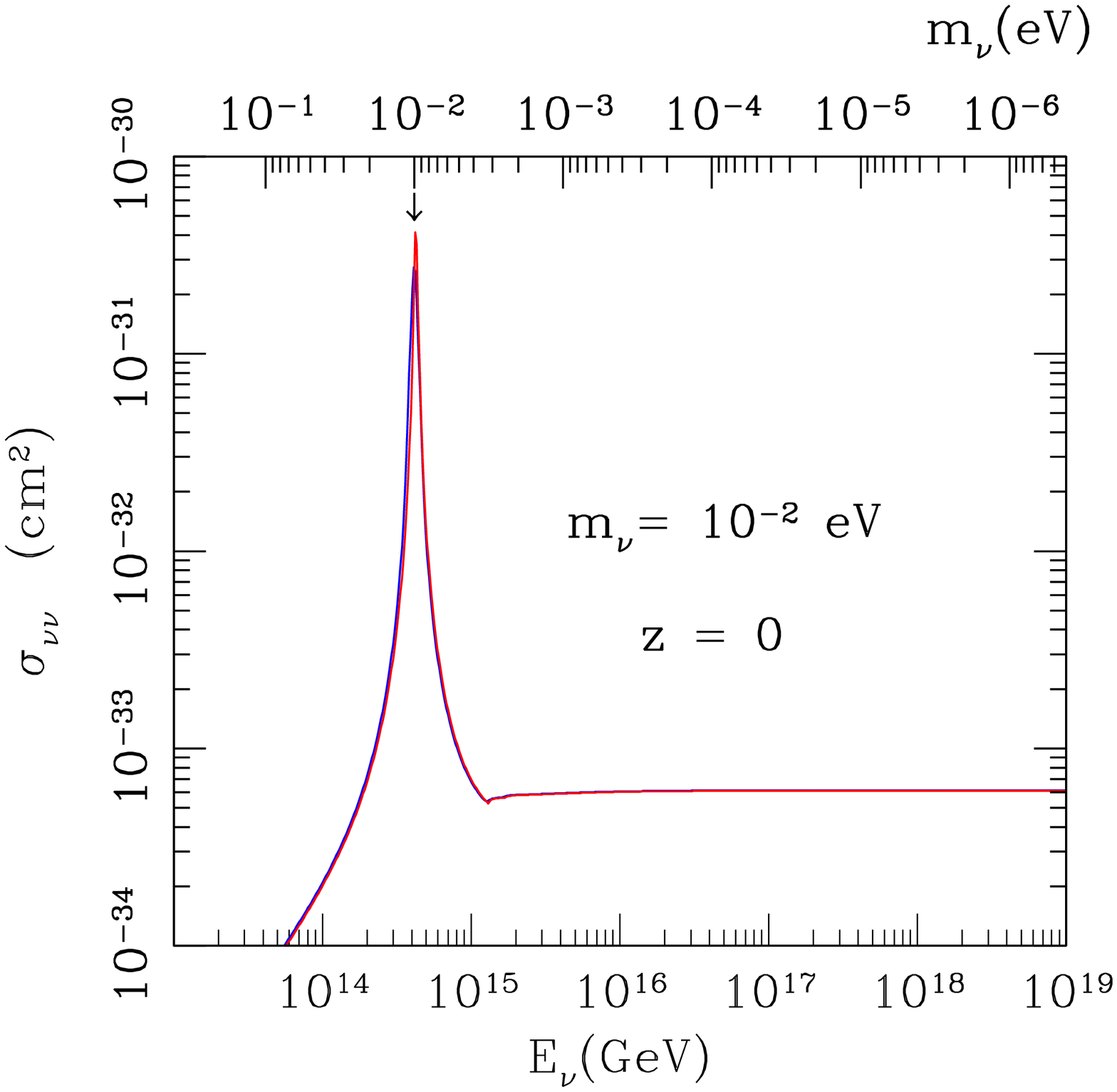}%
\includegraphics[width=4.5cm]{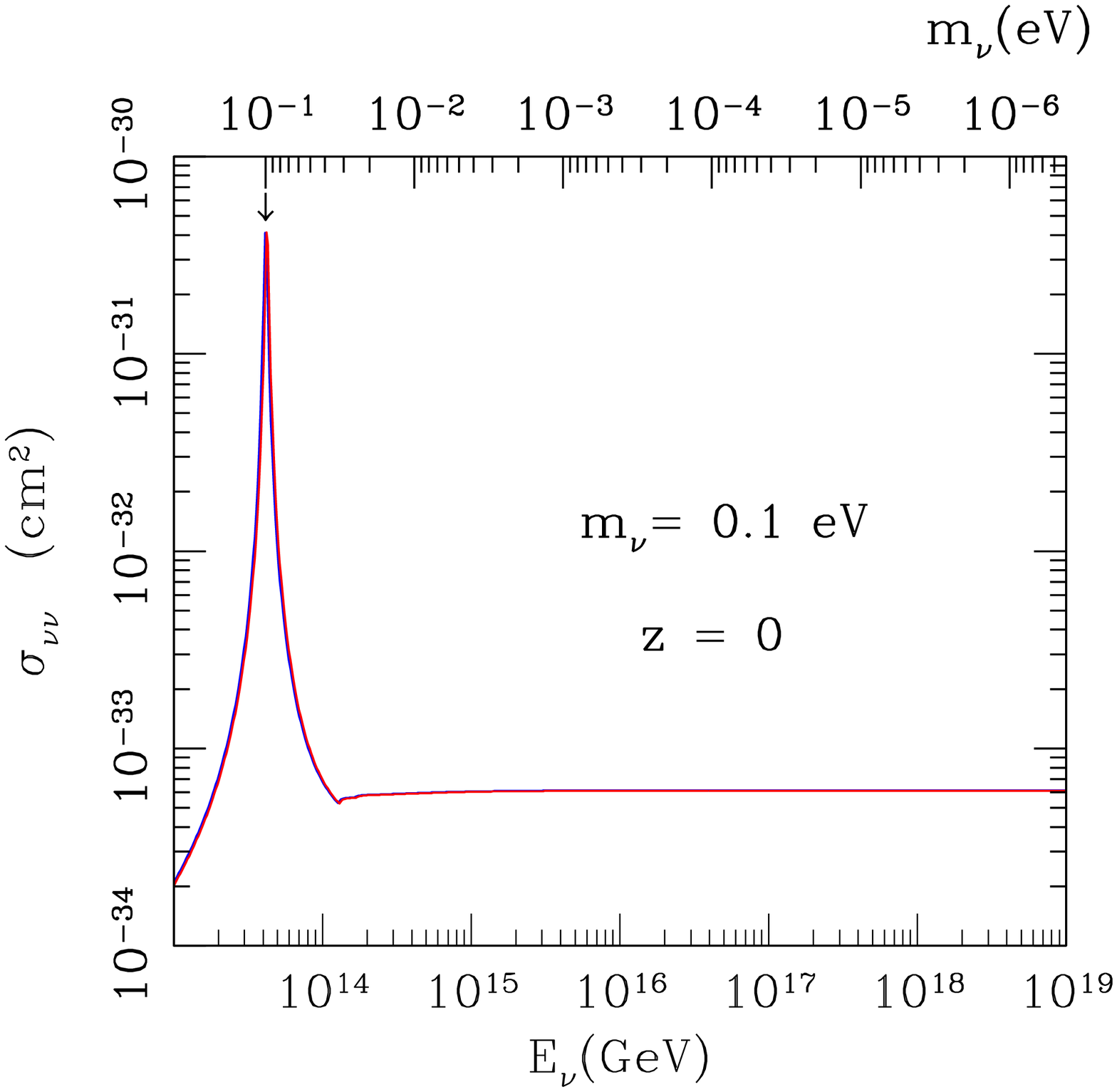} \\
\includegraphics[width=4.5cm]{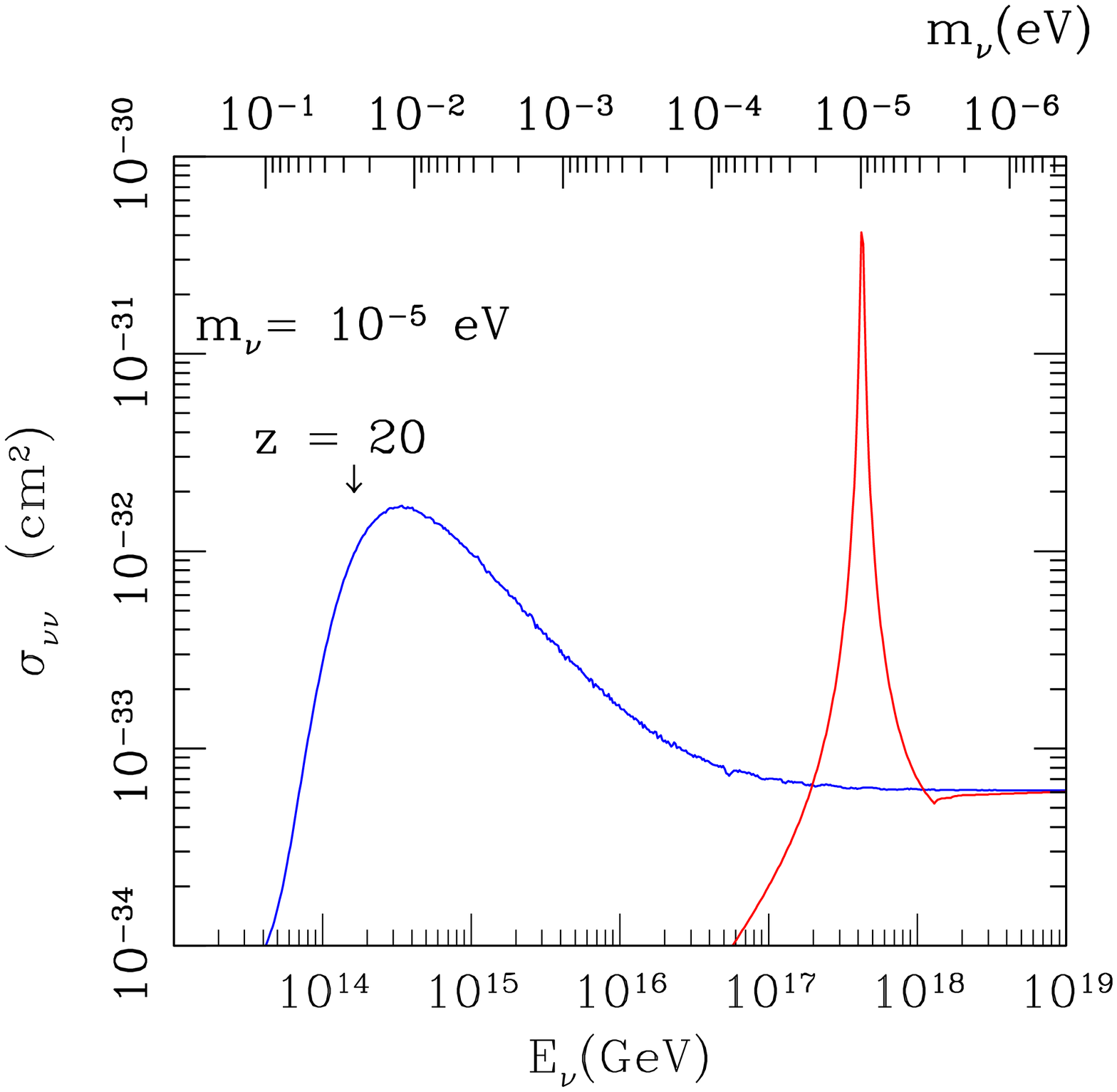}%
\includegraphics[width=4.5cm]{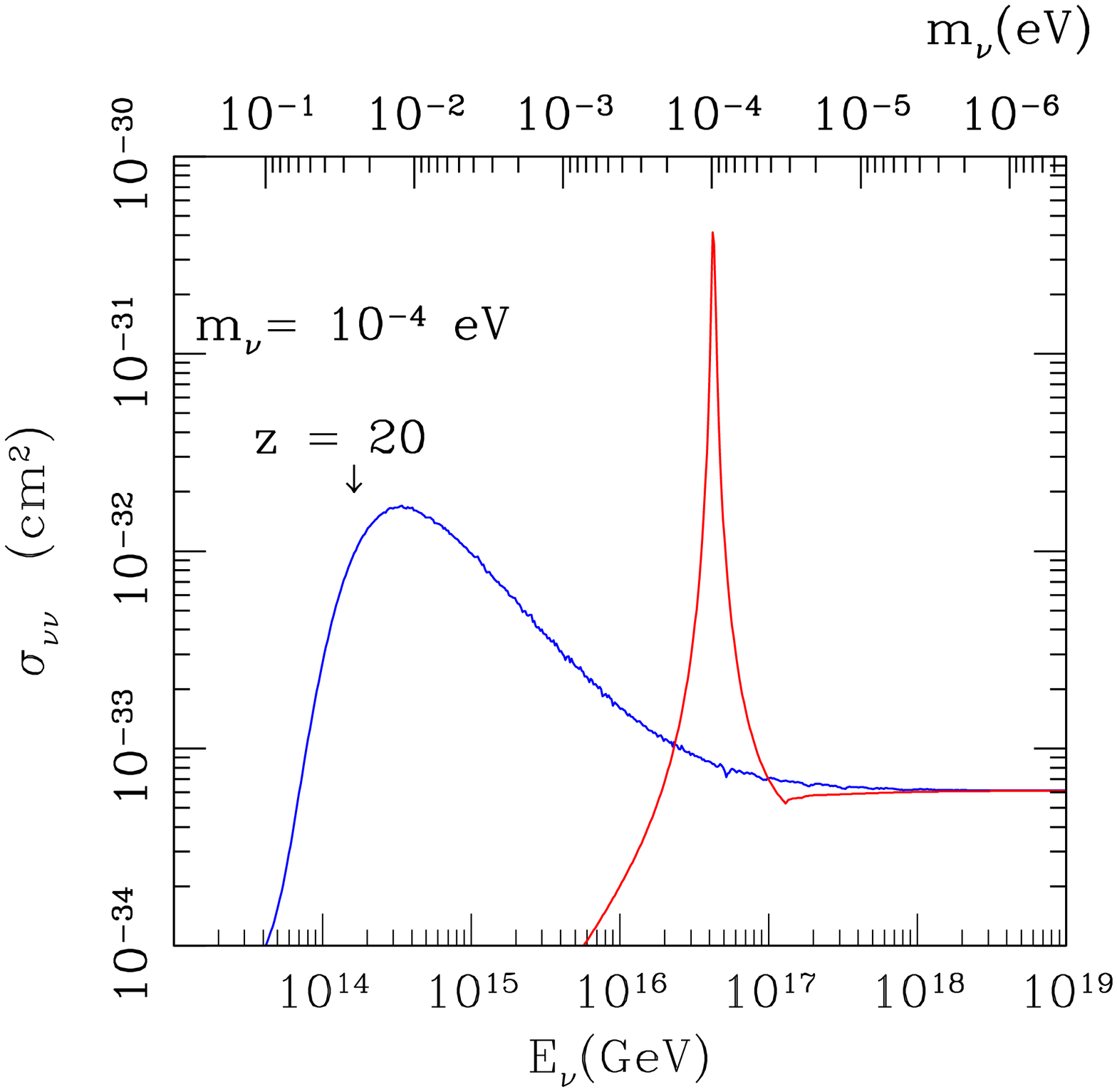}%
\includegraphics[width=4.5cm]{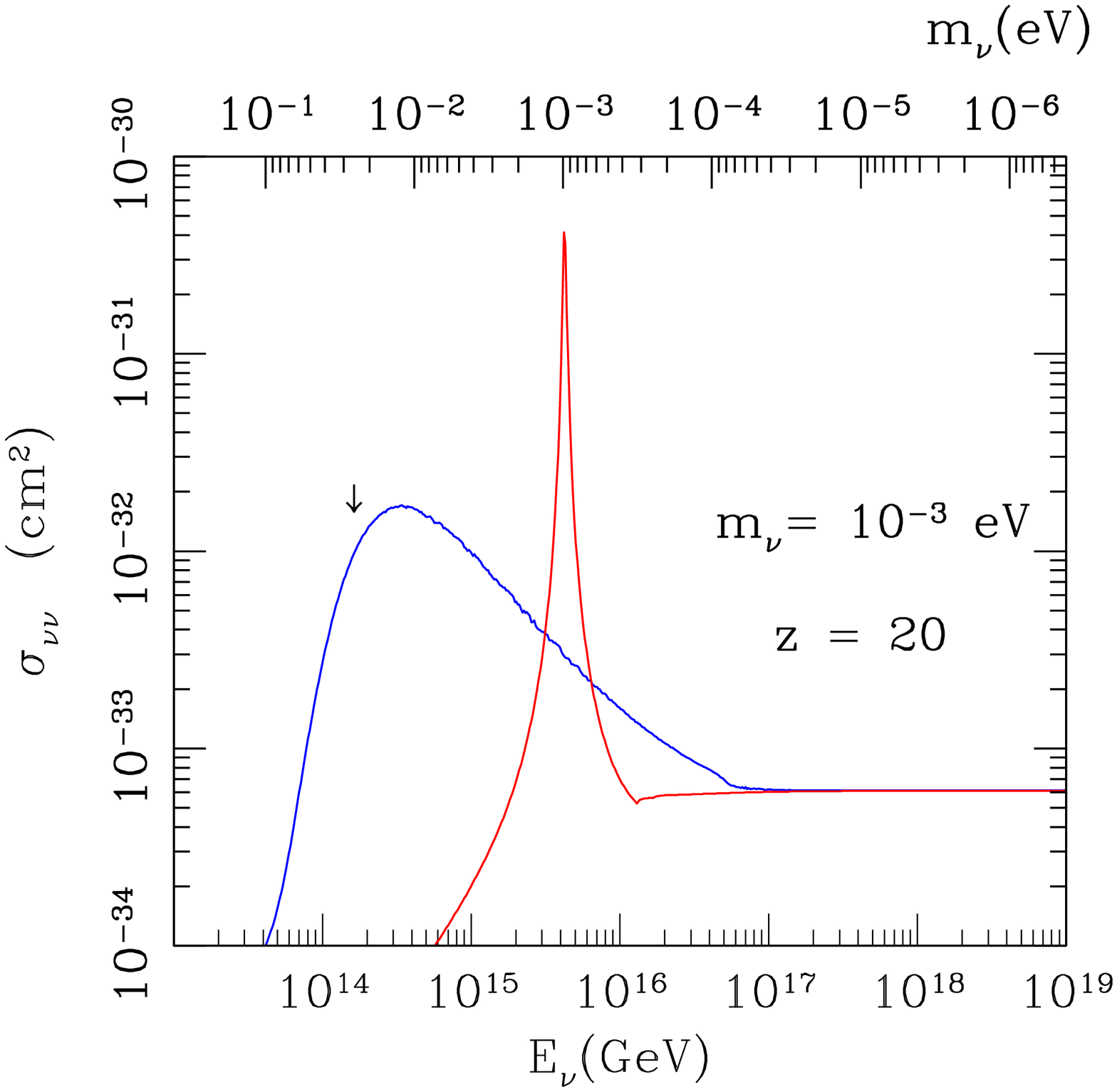}%
\includegraphics[width=4.5cm]{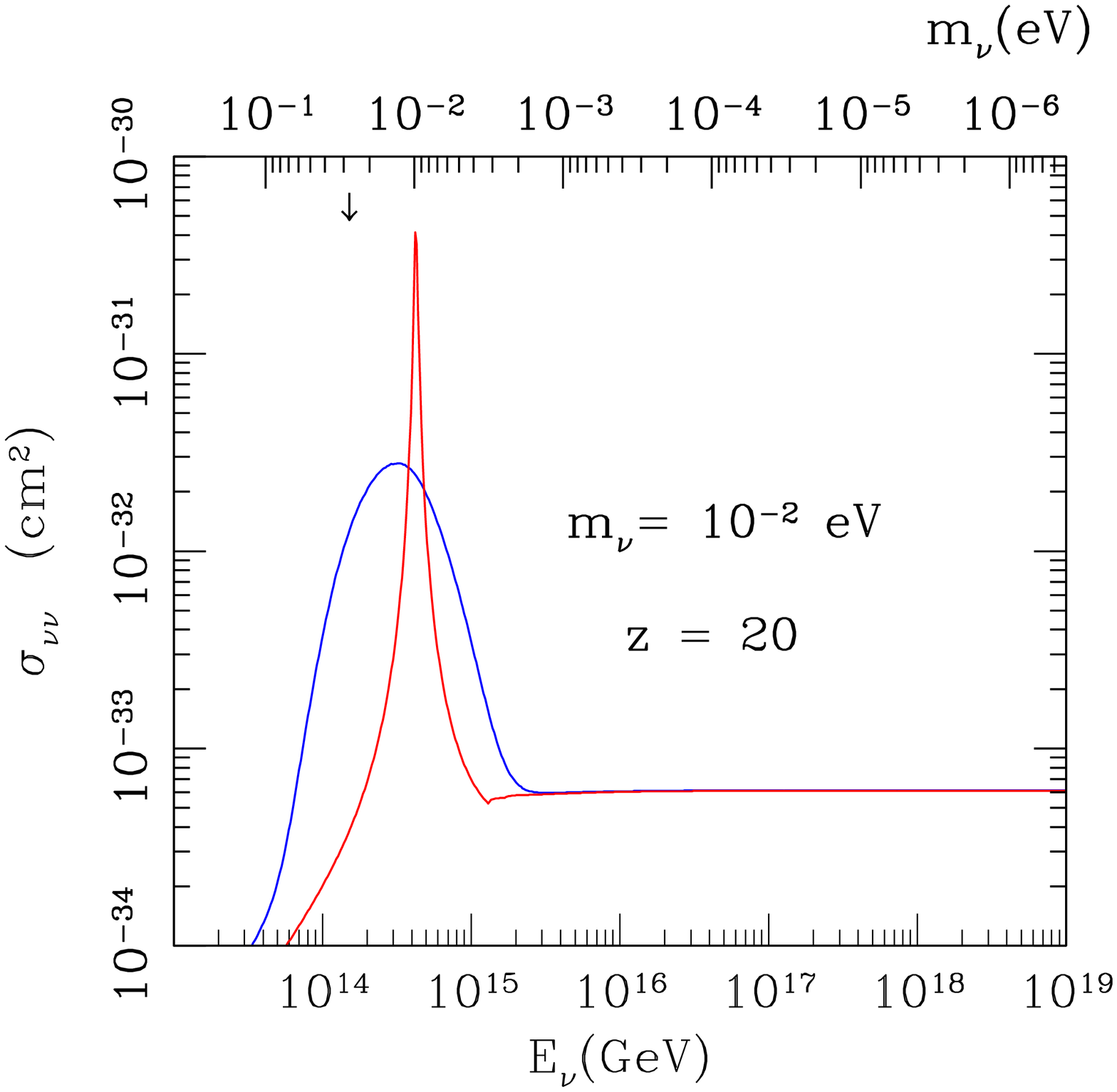}%
\includegraphics[width=4.5cm]{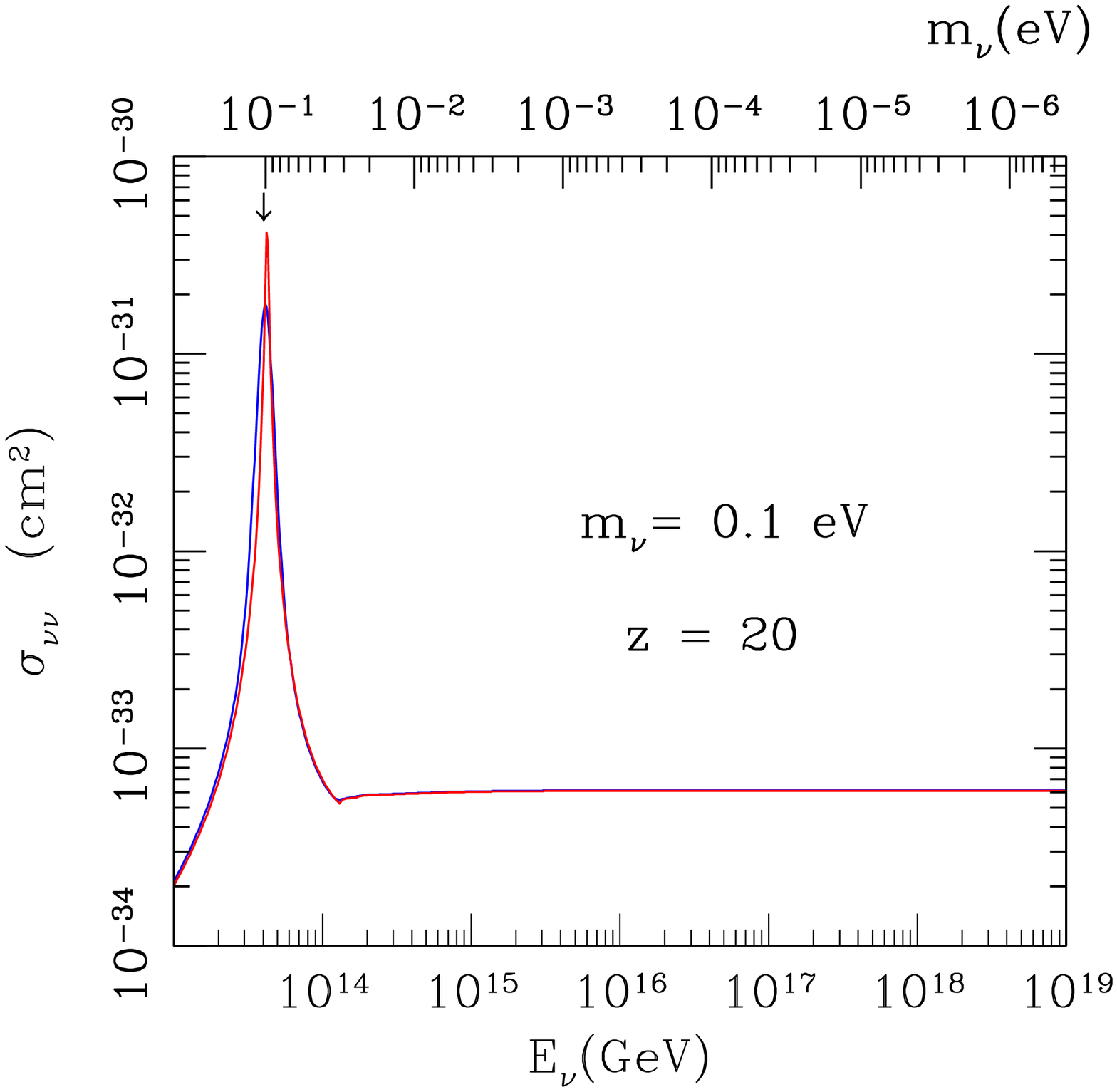}%                                             
\caption{Neutrino-antineutrino (annihilation) cross section as a 
function of the incident neutrino energy. In each panel, the narrow 
peak (red curve) applies for a relic neutrino at rest. The broader 
(blue) curves, shifted to lower energies result when the Fermi motion 
of the thermal distribution of relic neutrinos is taken into account. 
The upper series corresponds to relics in the present Universe ($z = 
0$); the lower series corresponds to redshift $z = 20$. From left to 
right, the panels depict relic neutrino masses $10^{-5}, 10^{-4}, 
10^{-3}, 10^{-2}$, and $10^{-1}\ev$. Arrows designate the thermally 
displaced resonant energy $\widetilde{E}_{\nu}^{Z \mathrm{res}}$ 
given by Eq.~(\ref{eq:smresavg}). \label{fig:fermimo}}
\end{figure*}
\end{turnpage}

We display the effect of Fermi motion on the $Z^{0}$-formation cross 
section in Figure~\ref{fig:fermimo}. 
The annihilation cross section depends on the relic neutrino energy and 
momentum. We have integrated  the exact expression for the cross section weighted  with 
the  Fermi--Dirac momentum distribution of the relic neutrinos over the relic neutrino momentum phase 
space.  We show two series corresponding to redshifts $z = 0$ and $20$ and
relic neutrino masses $10^{-5}, 10^{-4}, 10^{-3}, 10^{-2}$, and
$10^{-1}\ev$. In each panel, the narrow peak (red curve) shows the 
annihilation cross section as a function of incident neutrino energy 
for a stationary relic-neutrino target; the broad peak (blue curve) 
shows the annihilation cross section averaged over the thermal 
distribution of relic-neutrino momenta. In every case, the thermally 
averaged cross section peaks near $\widetilde{E}_{\nu}^{Z 
\mathrm{res}}$, indicated by a downward arrow. Consequently, the 
neutrino mass inferred from the location of an absorption line in a 
hypothetical experiment at fixed redshift would 
be approximately $\langle\varepsilon_{\nu}(z)\rangle$, rather than the 
true neutrino mass, whenever $\langle\varepsilon_{\nu}(z)\rangle \gtrsim m_{\nu}$.
%%%%%%%%%%%%%%%%%%%%%%%%%%%%%%%%%%%%%%%%%%%%%%%%%%%%%%%%%%%%%%%%%%%%%%%%%%%%%%%%%

\begin{figure*}
     \includegraphics[width=8.75cm]{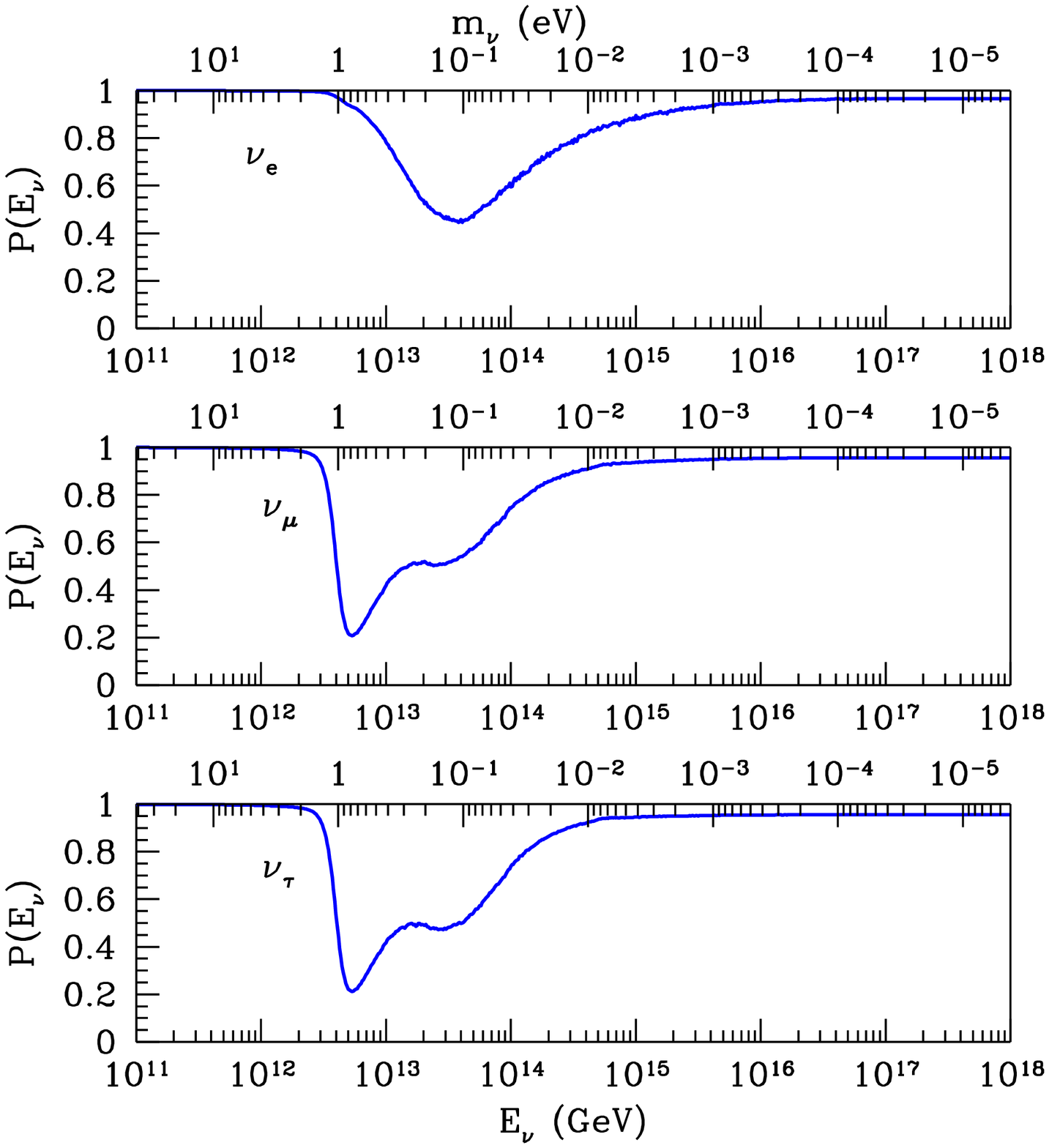}%{figs/lightweighted.eps}%                                             
     \includegraphics[width=8.75cm]{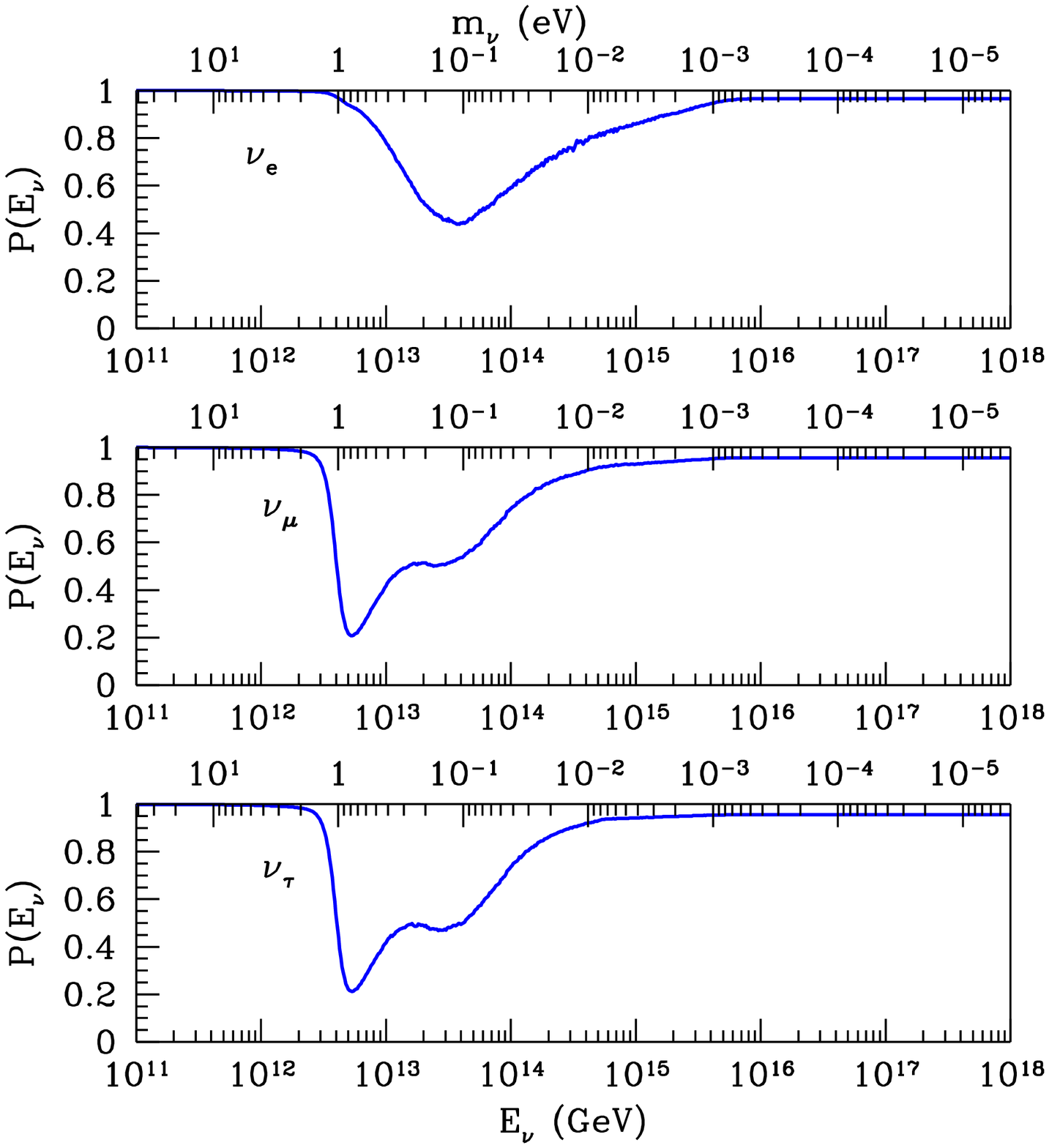}%{figs/mediumweighted.eps}%                                             
 \caption{Survival probabilities for $\nu_e$, $\nu_\mu$, and 
 $\nu_\tau$  as a function of the neutrino energy,
 after integration back to redshift $z = 20$, taking into account the 
 Fermi smearing induced by the thermal motion of the relic neutrinos.
The results apply for a normal hierarchy with lightest neutrino mass
 $m_{\ell}=10^{-5}\ev$ (left panel) or $m_{\ell}=10^{-3}\ev$ (right panel).}
 \label{fig:5zcnormal}
 \end{figure*}

Now we consider the influence of thermal motion on the structure of 
neutrino absorption lines in an evolving Universe, integrating over 
redshift as we did in \S\ref{subsec:expred}. We show in 
Figure~\ref{fig:5zcnormal} the survival probabilities for 
$\nu_{e},\nu_{\mu},\nu_{\tau}$ integrated from the present back to 
redshift $z = 20$, for $m_{\ell} = 10^{-5}\hbox{ and }10^{-3}\ev$. 
Comparing with the analogous calculation neglecting the Fermi motion 
of the relic neutrinos, whose outcome was depicted in 
Figure~\ref{fig:5znormal}, we find that the three distinct dips have 
been merged into a single complex dip. As we could anticipate from the 
displacement of the resonant energies displayed in Figure~\ref{fig:fermimo}, 
the prospect of determining distinct absolute neutrino masses has 
faded appreciably.

\begin{figure}
     \includegraphics[width=8.75cm]{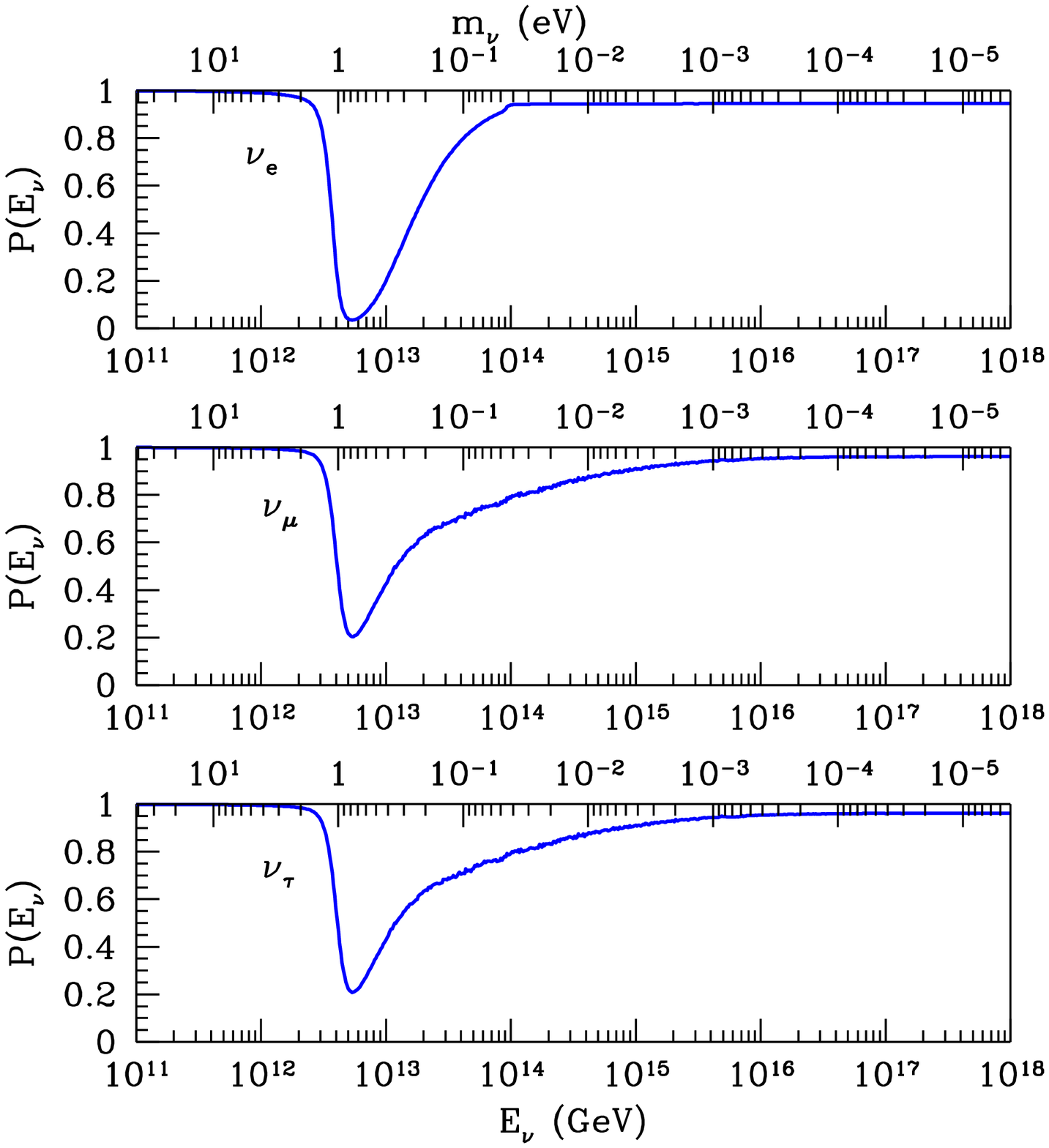}%{figs/lightweightedi.eps}%                                             
 \caption{Survival probabilities for $\nu_e$, $\nu_\mu$, and 
 $\nu_\tau$  as a function of the neutrino energy,
 after integration back to redshift $z = 20$, taking into account the 
 Fermi smearing induced by the thermal motion of the relic neutrinos.
The results apply for an inverted hierarchy with lightest neutrino mass
 $m_{\ell}=10^{-5}\ev$.}
 \label{fig:5zcinv}
 \end{figure}

\begin{figure*}
   \includegraphics[width=9.125cm]{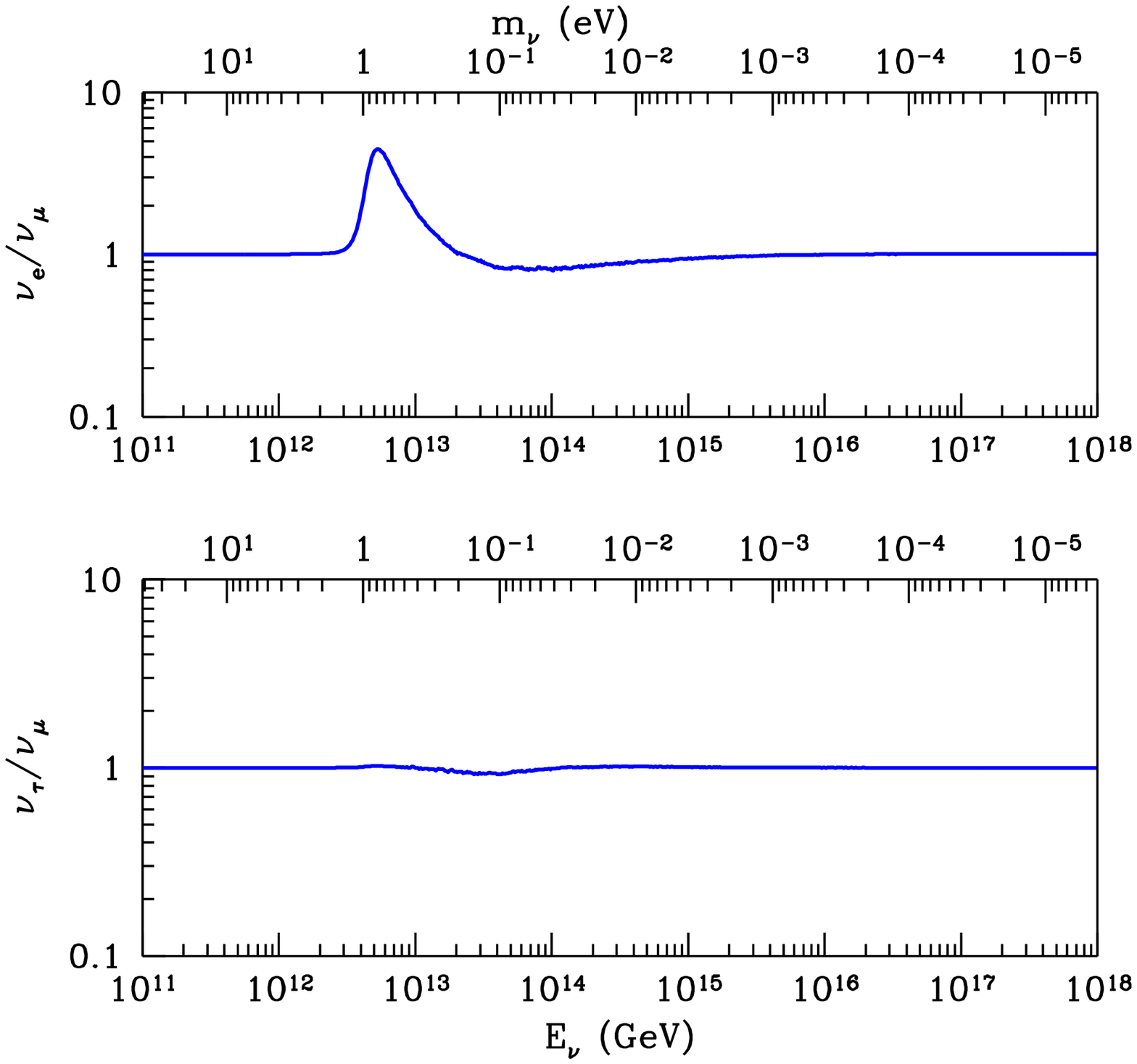}%{figs/ratios_test_weighted.eps}  
   \includegraphics[width=9.125cm]{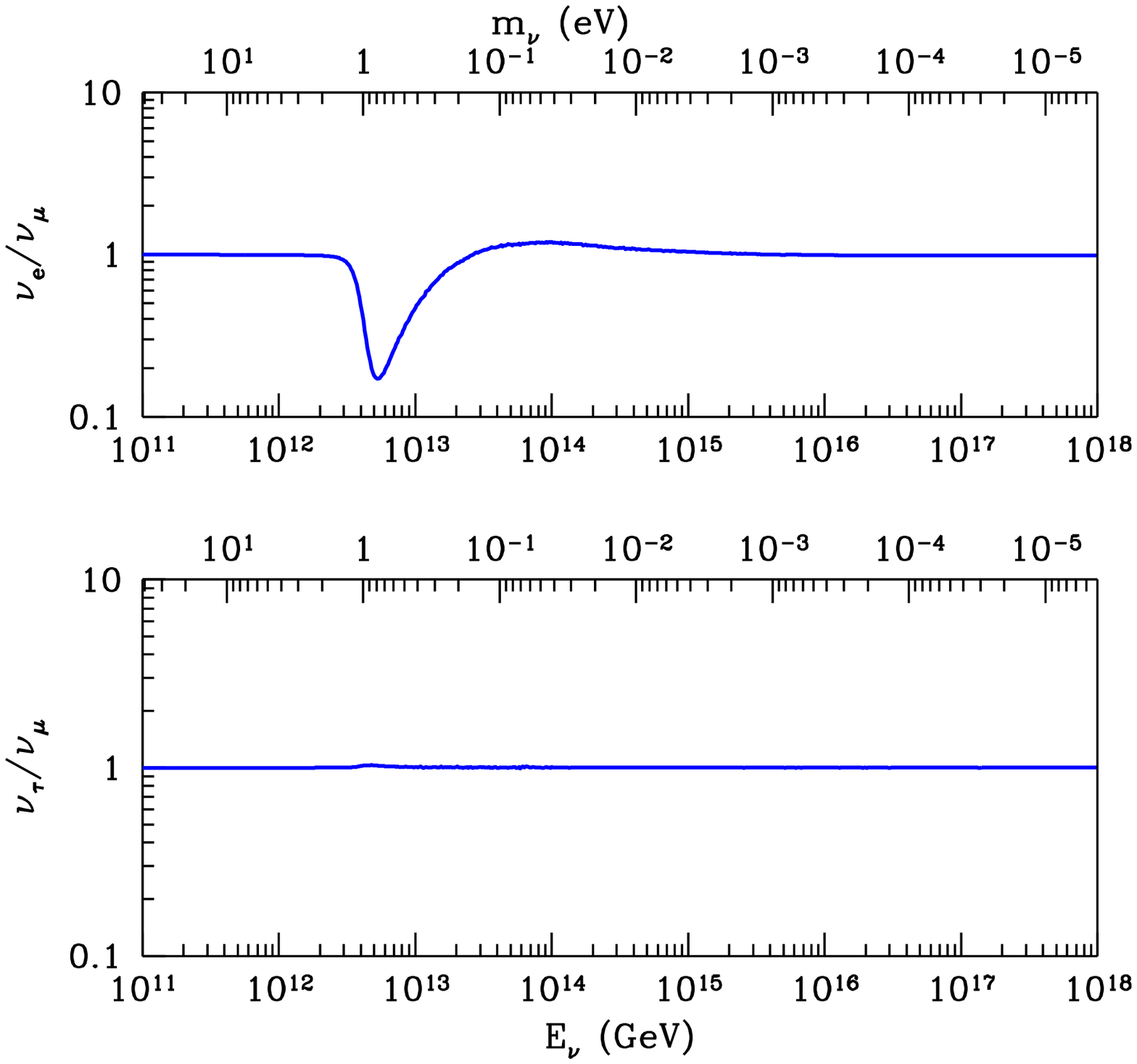}%{figs/ratios_testi_weighted.eps}  %ratios_itestzn.eps 
   \vspace*{-24pt}
    \caption{Flux ratios $\nu_{e}/\nu_{\mu}$ and 
    $\nu_{\tau}/\nu_{\mu}$ at Earth, for normal (left panel) and 
    inverted (right panel) mass hierarchies with $m_{\ell} = 
    10^{-5}\ev$, after integration back to redshift $z=20$ and a 
    thermal averaging over the relic-neutrino momentum distribution.
    The scale at the top shows the neutrino mass 
    defined as $m_{\nu} = M_{Z}^{2}/2E_{\nu}$ that would be inferred if 
    neutrino energies were not redshifted.     \label{fig:ratsztherm20}}
\end{figure*}

The distinction between the normal and inverted hierarchy remains, 
however.  We plot in Figure~\ref{fig:5zcinv}
the survival probabilities in the case of an inverted hierarchy with 
$m_{\ell} = 10^{-5}\ev$. Again, Fermi motion has merged the two 
distinct absorption lines (compare Figure~\ref{fig:5zinvert})
into one. The pattern of $\nu_{e}, \nu_{\mu}$ 
attenuation is different from the normal-hierarchy case, as we show 
in Figure~\ref{fig:ratsztherm20}.
In the normal-hierarchy case, $\nu_{e}/\nu_{\mu} > 1$ for the 
prominent dip at the lowest energy, whereas for the inverted 
hierarchy $\nu_{e}/\nu_{\mu} < 1$.

\begin{figure*}
     \includegraphics[width=9.125cm]{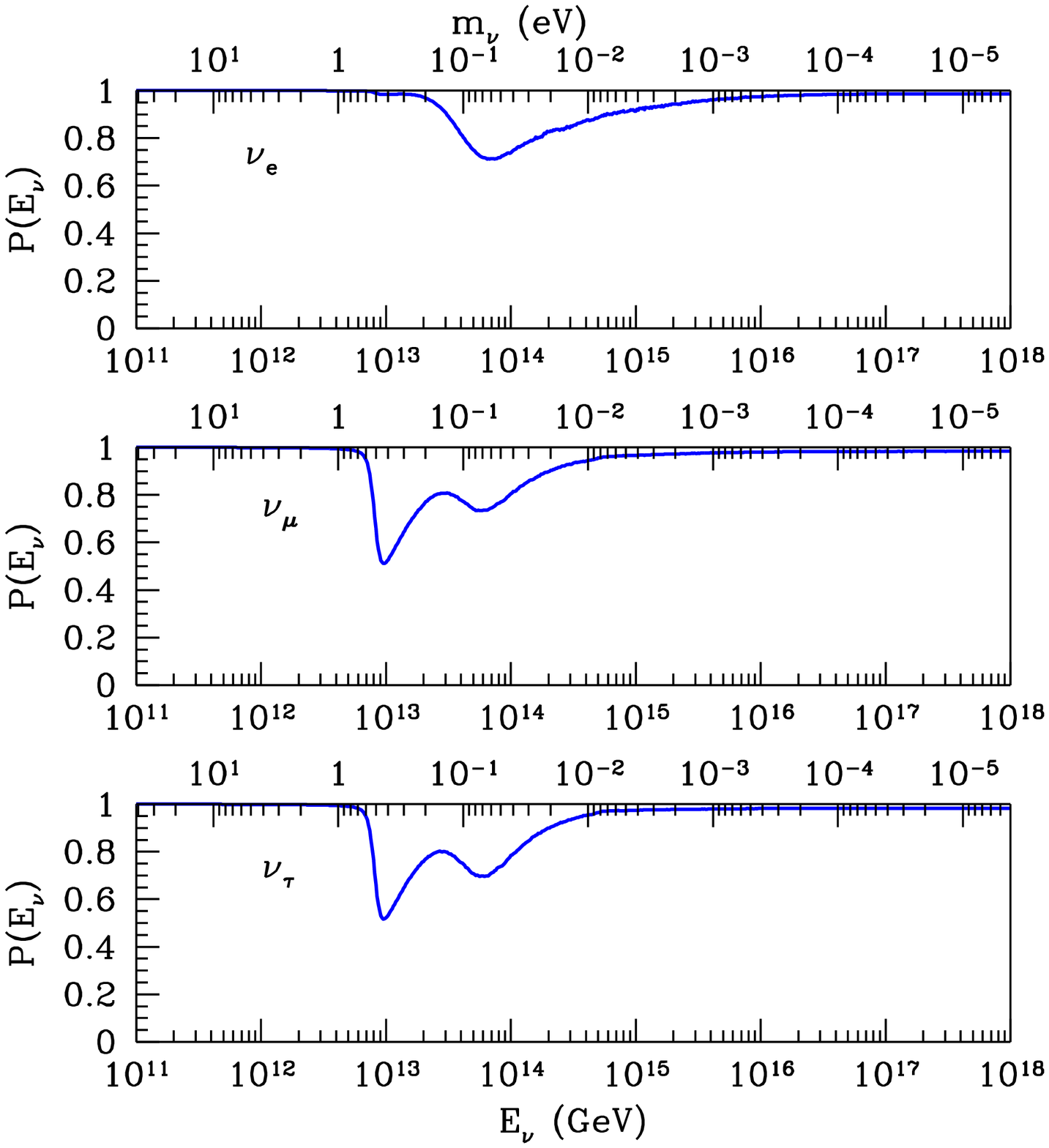}%{figs/lightweighted_10.eps}%                                             
     \includegraphics[width=9.125cm]{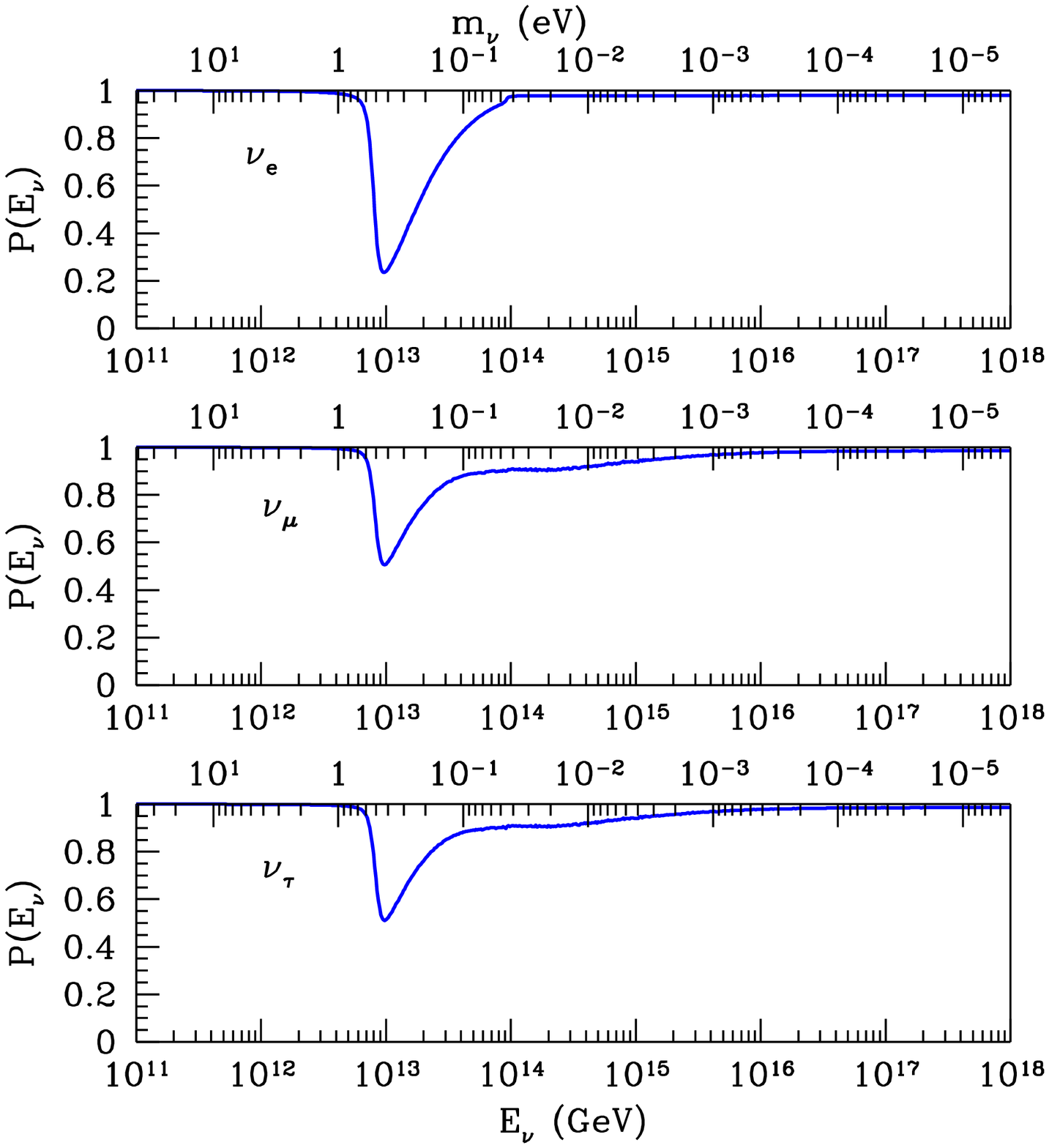}%{figs/lightweighted_10i.eps}%                                             
\vspace*{-12pt}
\caption{Survival probabilities for $\nu_e$, $\nu_\mu$, and 
 $\nu_\tau$  as a function of the neutrino energy,
 after integration back to redshift $z = 10$, taking into account the 
 Fermi smearing induced by the thermal motion of the relic neutrinos.
The results apply for a normal hierarchy (left panel) or inverted 
hierarchy (right panel), with lightest neutrino mass
 $m_{\ell}=10^{-5}\ev$.}
 \label{fig:5z10}
 \end{figure*}

 Although the thermal motion of the relics is largely responsible for 
obliterating the three-dip structure, we might inquire whether some 
of the structure would survive if the sources of ultrahigh-energy 
neutrinos came into existence more recently than redshift $z = 20$. 
In Figure~\ref{fig:5z10} we show the survival probabilities for normal 
and inverted hierarchies, with the lightest neutrino mass $m_{\ell} = 
10^{-5}\ev$, integrating from the present back to redshift $z = 10$.
The multiple dip structure is somewhat more pronounced than in the 
$z = 20$ scenario, but still indistinct. Because the Universe is more 
transparent to neutrinos when evolved back through $z = 10$ than 
evolved back through $z = 20$, the dips are less prominent than in 
Figures~\ref{fig:5zcnormal} and \ref{fig:5zcinv}. Flux ratios again 
distinguish the normal and inverted hierarchies.

\section{Unconventional Neutrino Histories \label{sec:loose}}
\subsection{Lepton asymmetry in the early Universe}
If the number of leptons differed from the number of antileptons in the
early universe, the present density of relic neutrinos might be very
different from the expected value of $56\cm^{-3}$.  In the standard
big-bang model, the neutrino asymmetry $\eta_{\nu_{\alpha}} \equiv
(n_{\nu_{\alpha}} - n_{\bar{\nu}_{\alpha}})/n_{\gamma}$, is assumed to
be negligibly small, comparable to the tiny-but-crucial baryon asymetry
$\eta_{B} \equiv (n_{B} - n_{\bar{B}})/n_{\gamma} = (6.1 \pm 0.2)
\times 10^{-10}$~\cite{Eidelman:2004wy,Spergel:2003cb}.\footnote{The
apparent electrical neutrality of the universe implies a similarly
small electron-positron asymmetry.} Such tracking is typical of
scenarios in which $B-L$ is a conserved quantum number and leptogenesis
is the forerunner of baryogenesis.  However, several models are known
in which a small $\eta_{B}$ is accompanied by a large
$\eta_{\nu}$~\cite{Affleck:1984fy,Casas:1997gx}. Consequently, it is 
worth our while to examine the observational constraints on the 
neutrino asymmetry.

A significant lepton asymmetry in the early universe, represented by a
nonvanishing chemical potential $\xi_{\alpha}$, would increase the
expansion rate of the universe, increasing the amplitude of the
acoustic peaks observed in the cosmic microwave background anisotropy.
A combined analysis of CMB and SN Ia data~\cite{Hansen:2001hi} finds
$-0.01\le \xi_e \le 0.22$ and $|\xi_{\mu,\tau}|\le 2.6$ at $95\%$ CL.
(The $\nu_{e}$ asymmetry is more tightly constrained because it
influences the neutron-to-proton ratio at the epoch of big-bang
nucleosynthesis, and hence the helium fraction.)  If neutrino mixing
equilibrates all the chemical potentials before BBN, then the bound for
$\xi_e$ applies to all species~\cite{Lunardini:2000fy}, and
$|\xi_{\nu{\alpha}}|<0.1$~\cite{Dolgov:2002ab,Abazajian:2002qx},
whereupon in the current universe $n_{\nu_{\alpha}} \approx
56\cm^{-3}$.  However, if a majoron field should block neutrino
oscillations in the primordial
plasma~\cite{Babu:1991at,Bento:2001xi,Dolgov:2004jw}, then $\xi_{\mu}$
and $\xi_{\tau}$ do not equilibrate with $\xi_e$.  If we take the 
largest allowed values, $|\xi_{\mu,\tau}| \simeq 3 $, then a large lepton
asymmetry appears, $n_{\nu_{\alpha}}-n_{\bar{\nu}_\alpha}\simeq
n_{\nu_{\alpha}}\simeq 1050\cm^{-3}$~\cite{gelmini2}. The increased 
neutrino density would reduce the interaction length for annihilation 
on the favored relic neutrino species.

\subsection{Neutrinoless Universe}
It is conceivable that neutrinos may interact in ways not foreseen in 
the standard electroweak theory. If, for example, the neutrinos couple 
to an extremely light boson, relic neutrinos might annihilate at late 
times, greatly reducing---indeed, likely erasing---the neutrino 
density expected in the standard cosmology with standard-model 
interactions~\cite{Beacom:2004yd}. If no neutrino relics remain in the 
recent ($z \lesssim 20$) Universe, then no neutrino absorption lines 
will be observed. To demonstrate that the search for absorption lines 
has covered the right terrain, it will be extremely helpful to 
establish the absolute scale of neutrino masses.

\subsection{Neutrino Clustering}
One early inspiration for the study of ultrahigh-energy neutrino 
absorption was the prospect that massive neutrinos, with masses in 
the range $30$--$100\ev$, might account for the dark matter in the 
Universe~\cite{Roulet:1992pz}. Had that been the case, neutrinos might have 
been absorbed not only on relics at large redshift, but also on 
relics clustered in galactic halos. For neutrino masses in the range 
we now consider likely, $m_{\nu} \lesssim 0.1\ev$, gravitational 
clustering of neutrinos appears to be insignificant~\cite{Singh:2002de}. 
Moreover, since light neutrinos would cluster---or accrete onto other 
structures---only at late times, absorption spectroscopy is unlikely 
to record neutrino density contrasts~\cite{Ringwald:2004np}. The 
question of detecting neutrino clumps by other techniques remains 
open.  

\section{Summation and Outlook \label{sec:summa}}
Let us assess the prospects for developing cosmic-neutrino absorption
spectroscopy as a new window on Nature.  First, we require that
neutrino observatories establish the existence of cosmic-neutrino
fluxes that extend to energies in the Greisen--Zatsepin--Kuzmin regime
and beyond.  If the requisite neutrino fluxes exist, then the essential
task will be to observe absorption lines at energies corresponding to
the masses of one or more neutrino species.  Making those observations
will require detectors with vast effective volumes, or long exposure
times, or both.  Detecting the relic-neutrino background would, by
itself, be an enormously satisfying observational accomplishment.  A
survey of other techniques that might detect the relics is given in
Ref.~\cite{gelmini2}. Short of direct detection, it may be possible 
to quantify the influence of primordial anisotropies in the relic 
neutrinos upon the cosmic microwave 
background~\cite{Hu:1995fq,Trotta:2004ty}.

The observation of cosmic-neutrino absorption lines will open the
way---at least in principle---to new insights about neutrino properties
and the thermal history of the universe.  Our calculations, with their
successive inclusion of potentially significant effects, show that how
the tale unfolds will depend on factors we cannot foresee.  The earlier
in redshift the relevant cosmic-neutrino sources appear, the lower the
present-day energy of the absorption lines and the denser the column of
relics the super-high-energy neutrinos must traverse.  In particular,
the appearance of dips at energies much lower that we expect points to
early---presumably nonacceleration---sources, that could give us insight
into fundamental physics at early times and high energy scales.  On the
other hand, integration over a longer range in redshift means more
smearing and distortion of the absorption lines.

If the lightest neutrino mass is small, then Fermi smearing due to 
the thermal motion of the relics sets an effective lower bound on the 
neutrino mass, as reflected in the absorption lines. That moves the 
absorption lines to lower, more accessible, energies, but reduces the 
power of absorption-line spectroscopy to distinguish the neutrino mass 
eigenstates. Once the age of the cosmic-neutrino sources is clear, 
analyzing the thermal smearing may provide at least a rough 
measurement of the current temperature of the relic neutrinos.
In the best (imaginable) circumstance, discriminating the 
interactions of electron, muon, and tau neutrinos might enable future 
experiments to determine, or verify, the flavor content of the mass 
eigenstates. 

It appears relatively secure to conclude that future observations of 
the $\nu_{e}/\nu_{\mu}$ ratio at the first absorption dip can decide 
whether the neutrino mass hierarchy is normal or inverted. 
Accelerator-based long-baseline neutrino-oscillation experiments 
should be the first to measure the sign of the ``atmospheric'' mass-squared 
difference.  This is not necessarily a losing proposition for 
cosmic-neutrino absorption spectroscopy: the more 
information---neutrino masses, flavor composition, etc.---other 
experiments provide to neutrino observatories, the more perceptive 
cosmic-neutrino absorption spectroscopy might become about the 
thermal history of the universe.

The experiments we describe in this study will not be done very soon,
and their interpretation is likely to require many waves of observation
and analysis.  Nevertheless, they offer the possibility to establish
the existence of another relic from the big bang and, conceivably, they
may open a window on periods of the thermal history of the universe not
readily accessible by other means.

\begin{acknowledgments}
Fermilab is operated by Universities Research Association Inc.\ under
Contract No.\ DE-AC02-76CH03000 with the U.S.\ Department of Energy.
One of us (C.Q.) is grateful for the hospitality of the Kavli Institute
for Theoretical Physics during the program, \textit{Neutrinos: Data,
Cosmos, and Planck Scale.} This research was supported in part by the
National Science Foundation under Grant No.\ PHY99-07949. G.B. and O.M. 
acknowledge the stimulating environment of the Aspen Center for 
Physics. G.B. thanks the Fermilab Theory Group for the support of its 
summer visitors program.
We thank Nicole Bell, Andreas Ringwald, and Tom Weiler for informative 
discussions. Our calculations made extensive use of the Fermilab 
General-Purpose Computing Farms~\cite{Albert:2003vv}. 
\end{acknowledgments}


\end{document}